\documentclass[aps,pre,onecolumn,longbibliography,nofootinbib,showkeys,showpacs,amssymb,floatfix]{revtex4-1}

\usepackage{latexsym}
\usepackage{amsfonts}
\usepackage{color}
\usepackage{graphicx}
\usepackage{mathptmx}      
\usepackage{bm}
\usepackage{enumerate}
\usepackage{subfigure}
\usepackage{listings}
\usepackage{grffile}

\newcommand\sumprime{\mathop{{\sum}'}}

\begin{document}
\title{Relaxation, thermalization and Markovian dynamics of two spins coupled to a spin bath}

\author{H. De Raedt}
\email{h.a.de.raedt@rug.nl}
\thanks{Corresponding author}
\affiliation{Zernike Institute for Advanced Materials,\\
University of Groningen, Nijenborgh 4, NL-9747AG Groningen, The Netherlands}
\author{F. Jin}
\affiliation{Institute for Advanced Simulation, J\"ulich Supercomputing Centre,\\
Forschungszentrum J\"ulich, D-52425 J\"ulich, Germany}
\author{M.I. Katsnelson}
\affiliation{Institute for Molecules and Materials, Radboud University, \\
Heyendaalseweg 135, NL-6525AJ Nijmegen, The Netherlands}
\author{K. Michielsen}
\affiliation{Institute for Advanced Simulation, J\"ulich Supercomputing Centre,\\
Forschungszentrum J\"ulich, D-52425 J\"ulich, Germany}
\affiliation{RWTH Aachen University, D-52056 Aachen, Germany}

\date{\today}

\begin{abstract}
It is shown that by fitting a Markovian quantum master equation
to the numerical solution of the time-dependent Schr\"odinger equation
of a system of two spin-1/2 particles interacting with a bath of up to 34 spin-1/2 particles,
the former can describe the dynamics of the two-spin system rather well.
The fitting procedure that yields this Markovian quantum master equation
accounts for all non-Markovian effects in as much the general structure of this equation allows
and yields a description that is incompatible with the Lindblad equation.
\end{abstract}


\maketitle

\section{Introduction}\label{section1}

In laboratory experiments, a physical system of interest can never be considered as being
completely isolated from its environment.
Therefore, in their theoretical description, the quantum system of interest (henceforth called system)
should be considered as an open quantum system, that is a system interacting with its environment.
As most open quantum systems are way too complicated to be treated without making approximations,
the standard procedure in theoretical treatments of open quantum systems
is to derive closed approximate equations of motion of the system operators, a quantum master equation (QMEQ)  from
the underlying time-dependent Schr\"odinger equation (TDSE)
by eliminating the environmental degrees of freedom~\cite{REDF57,NAKA58,ZWAN60,BREU02}.
Generically, such derivations involve the so-called Markov approximation,
which is based on the assumption that the correlations of the bath degrees of freedom vanish on a short time span, short
compared to the characteristic time scale of the system dynamics.
When the time scale of the system is comparable to that of the decay of the bath correlations
the Markovian approximation may no longer be adequate~\cite{SASS90,WEIS99,BREU02,TANI06,BREU06,MORI08,SAEK08,UCHI09,MORI14,CHEN15}.

Alternatively, without reference to any particular model,
one may postulate a Markovian QMEQ for the density matrix
which preserves positivity during the time evolution (i.e. a non-negative definite density matrix at all times),
as Lindblad did~\cite{LIND76,BREU02}.
In this approach, the key question is then how to extract the parameters that enter
the Lindblad QMEQ from the microscopic model of interest.
In this paper, we adopt a similar strategy and use a least-square minimization procedure to
extract the parameters of a Markovian QMEQ from data obtained by numerical solution of the TDSE of the system + bath.
As shown later in this paper, this Markovian QMEQ is not of the Lindblad form.

In the mathematically strict sense, the unitary Schr\"odinger dynamics of the system + bath is incompatible with
the statement that one or more system operators exhibit exponential decay, the signature of Markovian behavior~\cite{FOND78}.
Therefore, even though the system + bath satisfies all the requirements for justifying
a Markovian QMEQ description, when looked at in detail, the numerical solution of the TDSE of the system + bath
may still reveal non-Markovian behavior (different from Poincar\'e cycles which, for the quantum systems
of interest, have astronomically large time scales).
Indeed, such features are observed when solving the TDSE of spin-1/2 models~\cite{JIN10x,DONK17},
see also later in this paper.
Therefore, the central issue is not whether the dynamics of the system is described by a
Markovian QMEQ because in a strict sense is not, but rather to what extent
the Markovian QMEQ provides an accurate description of the system dynamics.

An earlier paper~\cite{ZHAO16} addressed the question to what extent a QMEQ
captures the salient features of the exact Schr\"odinger equation dynamics
of a single spin coupled to a bath of spins.
This question was answered by solving the TDSE of the whole system and subsequently fitting
the data of the expectation values of the spin components to those of a Markovian QMEQ.
The main finding of that paper was that in all cases in which the approximations
used to derive a Markovian QMEQ seem justified~\cite{BREU02},
the Markovian QMEQ obtained by least-square fitting to the data obtained by solving the TDSE of the whole system
describes the dynamics of the single spin in contact with the spin bath rather well.
In this case, the mathematical structure of the Markovian QMEQ is the same as that of the Bloch equation~\cite{BLOC46}
and as a phenomenological description, the Markovian QMEQ offers no advantages over the latter.
Of course, when the system contains more than one spin, the Bloch equation can no longer be used
whereas a Markovian QMEQ still has the potential to describe the system dynamics.

The main aim of this paper is to present a quantitative assessment of the Markovian QMEQ description
in the case where the system consists of two spins instead of one and a Bloch-type description can no longer be used.
Such system-bath spin models are relevant for the description of relaxation processes
in nuclear magnetic and electron spin resonance~\cite{KUBO57,REDF57,ABRA61,SLIC90,ABRA70}
and have applications to quantum information processing~\cite{NIEL10,JOHN11}.
By using these resonance techniques one can probe the dynamics of an individual spin
but the two-spin dynamics is not directly accessible.
However, with the advent of small quantum information processors such as the
IBM Quantum Experience~\cite{IBMQE}, a cloud-based platform for gate-based quantum computing,
it may be possible to study the two-spin system dynamics in detail.

A second aim of this work is to use the two-spin system coupled to a heat bath
as an instance to test one of the underlying assumptions of statistical mechanics, namely
the assumption that a system interacting with a thermostat approaches thermal equilibrium.
To this end, we study in detail how the two-spin system relaxes to a stationary state and
scrutinize the conditions under which this stationary state approaches its thermal equilibrium state.
Here and in the following, we use the term ``the system thermalizes'' if and only if there is evidence that the
density matrix of the system relaxes to the thermal equilibrium state.
In other words, it is not sufficient to show that the system energy relaxes to its thermal equilibrium value:
all the expectation values of a complete set of system operators should relax to their respective thermal equilibrium values.

The paper is organized as follows.
In section~\ref{section2}, we specify the Hamiltonians of the system, bath and system-bath interaction.
Section~\ref{section3} briefly reviews the numerical techniques that we use to solve the
TDSE of the whole system, to compute the reduced density matrix,
and to prepare the bath in the thermal state at a given temperature.
We present simulation results that demonstrate that the method of preparation
yields the correct thermal averages, study the relaxation to the stationary state
and address the effects of the finite size of the bath on the thermalization.
Section~\ref{section4} recapitulates the steps in the numerical procedure
to extract a Markovian QMEQ from the data of the reduced density matrix
obtained from the solution of the TDSE and presents some representative results.
Rewriting the fitted Markovian QMEQ as a dynamical map~\cite{BREU02},
the matrix of coefficients that defines this map can be calculated numerically
and is found to be indefinite instead of non-negative definite,
ruling out that the fitted Markovian QMEQ is of the Lindblad form.
The paper concludes with the summary, given in section~\ref{section5}.


\section{System coupled to a bath: Model}\label{section2}

The Hamiltonian of the system (S) + bath (B) takes the generic form
\begin{eqnarray}
H&=& H_\mathrm{S} + H_\mathrm{B} + \lambda H_\mathrm{SB}
.
\label{s40}
\end{eqnarray}
The overall strength of the system-bath (SB) interaction is controlled by the parameter $\lambda$.
In the present work, we limit ourselves to a system which consists of two spin-1/2 particles described by the two-site
XXZ Hamiltonian
\begin{eqnarray}
H_\mathrm{S}&=&
-J_{\bot}\left(\sigma^x_{1}\sigma^x_{2}+\sigma^y_{1}\sigma^y_{2}\right)-J_{\parallel}\sigma^z_{1}\sigma^z_{2}
,
\label{s41}
\end{eqnarray}
where $\bm{\sigma}_{n}=(\sigma^x_{n},\sigma^y_{n},\sigma^z_{n})$
denote the Pauli-spin matrices for spin-1/2 particle $n$.
Throughout the present paper, we adopt units such that $\hbar=1$, express time in units of $1/4|J_{\parallel}|$,
and to limit the amount of data, we confine ourselves to the case $J_{\bot}=J_{\parallel}=-1/4$,
i.e. the system is described by the isotropic antiferromagnetic Heisenberg model.
For later reference, it is useful to recall here that the ground state of the latter model
is the singlet state defined by
\begin{eqnarray}
|S\rangle&=&\frac{1}{\sqrt{2}}\left( |\uparrow \downarrow\rangle - |\downarrow\uparrow \rangle\right)
,
\label{singletstate0}
\end{eqnarray}%
and that, in the units adopted in this paper,  the ground state energy is $E_\mathrm{singlet}=-3/4$.

We consider two extreme cases for the interaction of the two-spin system with the spin bath.
In the first case, each system spin is connected to one, different bath spin.
The Hamiltonian of the system-bath interaction reads
\begin{eqnarray}
H_\mathrm{SB}&=&
-J^x_{n,1}\sigma^x_{n}\sigma^x_{1}
-J^y_{n,1}\sigma^y_{n}\sigma^y_{1}
-J^z_{n,1}\sigma^z_{n}\sigma^z_{1}
-J^x_{m,2}\sigma^x_{m}\sigma^x_{2}
-J^y_{m,2}\sigma^y_{m}\sigma^y_{2}
-J^z_{m,2}\sigma^z_{m}\sigma^z_{2}
,
\label{s43b}
\end{eqnarray}
where $n$ and $m$ are chosen randomly from the set $\{1,\ldots,N_{\mathrm{B}}\}$ such that $n\not=m$.
Here and in the following $N_{\mathrm{B}}$ denotes the number of bath spins.
The $J^\alpha_{n,1}$ and $J^\alpha_{m,2}$ are real-valued random numbers in the range $[-J,+J]$.
As the system-bath interaction strength is controlled by $\lambda$, we may set $J=1/4$
without loss of generality.

In the second case,
each system spin is connected to all the bath spins.
The Hamiltonian for the system-bath interaction reads
\begin{eqnarray}
H_\mathrm{SB}&=& -\sum_{j=1}^{2}\sum_{n=3}^{N_{\mathrm{B}}+2}\left( 
J^x_{n,j}\sigma^x_{n}\sigma^x_{j}+
J^y_{n,j}\sigma^y_{n}\sigma^y_{j}+
J^z_{n,j}\sigma^z_{n}\sigma^z_{j}
\right)
.
\label{s43}
\end{eqnarray}
In this case, $\Vert H_\mathrm{SB} \Vert = {\cal O}(N_{\mathrm{B}})$.
Hence, unlike for Eq.~(\ref{s43b}) for which the system-bath interaction does not
depend on the number of bath spins, for Eq.~(\ref{s43})
the system-bath interaction increases as the number of bath spins increases~\cite{ZHAO16}.

For the spin bath we also consider two extreme alternatives.
The first is a ring with Hamiltonian 
\begin{eqnarray}
H_\mathrm{B}&=& -\sum_{n=3}^{N_{\mathrm{B}}+2}\left( 
K^x_n\sigma^x_{n}\sigma^x_{n+1}
+K^y_n\sigma^y_{n}\sigma^y_{n+1}
+K^z_n\sigma^z_{n}\sigma^z_{n+1}
\right)
-\sum_{n=3}^{N_{\mathrm{B}}+2}\left( h^x_{n} \sigma^x_{n}+h^z_{n} \sigma^z_{n}\right) 
.
\label{s42a}
\end{eqnarray}
We use Eq.~(\ref{s42a}) in two very different forms.
In one form, we take all the $K^x_n$'s, $K^y_n$'s, and $K^z_n$'s to be uniform random numbers in the range $[-K,K]$
and the fields $h^x_{n}$ and $h^z_{n}$ to be uniform random numbers in the range
$[-h^x_\mathrm{B},+h^x_\mathrm{B}]$ and $[-h^z_\mathrm{B},+h^z_\mathrm{B}]$, respectively.
For random couplings and random fields,
it is unlikely that the model Eq.~(\ref{s42a}) is integrable (in the Bethe-Ansatz~\cite{BETH31,HULT38,GAUD83} sense)
or has any other special features such as conserved magnetization etc.
In the other form, we take $K^x_n=K^y_n=K^z_n=K$ and $h^x_{n}=h^z_{n}=0$ for all $n$.
Then, Eq.~(\ref{s42a}) is just the Hamiltonian of the isotropic Heisenberg ring
which is known to be integrable (in the Bethe-Ansatz sense).
Thus, a comparison of the results obtained by using these two extreme forms
allows us to gauge the importance of integrablility for the relaxation/thermalization
processes of interest.

As the second model for the spin bath, we consider a spin-glass defined by the Hamiltonian
\begin{eqnarray}
H_\mathrm{B}&=& -\sumprime_{m,n=3}^{N_{\mathrm{B}}+2}\left( 
K^x_{m,n}\sigma^x_{m}\sigma^x_{n} +K^y_{m,n}\sigma^y_{m}\sigma^y_{n} + K^z_{m,n}\sigma^z_{m}\sigma^z_{n}
\right)
-\sum_{n=3}^{N_{\mathrm{B}}+2}\left( h^x_{n} \sigma^x_{n}+h^z_{n} \sigma^z_{n}\right) 
,
\label{s42b}
\end{eqnarray}
where the $K^x_{m,n}$'s, $K^y_{m,n}$'s, and $K^z_{m,n}$'s are uniform random numbers in the range $[-K,K]$
and the prime on the summation sign indicates that contributions with $m=n$ are excluded.
Because Eq.~(\ref{s42b}) contains $(N_{\mathrm{B}}-1)N_{\mathrm{B}}$
spin-spin coupling terms instead of the $N_{\mathrm{B}}$ coupling terms in Eq.~(\ref{s42a}),
it takes a factor $N_{\mathrm{B}}-1$ more CPU time
to solve the TDSE for the same length of time interval.
Therefore, in particular for $N_{\mathrm{B}}>28$, we use Eq.~(\ref{s42b}) judiciously.

The bath Hamiltonian Eq.~(\ref{s42a}) with random couplings and fields has the property that
the distribution of nearest-neighbor energy levels is Wigner-Dyson-like,
suggesting that the corresponding classical baths exhibit chaos.
Earlier work along the lines presented in the present paper has shown that
spin baths with a Wigner-Dyson-like distribution are more effective as
sources for fast decoherence than spin baths with Poisson-like distribution~\cite{LAGE05}.
Fast decoherence is a prerequisite for a system to exhibit fast relaxation to the
thermal equilibrium state~\cite{YUAN09,JIN10x}.
Extensive simulation work on spin-baths with very different degrees of connectivity~\cite{YUAN06,YUAN07,YUAN08,JIN13a,NOVO16}
suggests that as long as there is randomness in the system-bath coupling and
randomness in the intra-bath coupling, the simple model Eq.~(\ref{s42a}) may be considered as a generic spin bath.
However, as we show below, the details of the relaxation process change if we use as a model
of the bath  Eq.~(\ref{s42b}) instead of Eq.~(\ref{s42a}).

\section{Quantum dynamics of the whole system}\label{section3}

The time evolution of a closed quantum system defined
by Hamiltonian Eq.~(\ref{s40}) is governed by the TDSE
\begin{eqnarray}
i\frac{\partial}{\partial t} |\Psi(t)\rangle &=& H|\Psi(t)\rangle
.
\label{s44}
\end{eqnarray}%
The pure state $|\Psi(t)\rangle$ of the whole system $\mathrm{S}+\mathrm{B}$ evolves in time according to
\begin{eqnarray}
|\Psi(t)\rangle
&=&
e^{-itH}|\Psi(0)\rangle=\sum_{i=1}^{D_{\mathbf{S}}} \sum_{p=1}^{D_{\mathrm{B}}} c(i,p,t)|i,p\rangle
,
\label{s45}
\end{eqnarray}%
where $D_{\mathrm{S}}=4$ and $D_{\mathrm{B}}=2^{{N_\mathrm{B}}}$ are the dimensions of the Hilbert space
of the system and bath, respectively.
The coefficients $\{c(i,p,t)\}$ are the complex-valued amplitudes of
the corresponding elements of the set $\{ |i,p\rangle \}$ which denotes
the complete set of orthonormal states in the up--down basis of the system and bath spins.

The size of the quantum systems that can be simulated, that is the
size for which Eq.~(\ref{s45}) can actually be computed, is primarily
limited by the memory required to store the pure state.

Solving the TDSE requires storage of all the complex numbers $\{ c(i,p,t)|i=1,\ldots,4\;,p=1,\ldots,2^{N_\mathrm{B}}\}$.
Clearly, the amount of memory that is required is proportional to $2^{N_\mathrm{B}+2}$,
which increases exponentially with the number of spins of the bath.
Using 64-bit floating-point arithmetic (corresponding to $16=2^4$ bytes for each complex number),
representing a pure state of $N_\mathrm{B}+2$ spin-$1/2$ particles
on a digital computer requires at least $2^{N_{\mathrm{B}}+6}$ bytes.
For example, for $N_\mathrm{B}=22$ ($N_\mathrm{B}=34$) we need at least 256~MB (1~TB)
of memory to store a single state $|\Psi(t)\rangle$.
In practice we need storage for three vectors, and memory for communication buffers, local variables and the code itself.

From a numerical-analysis viewpoint,
the real-time propagation by $e^{-itH}$ is best carried out by means
of the Chebyshev polynomial algorithm~\cite{TALE84,LEFO91,IITA97,DOBR03}.
This algorithm is known to yield results that are very accurate
(close to machine precision), independent of the time step used~\cite{RAED06}.
A disadvantage of this algorithm is that, especially when
the number of spins exceeds 28, it consumes significantly
more CPU and memory resources than a Suzuki-Trotter product-formula based algorithm~\cite{RAED06}.

Advancing a pure state by one time step $\tau$ by a Suzuki-Trotter product-formula based algorithm
can symbolically be written as $|\Psi \rangle \leftarrow U_K\ldots U_1 |\Psi\rangle$
where the $U$'s are sparse unitary matrices with a relatively complicated structure.
A characteristic feature of the problem at hand is that for most of the $U$'s,
all elements of the set $\{c(i,p,t)|i=1,\ldots,4\;,p=1,2^{N_\mathrm{B}}\}$ are involved in the operation.
This translates into a complicated scheme for efficiently accessing memory,
which in turn requires a sophisticated MPI communication scheme on a distributed memory system~\cite{RAED07X}.
The CPU time required for one such typical $U$-operation also increases exponentially with the number of spins.

Using the latter to solve the TDSE for $0\le t \le 200$ and ${N_\mathrm{B}}+2=36$
requires somewhat more than 1 TB of memory and takes about 15 hours of elapsed time, using
131072 IBM BlueGene/Q cores.
The Chebyshev polynomial algorithm takes about 3 times this amount of resources.
Therefore, we only use the latter to verify that the numerical results
of the product-formula based algorithm are, for practical purposes, as good
as the numerically exact results and then use the product-formula based for
the production runs.

We end this section by addressing an important aspect of the simulation procedure.
As is clear from the presentation of the various Hamiltonians, we often use randomly chosen couplings.
Likewise, to prepare the initial state of the bath (see section~\ref{section3b}), we also use random numbers.
In practice, all the random numbers that are required to
define the interactions and to construct the initial state are generated afresh for each simulation run.
In other words, we may expect that our numerical results show fluctuations
due to that the interactions or initial states are unlikely to be the same.
However, as the data presented in this paper show, the conclusions that can be drawn from the data
are robust in the sense that they do not seem to depend on different random choices of couplings and initial states.

\subsection{Density matrix}\label{section3a}

According to quantum theory, observables are represented by Hermitian matrices
and the correspondence with measurable quantities is through their averages defined as~\cite{NEUM55,BALL03}
\begin{equation}
\langle  {\cal A}(t) \rangle = \mathbf{Tr\;} \rho(t) {\cal A} = \mathbf{Tr\;} \rho {\cal A}(t)
,
\label{s46a}
\end{equation}
where ${\cal A}$ denotes a Hermitian matrix representing the observable,
$\rho(t)$ is the density matrix of the whole system $S+B$ at time $t$
and $\mathbf{Tr\;}$ denotes the trace over all states of the whole system $\mathrm{S}+\mathrm{B}$.

The state of the system $S$ is completely described by the reduced density matrix
\begin{equation}
\rho_{\mathrm{S}}(t)\equiv\mathbf{Tr}_{\mathrm{B}}\rho(t)
,
\label{s46}
\end{equation}
where $\rho(t)$ is the density matrix of the whole system $S+B$ at time $t$,
$\mathbf{Tr}_{\mathrm{B}}$ denotes the trace over the degrees of freedom of the bath,
and $\mathbf{Tr}_{\mathrm{S}}\rho_{\mathrm{S}}(t)=\mathbf{Tr\;}\rho(t)=1$.

For numerical purposes it is convenient to express $4\times4$ matrices
in terms of the sixteen $4\times4$ matrices defined by
$\{\mathbf{e}_0,\ldots,\mathbf{e}_{15} \}\equiv
\{\openone_1\otimes\openone_2,
\sigma_1^x\otimes\openone_2,
\sigma_1^y\otimes\openone_2,
\sigma_1^z\otimes\openone_2,
\openone_1\otimes\sigma_2^x,
\openone_1\otimes\sigma_2^y,
\openone_1\otimes\sigma_2^z,
\sigma_1^x\otimes\sigma_2^x,
\sigma_1^y\otimes\sigma_2^y,
\sigma_1^z\otimes\sigma_2^z,
\sigma_1^x\otimes\sigma_2^y,
\sigma_1^y\otimes\sigma_2^x,
\sigma_1^x\otimes\sigma_2^z,
\sigma_1^z\otimes\sigma_2^x,
\sigma_1^y\otimes\sigma_2^z,
\sigma_1^z\otimes\sigma_2^y
\}$.
These matrices span the vector space of $4\times4$ complex-values matrices
and are orthonormal with respect to the inner product $(X|Y)=(1/4)\mathbf{Tr}_{\mathrm{S}} X^\dagger Y$.
With the help of these basis vectors,
the reduced density matrix can, without loss of generality, be written as
\begin{equation}
\rho_{\mathrm{S}}(t)=\frac{1}{4}\sum_{i=0}^{15} \rho_i(t) \mathbf{e}_i
,
\label{s48}
\end{equation}
where all the $\rho_i(t)$ are real numbers.
From  Eq.~(\ref{s48}) it follows immediately  that
\begin{eqnarray}
\rho_i(t)&=&\mathbf{Tr}_{\mathrm{S}}\;\rho_{\mathrm{S}}(t)\mathbf{e}_i
=\mathbf{Tr}\;\rho(t)\mathbf{e}_i
,
\label{s49a}
\end{eqnarray}
and that $\rho_0(t)=1$ because $\mathbf{Tr}_{\mathrm{S}}\rho_{\mathrm{S}}(t)=1$.
Equation~(\ref{s49a}) shows that $\rho_i(t)$ is nothing but the expectation value
of the operator $\mathbf{e}_i$, as measured with respect to the whole system.

\subsection{Random state technology}\label{section3aa}

If the numerical solution of the TDSE for a pure state of
$N_\mathrm{B}+2$ spins already requires  resources that
increase exponentially with the number of spins of the bath,
computing Eq.~(\ref{s46a}) seems an even more daunting task.
Fortunately, we can make use of the ``random-state technology''
to reduce the computational cost to that of solving the TDSE for one pure state~\cite{HAMS00}.
The key is to note that if $|\Phi\rangle$ is a pure state, picked randomly
from the $D=2^{{N_\mathrm{B}+2}}$-dimensional unit hypersphere,
one can show in general that for Hermitian matrices $X$~\cite{HAMS00,REIM07,BART09,SUGI12,SUGI13,STEI14b}
\begin{equation}
\mathbf{Tr\;} X \approx D\langle\Phi|X|\Phi\rangle
.
\label{s46b}
\end{equation}
As shown in Appendix A, if $D$ is large
the statistical errors resulting from approximating
$\mathbf{Tr\;} X $ by $\langle\Phi| X |\Phi\rangle$ are small.
For large baths, this property makes the problem amenable to numerical simulation.
Therefore, from now on, we replace the ``$\mathbf{Tr\;}$'' by
a matrix element of a random pure state whenever the trace operation
involves a number of states that increases exponentially with the
number of spins (in the present case, bath spins only).
In practice, as the dimension of the Hilbert space of the bath may be assumed to be large,
we can, using this ``random-state technology'', replace the
trace operation in Eq.~(\ref{s46a}) by solving the TDSE with the initial state
\begin{equation}
| \Psi\rangle = \sqrt{D\rho} |\Phi\rangle
,
\label{s46z}
\end{equation}
such that
\begin{equation}
\langle  {\cal A}(t) \rangle \approx \langle \Psi|{\cal A}(t) | \Psi\rangle = \langle \Psi(t)|{\cal A} | \Psi(t)\rangle
.
\label{s46y}
\end{equation}
Similarly, we may  compute the trace over the bath degrees of freedom as
\begin{equation}
\left(\mathbf{Tr}_{\mathrm{B}} {\cal A}\right)_{i,j} \approx
\sum_{p=1}^{D_{\mathrm{B}}} c^{\ast}(i,p,t)c(j,p,t)\;\langle i,p|{\cal A} |j,p\rangle
,
\label{s46c}
\end{equation}
and the expectation values of the operators $\mathbf{e}_i$ are given by
\begin{eqnarray}
\rho_i(t)\approx  \langle\Psi(t)|\mathbf{e}_i|\Psi(t)\rangle,\quad i=1,\ldots,15
.
\label{s49}
\end{eqnarray}

\begin{table}[t]
\caption{
Simulation data for the system energy and the system-bath energy as obtained from
Eq.~(\ref{RST}) using the thermal random state Eq.~(\ref{s4k0}) with $\beta=5$.
The Hamiltonian of the system-bath interaction and spin bath
are given by Eq.~(\ref{s43b}) and Eq.~(\ref{s42a}), respectively.
The first row lists the exact result $E_\mathrm{S}(\beta=5)=-0.730$ of the isolated two-spin system,
the exact ground state energy being $E_0=-0.750$.
The data of columns (3,4) and (5,6) were obtained from different realizations
of the thermal random state and interaction parameters $J^\alpha_{n,1}$, $J^\alpha_{m,2}$ and $K^\alpha_n$,
see Eqs.~(\ref{s43b}) and (\ref{s42a}).
}
\begin{ruledtabular}
\begin{tabular*}{\textwidth}{@{\extracolsep{\fill}} cccccc}
$\lambda$ & $N_{\mathrm{B}}$ & \hfil $\langle H_{\mathrm{S}}\rangle$ \hfil  & \hfil $\lambda\langle H_{\mathrm{SB}}\rangle$
 \hfil & \hfil $\langle H_{\mathrm{S}}\rangle$ \hfil  & \hfil $\lambda\langle H_{\mathrm{SB}}\rangle$ \hfil  \\
\hline
0 & $-$ & $-0.730$ & $0$ & $-0.730$ & $0$ \\
\hline
$0.125 $&$ 16 $&$ -0.728 $&$  -0.378\times10^{-1} $&$ -0.731 $&$  -0.173\times10^{-1} $ \\
$0.125 $&$ 18 $&$ -0.728 $&$  -0.335\times10^{-1} $&$ -0.715 $&$  -0.098\times10^{-1} $ \\
$0.125 $&$ 20 $&$ -0.725 $&$  -0.216\times10^{-1} $&$ -0.742 $&$  -0.117\times10^{-1} $ \\
$0.125 $&$ 22 $&$ -0.732 $&$  -0.452\times10^{-1} $&$ -0.728 $&$  -0.399\times10^{-1} $ \\
$0.125 $&$ 24 $&$ -0.727 $&$  -0.259\times10^{-1} $&$ -0.720 $&$  -0.045\times10^{-1} $ \\
\hline
$0.250 $&$ 16 $&$ -0.737 $&$  -0.147\times10^{-1} $&$ -0.726 $&$  -0.432\times10^{-1} $ \\
$0.250 $&$ 18 $&$ -0.730 $&$  -0.075\times10^{-1} $&$ -0.718 $&$  -0.291\times10^{-1} $ \\
$0.250 $&$ 20 $&$ -0.727 $&$  -0.215\times10^{-1} $&$ -0.723 $&$  -0.395\times10^{-1} $ \\
$0.250 $&$ 22 $&$ -0.730 $&$  -0.234\times10^{-1} $&$ -0.728 $&$  -0.330\times10^{-1} $ \\
$0.250 $&$ 24 $&$ -0.732 $&$  -0.181\times10^{-1} $&$ -0.724 $&$  -0.310\times10^{-1} $ \\
\hline
$0.500 $&$ 16 $&$ -0.705 $&$  -0.612\times10^{-1} $&$ -0.719 $&$  -0.423\times10^{-1} $ \\
$0.500 $&$ 18 $&$ -0.689 $&$  -1.102\times10^{-1} $&$ -0.715 $&$  -0.665\times10^{-1} $ \\
$0.500 $&$ 20 $&$ -0.717 $&$  -0.473\times10^{-1} $&$ -0.716 $&$  -0.514\times10^{-1} $ \\
$0.500 $&$ 22 $&$ -0.719 $&$  -0.472\times10^{-1} $&$ -0.721 $&$  -0.661\times10^{-1} $ \\
$0.500 $&$ 24 $&$ -0.712 $&$  -0.496\times10^{-1} $&$ -0.711 $&$  -0.687\times10^{-1} $ \\
\end{tabular*}
\end{ruledtabular}
\label{tab1}
\end{table}

\subsection{Thermal equilibrium state}\label{section3b}

As a first check on the numerical method, it is of interest to simulate the case in which the system+bath is
initially in thermal equilibrium and study the effects of the bath size $N_\mathrm{B}$
and system-bath interaction strength $\lambda$ on the expectation values of the system spins.
The procedure is as follows.
First we generate a thermal random state of the whole system, meaning that
\begin{eqnarray}
|\Phi(\beta)\rangle&=& \frac{e^{-\beta H/2}|\Phi\rangle}{ \langle\Phi|e^{-\beta H}|\Phi\rangle^{1/2} }
,
\label{s4k0}
\end{eqnarray}
where $\beta$ denotes the inverse temperature.
As one can show that for any observable ${\cal A}(t)$~\cite{HAMS00}
\begin{equation}
\langle {\cal A}(t)\rangle=\frac{\mathbf{Tr\;} e^{-\beta H}{\cal A}(t)}{\mathbf{Tr\;} e^{-\beta H}}
\approx \langle\Phi(\beta)| {\cal A}(t)|\Phi(\beta)\rangle
,
\label{RST}
\end{equation}
we can use $\langle\Phi(\beta)| {\cal A}(t)|\Phi(\beta)\rangle$ to estimate $\langle {\cal A}(t)\rangle$.
As shown in Appendix~\ref{APP1}, in general we may expect the statistical errors
incurred by approximation Eq.~(\ref{RST}) to vanish exponentially with the number of spins.

As $e^{-\beta H}$ commutes with $e^{-it H}$,
$\langle {\cal A}(t)\rangle=\langle {\cal A}(t=0)\rangle$ is time independent.
Excluding the trivial case that $[H,{\cal A}(t)]=0$,
$\langle\Phi(\beta)| {\cal A}(t)|\Phi(\beta)\rangle=
\langle\Phi(\beta)|e^{+it H} {\cal A}e^{-it H}|\Phi(\beta)\rangle$ depends on time.
Indeed, in general the random state $|\Phi(\beta)\rangle$ is unlikely to be an eigenstate of $H$.
Therefore, the simulation data obtained by solving the TDSE with
$|\Phi(\beta)\rangle$ as the initial state should display some time dependence.
However, from Appendix~\ref{APP1}, it follows directly that the time-dependent fluctuations
will vanish very fast with the number of spins.
Hence this time dependence, an artifact of using ``random state technology'',
reveals itself as small statistical fluctuations and therefore can be ignored.

In Table~\ref{tab1} we present simulation results of the system energy $\langle H_{\mathrm{S}}\rangle$
and system-bath energy $\lambda\langle H_{\mathrm{SB}}\rangle$, calculated according to Eq.~(\ref{RST}).
The Hamiltonian of the system-bath interaction and spin bath
are given by Eq.~(\ref{s43b}) and Eq.~(\ref{s42b}), respectively,
$N_{\mathrm{B}}=16,18,20,22,24$ and $\lambda=0.125,0.25,0.5$.
For reference, we note that the ground state energy of the system in the singlet state is equal to $-3/4=0.75$.
As Table~\ref{tab1} shows, $\langle H_{\mathrm{S}}\rangle\approx-0.73$ hence, for the system being studied,
$\beta=5$ corresponds to a fairly low temperature.

The data of columns (3,4) and (5,6) were obtained for different realizations of the system-bath and bath interaction
parameters and different realizations of the thermal random states,
giving some indication of the statistical fluctuations
stemming from both the use of random couplings and different realizations of the thermal random states.

The results of the system-bath energy $\lambda\langle H_{\mathrm{SB}}\rangle$
for different $\lambda$ give an indication for the range
of $\lambda$ for which the system-bath interaction may be considered to be a perturbation.
Taking into account the statistical fluctuations, we conclude from the data of Table~\ref{tab1} that for $\beta=5$,
$\lambda=0.5$ may be outside the perturbative regime while $\lambda=0.125,0.25$ are not.
Disregarding statistical fluctuations, the data of Table~\ref{tab1} obtained with $H_{\mathrm{SB}}$ given by Eq.~(\ref{s43b})
do not show a clear signal of a dependence on the number of bath spins $N_{\mathrm{B}}$.
From a standard perturbation expansion, it follows that the perturbative regime grows as $\beta$ decreases.
Hence, the statement that $\lambda=0.5$ may be outside the perturbative regime does not necessarily hold for say $\beta=1$
and in fact it does not (data not shown).
From Table~\ref{tab1} it is also clear that the system-bath energy
$\lambda\langle H_{\mathrm{SB}}\rangle$ may vary considerably from
one realization to another, which in view of the random choices of the couplings is not a surprise.

In the case that we use system-bath interaction Hamiltonian Eq.~(\ref{s43}),
each system spin interacts with each of the $N_{\mathrm{B}}$ bath spins.
Therefore, the system-bath energy is proportional to $N_{\mathrm{B}}$,
in contrast to the case of Hamiltonian Eq.~(\ref{s43b}) in which the system-bath energy is of order one.
In this respect, the system-bath interaction Hamiltonian Eq.~(\ref{s43}) is not different from e.g. the
standard spin-boson model~\cite{BREU02}.
Taking into account that when using Eq.~(\ref{s43}), the effective system-bath interaction is
proportional to $\lambda N_{\mathrm{B}}$ instead of proportional to
$\lambda$, the simulation data obtained by using Eq.~(\ref{s43})
instead of Eq.~(\ref{s43b}) are similar to those shown in Table~\ref{tab1} and are therefore not shown.

In general, to determine whether the system-bath interaction is weak or not
we adopt a pragmatic approach: we simply compute the
averages and compare them with the theoretical results of the isolated system.
The coupling $\lambda$ is considered to be small enough if the averages and theoretical results
agree within a few percent.

\begin{figure}[ht]
\begin{center}
\includegraphics[width=0.30\hsize]{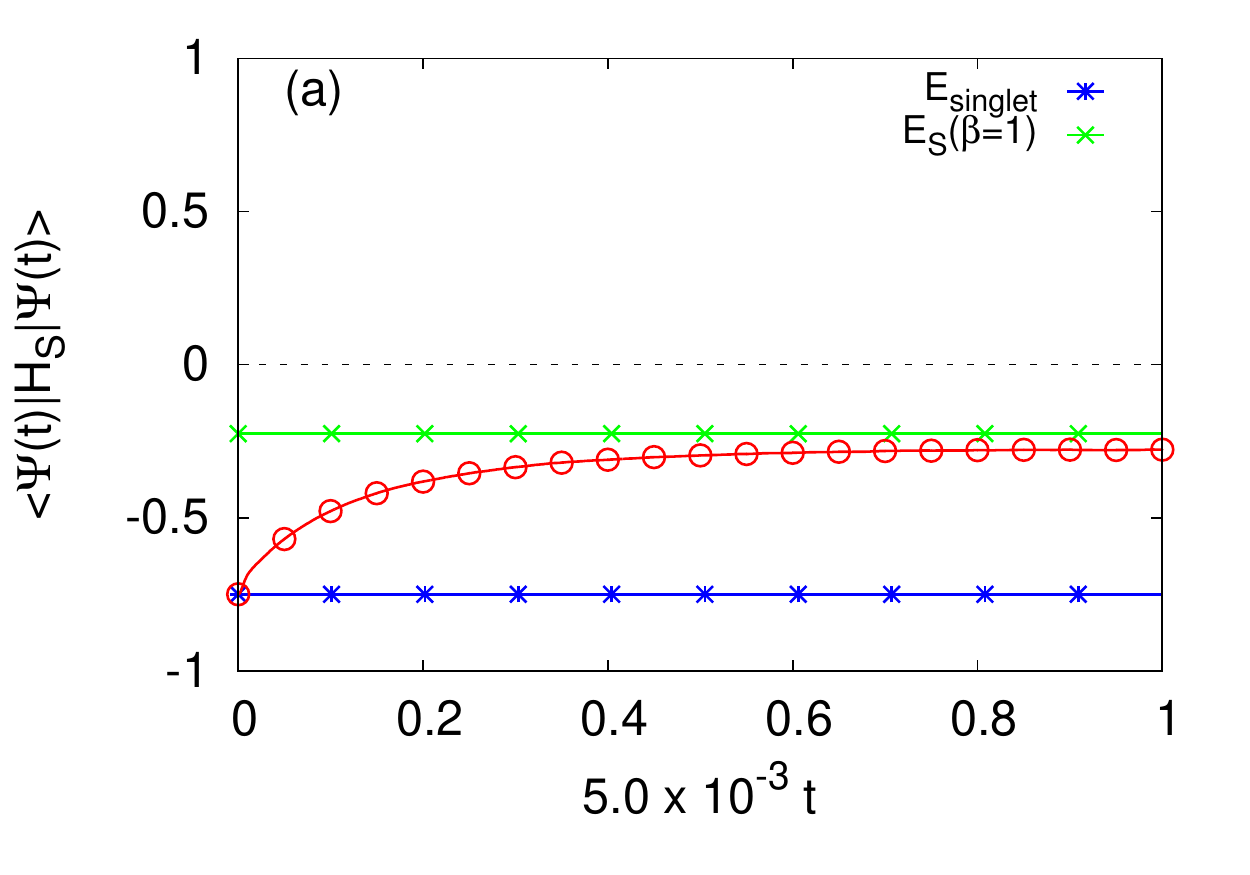}
\includegraphics[width=0.30\hsize]{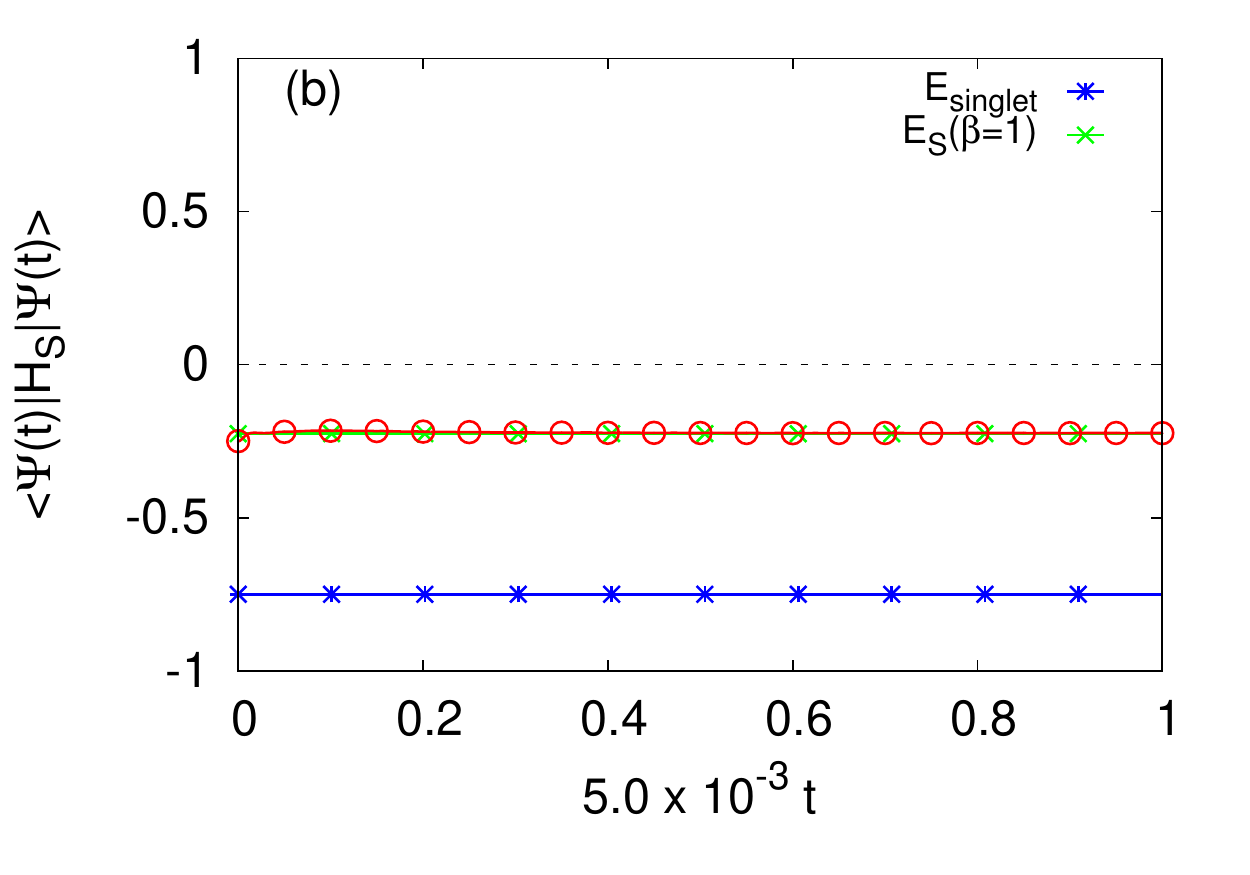}
\includegraphics[width=0.30\hsize]{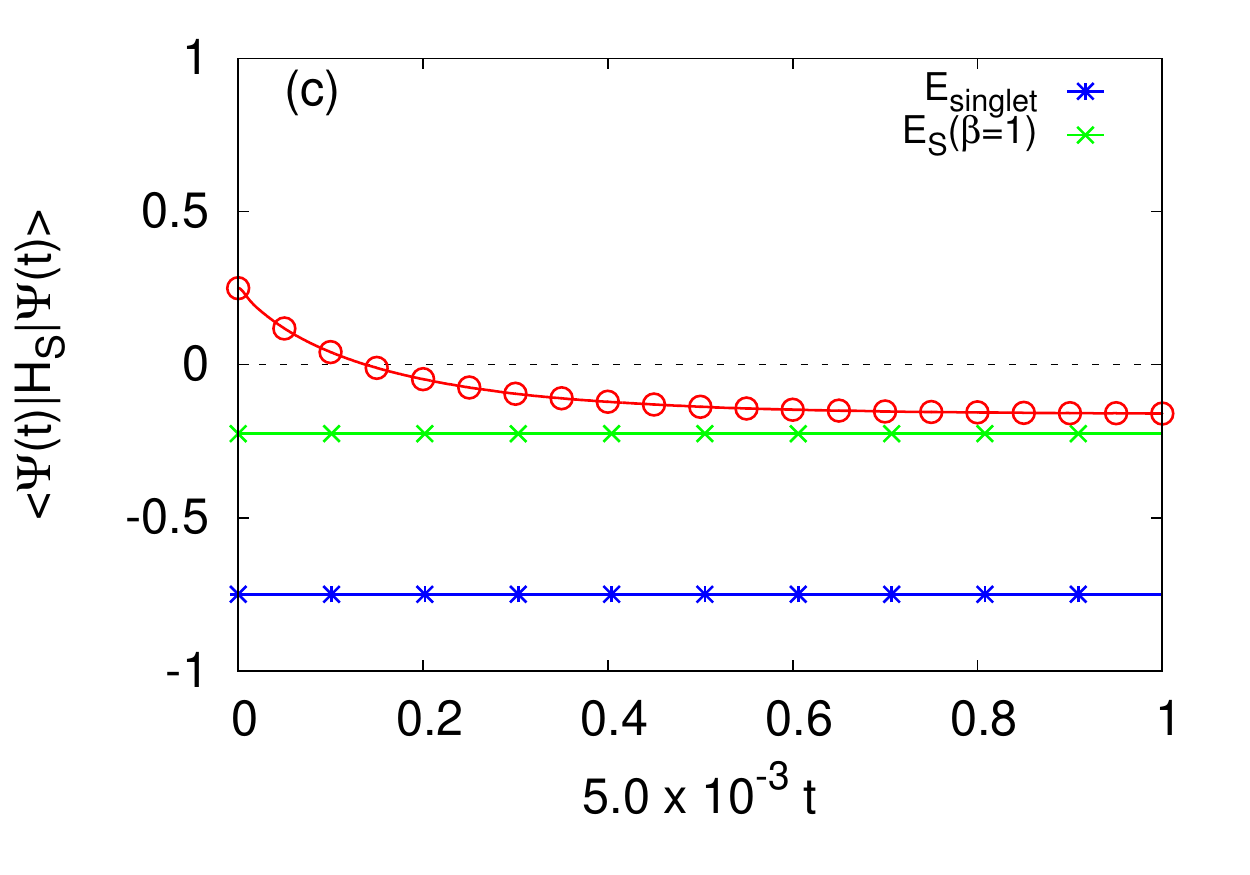}
\includegraphics[width=0.30\hsize]{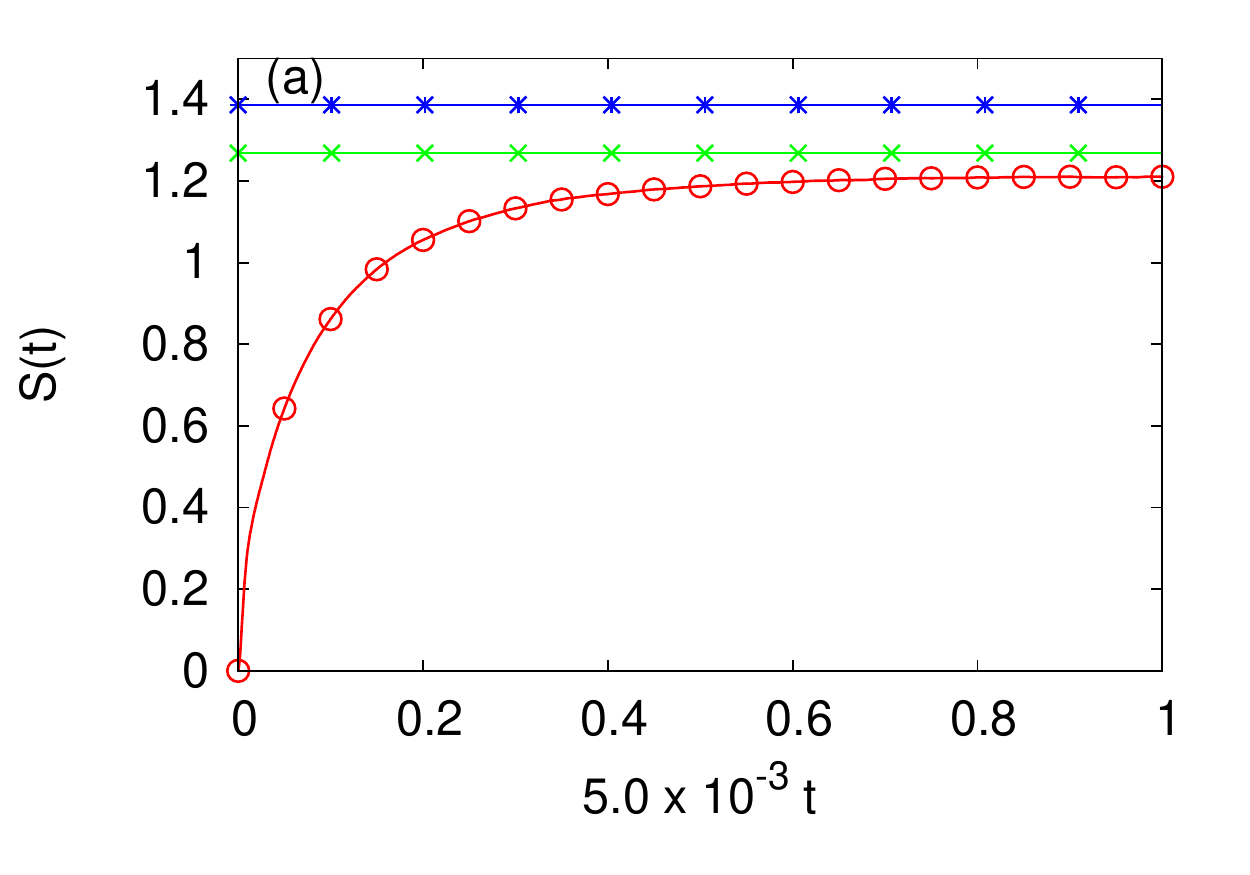}
\includegraphics[width=0.30\hsize]{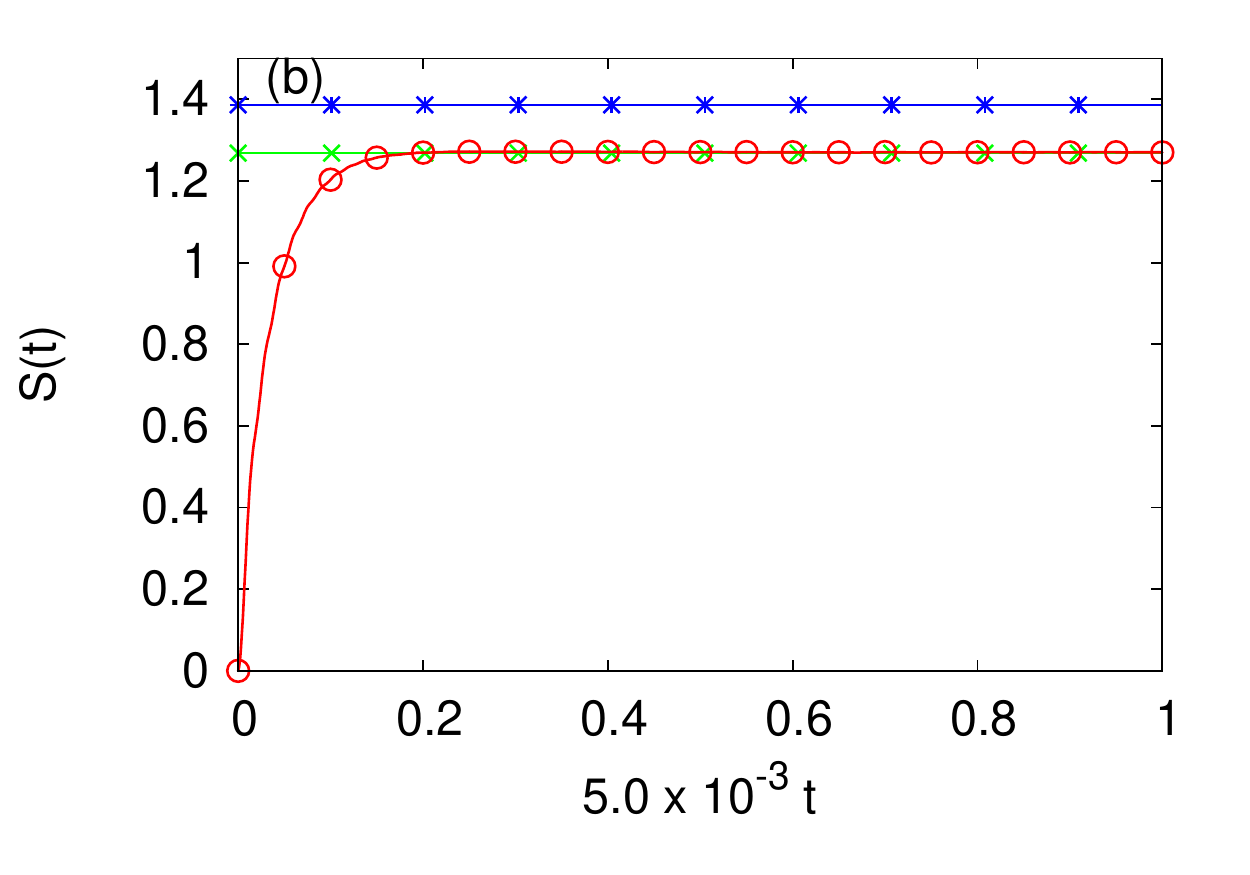}
\includegraphics[width=0.30\hsize]{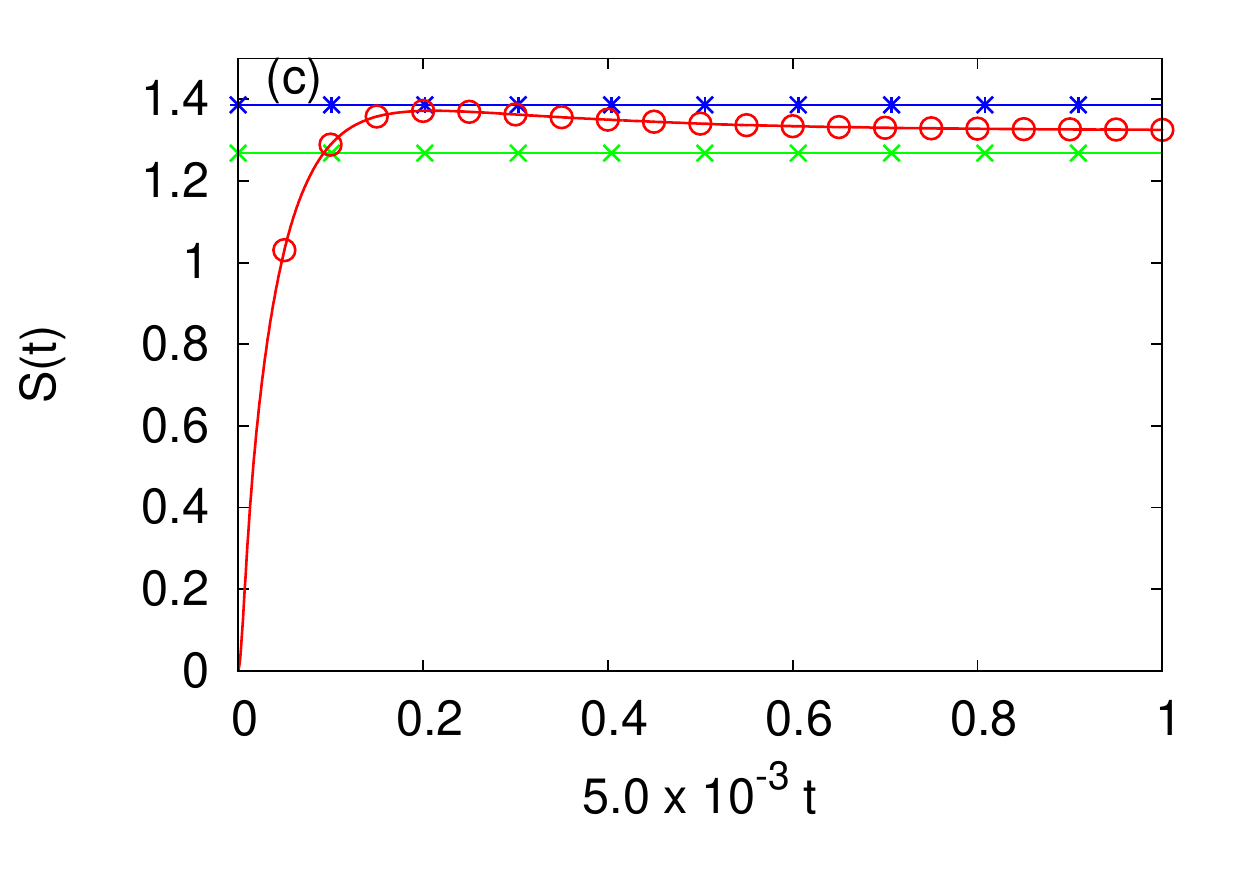}
\caption{(color online) %
Time dependence of the energy (open circles, top row) and entropy (open circles, bottom row) of the two-spin system in contact with a spin bath at moderate temperature,
as obtained from the solution of the TDSE for three different initial states $|\psi>\otimes|\Phi(\beta=1)\rangle$
where $|\Phi(\beta=1)\rangle$ denotes a thermal random state Eq.~(\ref{s4k0z}) of the bath only.
The system Hamiltonian is given by Eq.~(\ref{s41})
with $J_{\bot}=J_{\parallel}=1/4$ (antiferromagnetic Heisenberg model).
The Hamiltonian of the system-bath interaction and spin-glass bath
are given by Eq.~(\ref{s43b}) and Eq.~(\ref{s42b}), respectively.
The number of bath spins is $N_\mathrm{B}=20$, $K=1/2$, $h^x_\mathrm{B}=h^z_\mathrm{B}=0$.
The system-bath interaction strength is $\lambda=0.5$.
(a): $|\psi>=|S>$, where $|S>$ is the singlet state as given by Eq.~(\ref{singletstate0});
(b): $|\psi>=|\uparrow \downarrow>$;
(c): $|\psi>=|\uparrow \uparrow>$.
Crosses in the top row figures: ground state energy of the isolated two-spin system;
Stars in the top row figures: thermal energy of the isolated two-spin system at $\beta=1$;
Crosses in the bottom row figures: maximum entropy of the isolated two-spin system;
Stars in the bottom row figures: entropy of the isolated two-spin system at $\beta=1$;
}
\label{fig2b}
\end{center}
\end{figure}

\begin{figure}[ht]
\begin{center}
\includegraphics[width=0.30\hsize]{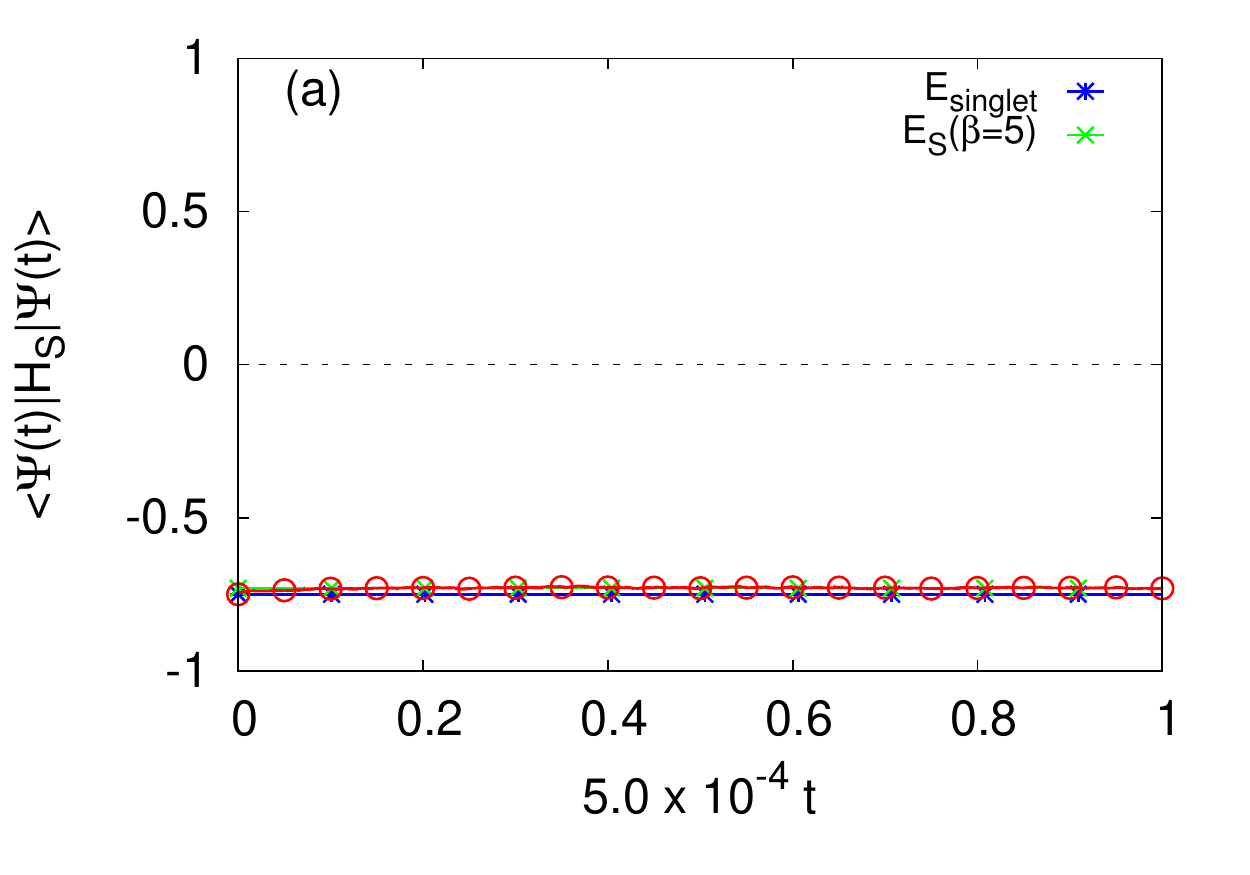}
\includegraphics[width=0.30\hsize]{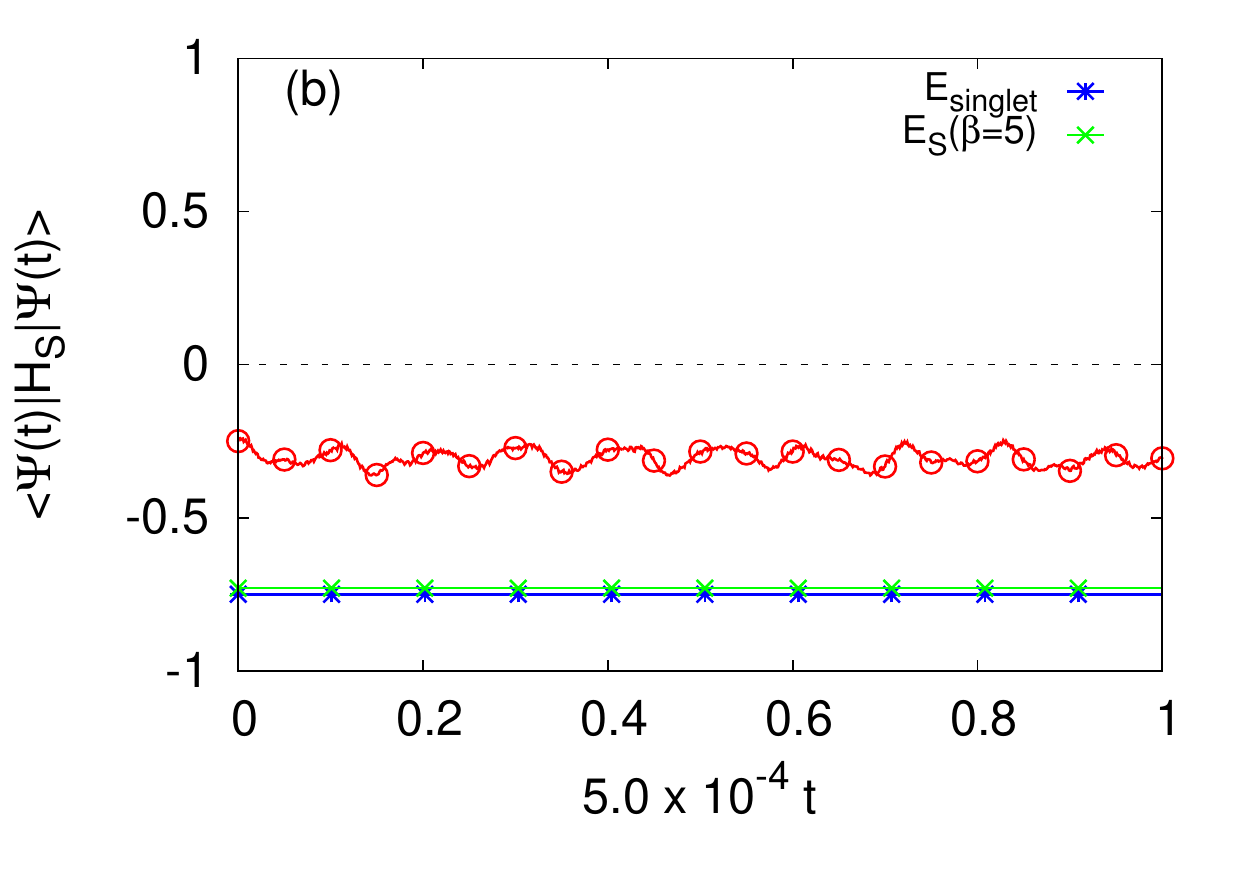}
\includegraphics[width=0.30\hsize]{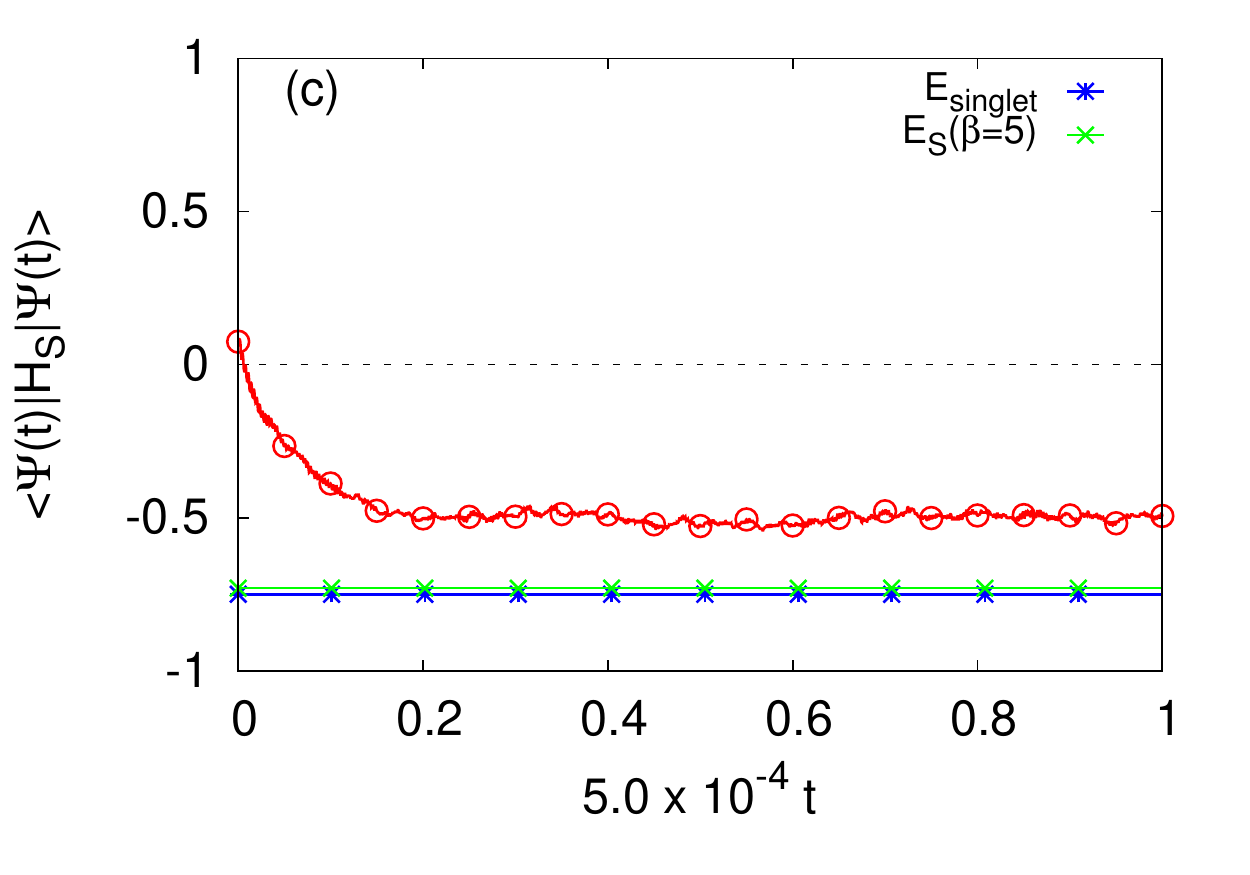}
\includegraphics[width=0.30\hsize]{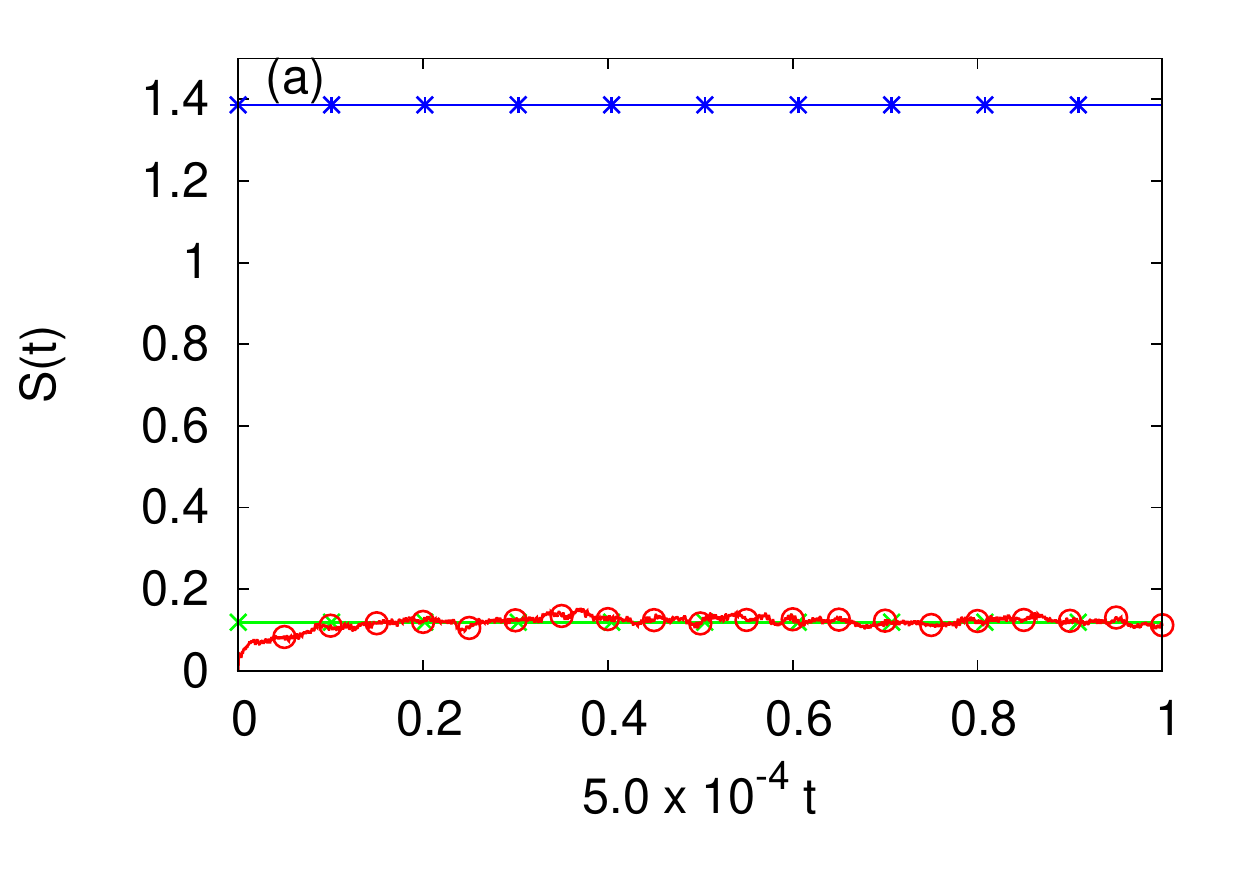}
\includegraphics[width=0.30\hsize]{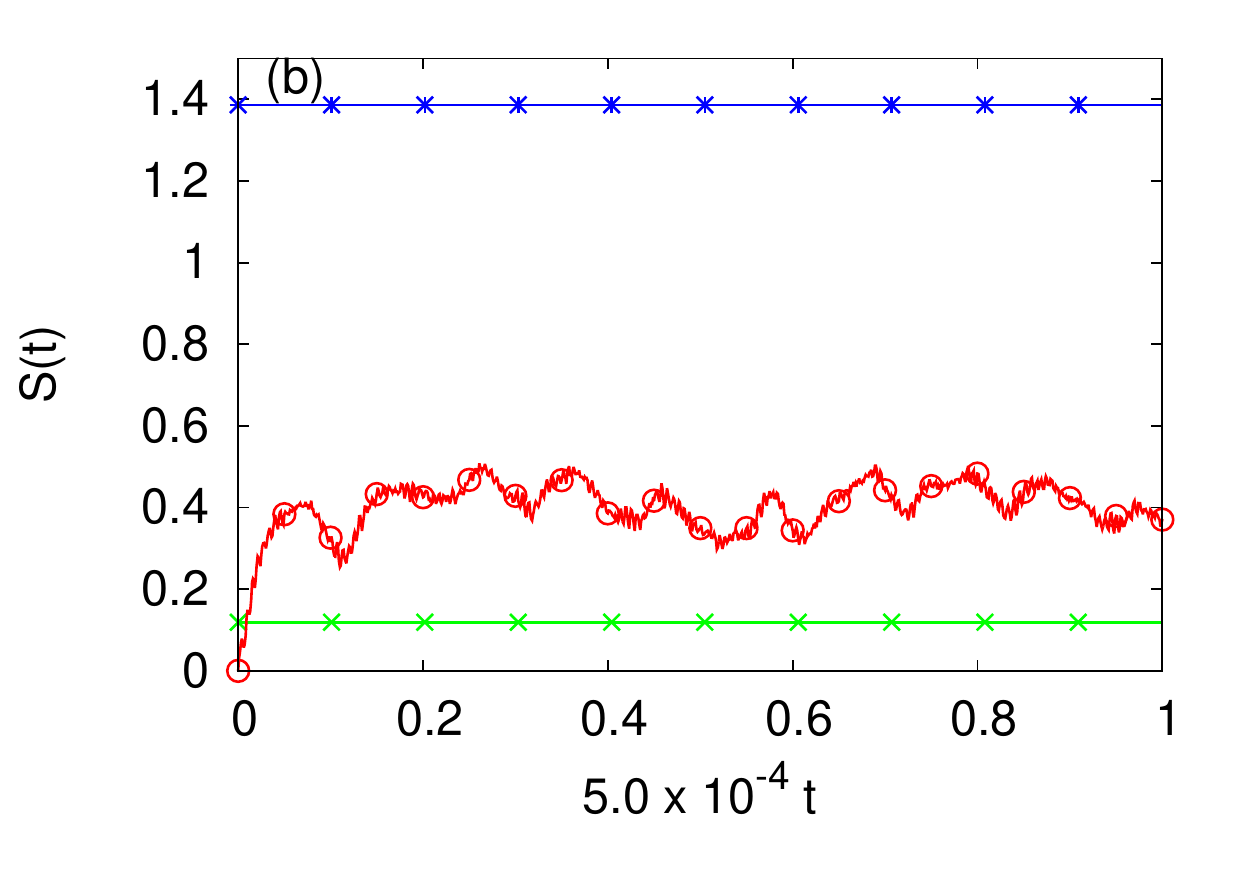}
\includegraphics[width=0.30\hsize]{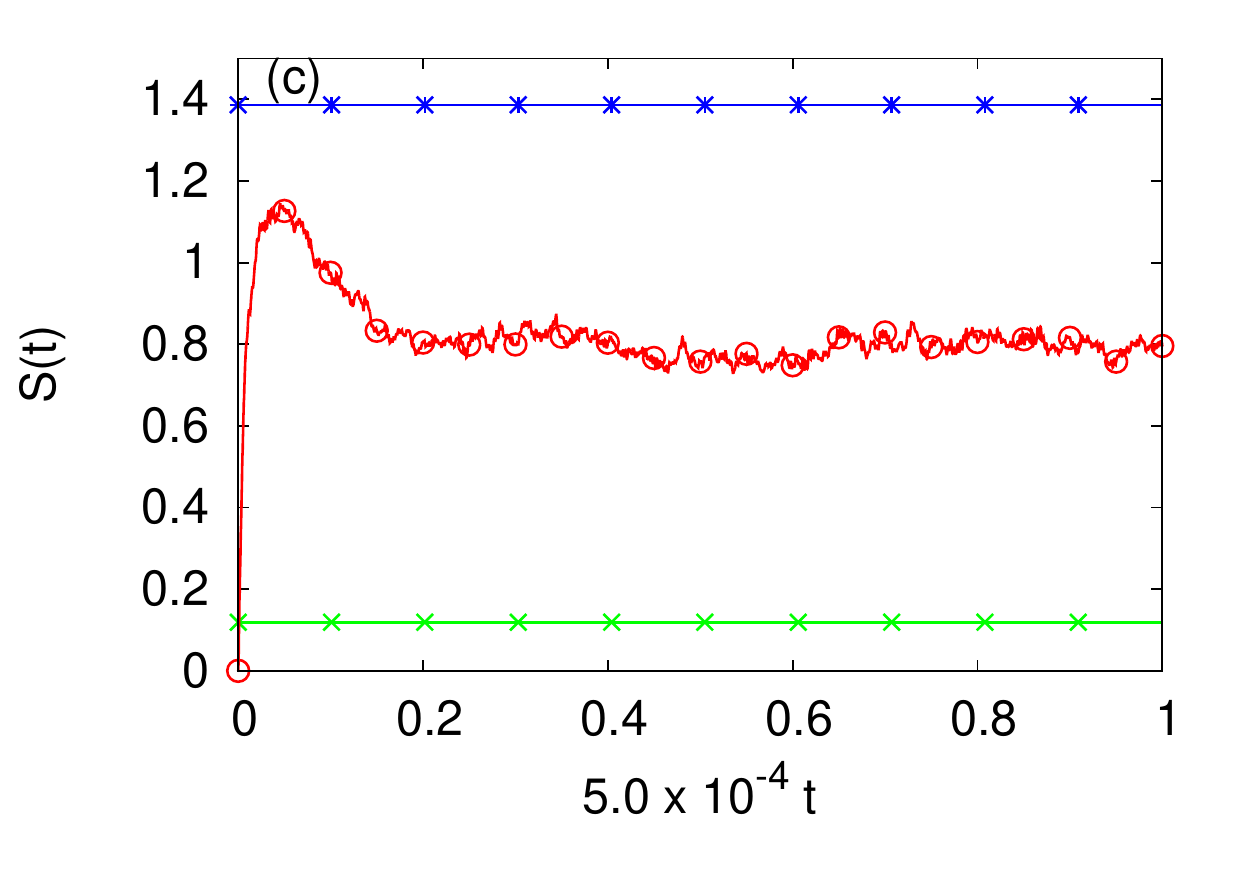}
\caption{(color online) %
Time dependence of the energy (open circles, top row) and entropy (open circles, bottom row) of the two-spin system in contact with a spin bath at low temperature,
as obtained from the solution of the TDSE with the initial state $|\psi>\otimes|\Phi(\beta=5)\rangle$.
The system Hamiltonian is given by Eq.~(\ref{s41})
with $J_{\bot}=J_{\parallel}=1/4$ (antiferromagnetic Heisenberg model).
The Hamiltonian of the system-bath interaction and spin bath
are given by Eq.~(\ref{s43b}) and Eq.~(\ref{s42a}), respectively.
The number of bath spins is $N_\mathrm{B}=20$, $K=1$, $h^x_\mathrm{B}=h^z_\mathrm{B}=1/4$.
The system-bath interaction strength is $\lambda=0.25$.
(a): $|\psi>=|S>$, where $|S>$ denotes the singlet state given by Eq.~(\ref{singletstate0});
(b): $|\psi>=|\uparrow \downarrow>$;
(c): $|\psi>=|R>\otimes |R>$ where $|R>$ denotes a random superposition of spin-up and spin-down states.
Crosses in the top row figures: ground state energy of the isolated two-spin system;
Stars in the top row figures: thermal energy of the isolated two-spin system at $\beta=1$;
Crosses in the bottom row figures: maximum entropy of the isolated two-spin system;
Stars in the bottom row figures: entropy of the isolated two-spin system at $\beta=1$;
}
\label{fig2c}
\end{center}
\end{figure}

\begin{figure}[ht]
\begin{center}
\includegraphics[width=0.45\hsize]{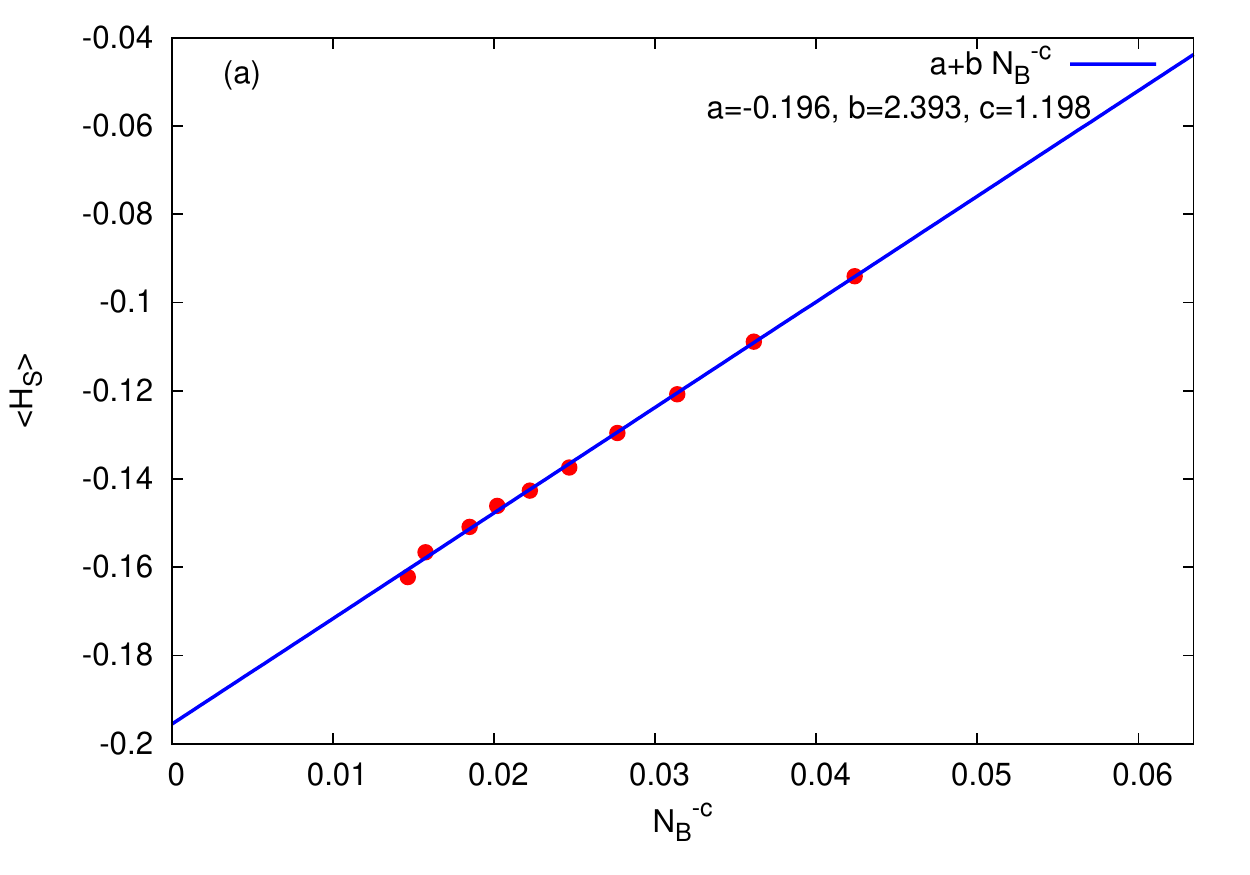}
\includegraphics[width=0.45\hsize]{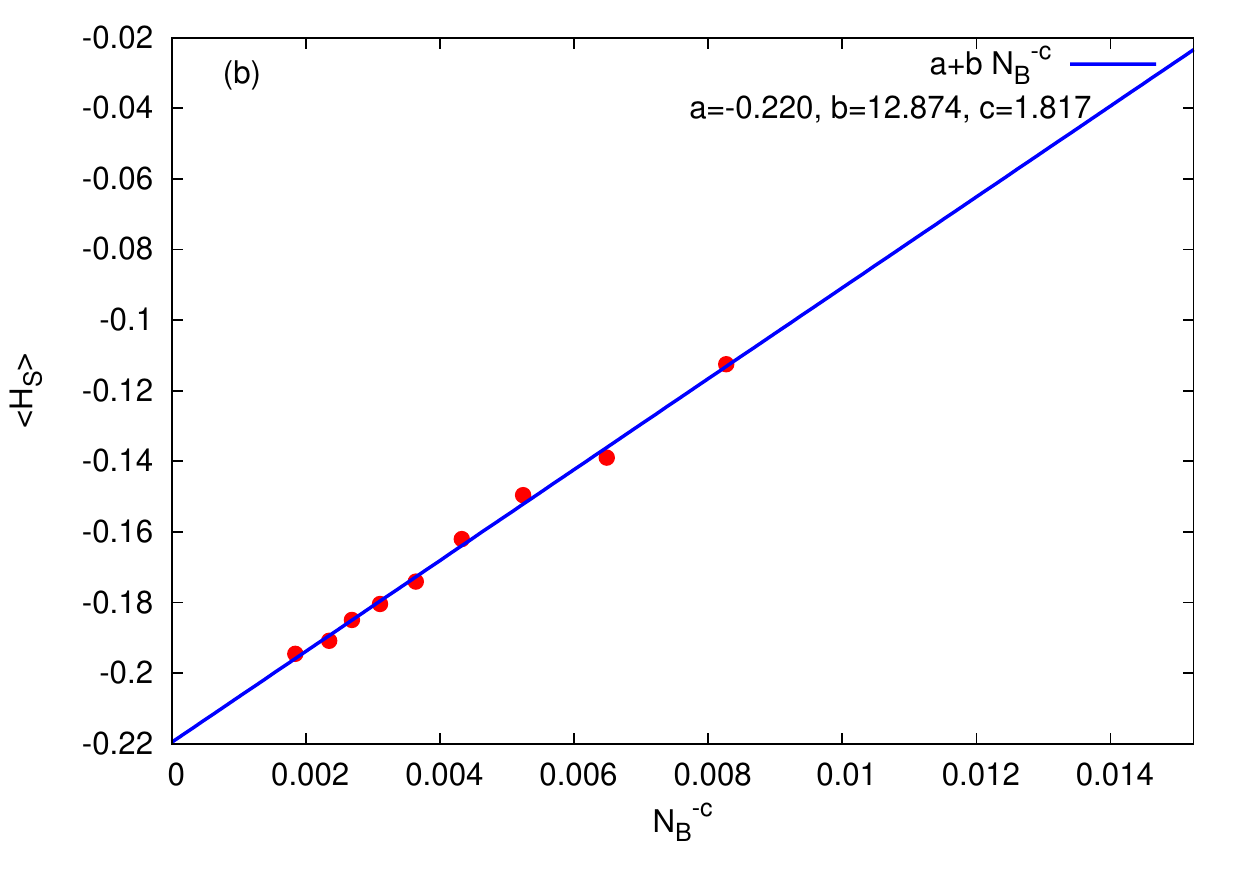}
\caption{(color online) %
Finite-size scaling of the system energy in the stationary state as a function of the number of bath spins
as obtained from the solution of the TDSE with the initial state $|\uparrow\uparrow>\otimes|\Phi(\beta=1)\rangle$.
Solid lines represent the function $a+bN_\mathrm{B}^{-c}$ with $a$ and $b$ determined by the best fit of the data.
The Hamiltonian of the system-bath interaction is given by Eq.~(\ref{s43b}), respectively.
(a): integrable spin bath Hamiltonian Eq.~(\ref{s42a}) with $K^x_n=K^y_n=K^z_n=K=1/2$ and $h^x_\mathrm{B}=h^z_\mathrm{B}=0$,
system-bath interaction $\lambda=1/2$ and $N_\mathrm{B}=14,16,18,20,22,24,28,32,34$;
(b): spin bath Hamiltonian Eq.~(\ref{s42b}) with all $K$'s random in the interval $[-1/2,+1/2]$, $h^x_\mathrm{B}=h^z_\mathrm{B}=0$,
system-bath interaction $\lambda=1/2$ and $N_\mathrm{B}=14,16,18,20,22,24,28,32$.
}
\label{fig2d}
\end{center}
\end{figure}

\subsection{Relaxation to a stationary state}\label{section3c}

In this and the sections that follow, the random state approach with $H$ replaced by
$H_{\mathrm{B}}$ is used to construct the thermal equilibrium state of the bath, that is
\begin{eqnarray}
|\Phi(\beta)\rangle&=& \frac{e^{-\beta H_{\mathrm{B}}/2}|\Phi\rangle}{ \langle\Phi|e^{-\beta H_{\mathrm{B}}}|\Phi\rangle^{1/2} }
,
\label{s4k0z}
\end{eqnarray}
where $|\Phi\rangle$ denotes a random state of the bath only.

In Fig.~\ref{fig2b}, we present typical simulation results of the system energy and system entropy
as a function of time and for three different initial states and $\beta=1$.
These results illustrate that
\begin{enumerate}[(i)]
\item
a bath of $N_\mathrm{B}=20$ is sufficiently large to let the system relax to a stationary state,
\item
if the energy of the initial state of the system ($\langle H_\mathrm{S}(t=0)\rangle=-0.75$)
is much smaller than the thermal energy of the isolated system,
the system energy in the stationary state is smaller than the latter
and the entropy is a monotonically increasing function of time,
indicating that the system only gains energy from the bath, see Fig.~\ref{fig2b}(a),
\item
if the energy of the initial state of the system ($\langle H_\mathrm{S}(t=0)\rangle=-0.25$)
is close to the system energy $E_\mathrm{S}(\beta=1)=-0.223$ in the thermal equilibrium state,
the stationary state is (very) close to the thermal equilibrium state at $\beta=1$, Fig.~\ref{fig2b}(b),
\item
if the energy of the initial state of the system ($\langle H_\mathrm{S}(t=0)\rangle=+0.25$)
is larger than the thermal energy of the isolated system,
the system energy in the stationary state is larger than the latter and
the entropy is not a monotonic function of time,
indicating that the system not only releases energy into the bath but also gains energy from the bath, Fig.~\ref{fig2b}(c).
\end{enumerate}
Qualitatively, these conclusions are corroborated by the results at low temperature $\beta=5$, shown in Fig.~\ref{fig2c}.
Note that the data presented in Fig.~\ref{fig2c} have been obtained for a spin bath, the Hamiltonian of which is very different
from the one used to produce the data shown in Fig.~\ref{fig2b}.

Both the results presented in Fig.~\ref{fig2b} and Fig.~\ref{fig2c} strongly suggest that for finite spin baths ($N_\mathrm{B}\le34$),
the stationary state depends on the initial state of the system.
From statistical mechanics we may expect that the stationary state
will approach the thermal state of the isolated system as $N_\mathrm{B}\rightarrow\infty$ because
the system-bath interaction is weak.
We scrutinize this expectation by performing simulations for different $N_\mathrm{B}$.

In Fig.~\ref{fig2d}, we show how the system energy in the stationary state, i.e. the value of the system energy at the end of the simulation run,
changes with the number of spins $N_\mathrm{B}$ of the spin bath.
Finite-size scaling results for the integrable spin bath (Heisenberg antiferromagnet) Eq.~(\ref{s42a}) are presented in Fig.~\ref{fig2d}(a).
The least-square fitting of $a+bN_\mathrm{B}^{-1}$ (with $N_\mathrm{B}=14,16,18,20,22,24,28,32,34$) to the data yields $a\approx-0.21$ and $b\approx1.58$,
with a RMSE of about 0.0031 (data not shown).
On the other hand, the least-square fitting of $a+bN_\mathrm{B}^{-c}$ to the data yields $a\approx-0.20$, $b\approx2.34$ and $c\approx1.12$
with a RMSE of about 0.0026 and is shown as the solid line in Fig.~\ref{fig2d}(a).
Apparently, in the case of the integrable baths with up to $N_\mathrm{B}=34$ spins, it is not easy to tell whether
the system energy will converge to the correct one of the isolated system as $N_\mathrm{B}\rightarrow\infty$
but at least the data do not indicate otherwise.

In Fig.~\ref{fig2d}(b) we show the results of the same analysis except that we used
the fully connected random-coupling spin bath (Eq.~(\ref{s42b})) instead of the integrable bath.
The least-square fitting of $a+bN_{\mathrm{B}}^{-3/2}$ (with $N_\mathrm{B}=14,16,18,20,22,24,28,32$) to the data yields $a\approx-0.23$ and $b\approx6.10$,
with a RMSE of about 0.006  (data not shown).
On the other hand, the least-square fitting of $a+bN_{\mathrm{B}}^{-c}$ to the data yields $a\approx-0.22$, $b\approx12.87$ and $c\approx1.82$
with a RMSE of about 0.005 and is shown as the solid line in Fig.~\ref{fig2d}(b).
For both fits, the extrapolated system energies (the values of $a$) are in good agreement
with the thermal energy of the isolated system ($E_\mathrm{S}(\beta=1)=-0.223$).

As mentioned earlier, to establish whether the system has evolved to its thermal equilibrium state
with a temperature that is close to the bath temperature, it is necessary to consider the
expectation values of a complete set of system operators.
Equivalently, we may also determine the effective Hamiltonian that describes the final state of the simulation run.
From the data of the expectation values of the complete set of system operators
$\{ \mathbf{e}_i\;|\;i=0,\ldots,15 \}$ at the final time $t_\mathrm{final}$ of a simulation,
we can extract the effective Hamiltonian ${\widehat H}_{\mathrm{S}}$ from the ansatz
\begin{eqnarray}
\langle\Psi(t_\mathrm{final})|\mathbf{e}_i|\Psi(t_\mathrm{final})\rangle &=&
\frac{\mathbf{Tr}\;e^{-\beta {\widehat H}_{\mathrm{S}} } \mathbf{e}_i }{ \mathbf{Tr}\;e^{-\beta {\widehat H}_{\mathrm{S}}} }
 \quad,\quad i=0,\ldots,15
.
\label{term0}
\end{eqnarray}
This is most conveniently done by expanding $e^{-\beta {\widehat H}_{\mathrm{S}}}$
in terms of $\{ \mathbf{e}_i\;|\;i=0,\ldots,15\}$ and using the orthogonality of the $\mathbf{e}_i$'s.
Then,  $-\beta {\widehat H}_{\mathrm{S}}$ follows by numerically computing the logarithm of $e^{-\beta {\widehat H}_{\mathrm{S}}}$.

If the final state $|\Psi(t_\mathrm{final})\rangle$ is a thermal random state
we must have ${\widehat H}_{\mathrm{S}}\approx {H}_{\mathrm{S}}$.
For instance, from the simulation run (with $N_\mathrm{B}=20$) of which some data is shown in Fig.~\ref{fig2b}(b), we find that
$-\beta {\widehat H}_{\mathrm{S}}=-1.00\sigma_1^x \sigma_2^x -0.99\sigma_1^y \sigma_2^y -1.00\sigma_1^z \sigma_2^z + {\cal R}(0.006)$
where ${\cal R}(\epsilon)$ denotes the sum of all remaining system operators, their prefactors being smaller than $\epsilon$.
Thus, in this case, the final state is indeed very close to a thermal random state at $\beta=1$.
In contrast, the simulation run (with $N_\mathrm{B}=32$) and initial state
$|\uparrow\uparrow>\otimes|\Phi(\beta=1)\rangle$ (data of the system energy is shown in Fig.~\ref{fig2d}(b)), yields
$-\beta{\widehat H}_{\mathrm{S}}=-0.88\sigma_1^x \sigma_2^x -0.88\sigma_1^y \sigma_2^y -0.88\sigma_1^z \sigma_2^z + {\cal R}(0.001)$,
indicating that the system is in thermal equilibrium at $\beta=0.88$ instead of $\beta=1$, in
agreement with the observation that for this choice of initial state of the system, the finite size of the
bath affects the final state (see Fig.~\ref{fig2d}(b)).

\subsection{Thermalization: relation to earlier work}\label{section3d}

The numerical results presented above are, for all practical purposes, exact and provide additional
insight into the question whether a (classical or quantum) system coupled to a heat bath thermalizes or not.
In fact, there are only a very few rigorous results about the thermalization of a classical system in contact with a thermostat,
i.e. interacting with a large bath.
Bogolyubov proved that the density matrix of an ensemble of classical oscillators relaxes to the
thermal equilibrium state under rather general conditions~\cite{BOGU70,BOGU94,CHIR86,STRO07}.
In Appendix~\ref{OSC}, we present some illustrative, numerically exact simulation results
of a system consisting of one classical oscillator which interacts (harmonically) with a small bath of classical oscillators.
We show that for a suitable choice of model parameters, an ensemble of such an integrable model indeed thermalizes but a single trajectory does not.
Most remarkably, simulations have shown that the canonical distribution
of a subsystem, containing several particles, of a closed classical system of a ring of coupled harmonic
oscillators (integrable system) or magnetic moments (nonintegrable system)
follows directly from the solution of the time-reversible Newtonian equation
of motion in which the total energy is strictly conserved~\cite{JIN13b}.
Regarding rigorous results for open quantum systems, the situation is not much brighter, even though to begin with,
the quantum theoretical description is statistical in nature~\cite{BREU02}.
For one harmonic oscillator weakly interacting with a bath of oscillators, the closed-form
equation of motion of the position and momentum of the system oscillator can be used
to prove analytically that the system energy relaxes to its thermal equilibrium value
if the number of bath oscillators is taken to be infinite and a suitable choice of model parameters is made~\cite{BREU02}.
We do not know of a rigorous proof that also the density matrix of the system relaxes to the thermal equilibrium (Gibbs) state, as in the case
of the classical system~\cite{BOGU94,CHIR86,STRO07}.
As discussed above, under certain conditions, our numerical results for a two-spin system coupled to a finite spin bath
show that the two-spin system thermalizes.

In the last two decades, theoretical research on the thermalization process of quantum systems interacting with a bath
has focused on the situation in which the whole system (S+B) is described by a single pure state.
The so-called ``canonical typicality''~\cite{GEMM03,GOLD06,POPE06,REIM07,BART09}
states that the reduced density matrix of a system is canonical if the state of the whole
system is one of the overwhelming majority of wave functions in the subspace corresponding to the energy interval encompassed by the
microcanonical ensemble, namely a random state of the confined Hilbert space.
Conceptually, canonical typically is very similar to the random-state method~\cite{IITA97,IITA97b,HAMS00,GELM03,GELM04,SUGI12,SUGI13,STEI14b}
employed in this paper.

Recently, a testable theoretical result regarding
the time scale of the relaxation process of closed quantum system has been obtained~\cite{REIM15}.
More specifically, a closed form expression for the relaxation process in terms of the function,
$F(t)=D\left(|\phi(t)|^2-1/D\right)/(D-1)$ where $\phi(t)$ is the Fourier transform of the
spectral density of the whole system was derived~\cite{REIM15}.
For special choices of the spectral density, this formula predicts a very short timescale of the relaxation
process of the order of the Boltzmann time ($h/k_BT$), in agreement with other results~\cite{GOLD15}.

\begin{figure}[ht]
\begin{center}
\includegraphics[width=0.33\hsize]{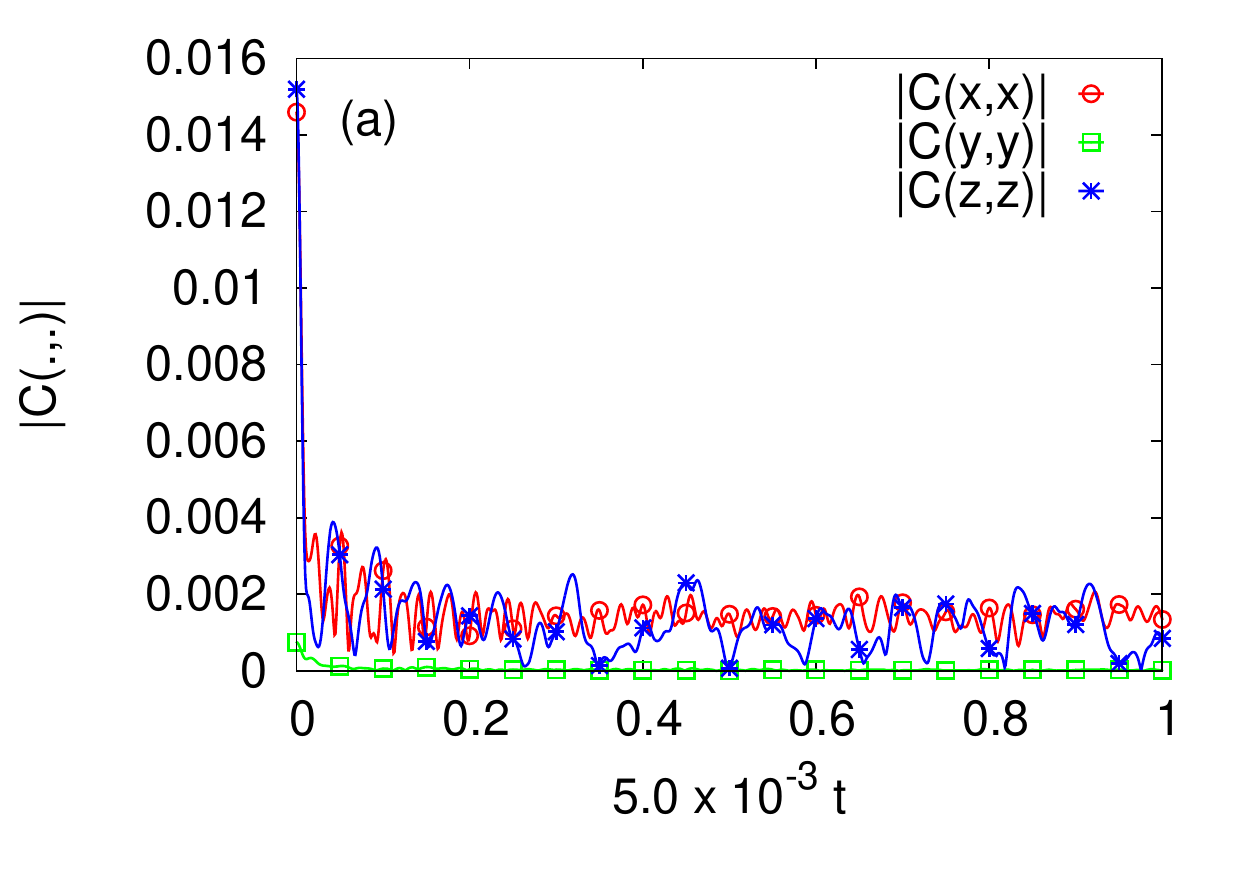}
\includegraphics[width=0.33\hsize]{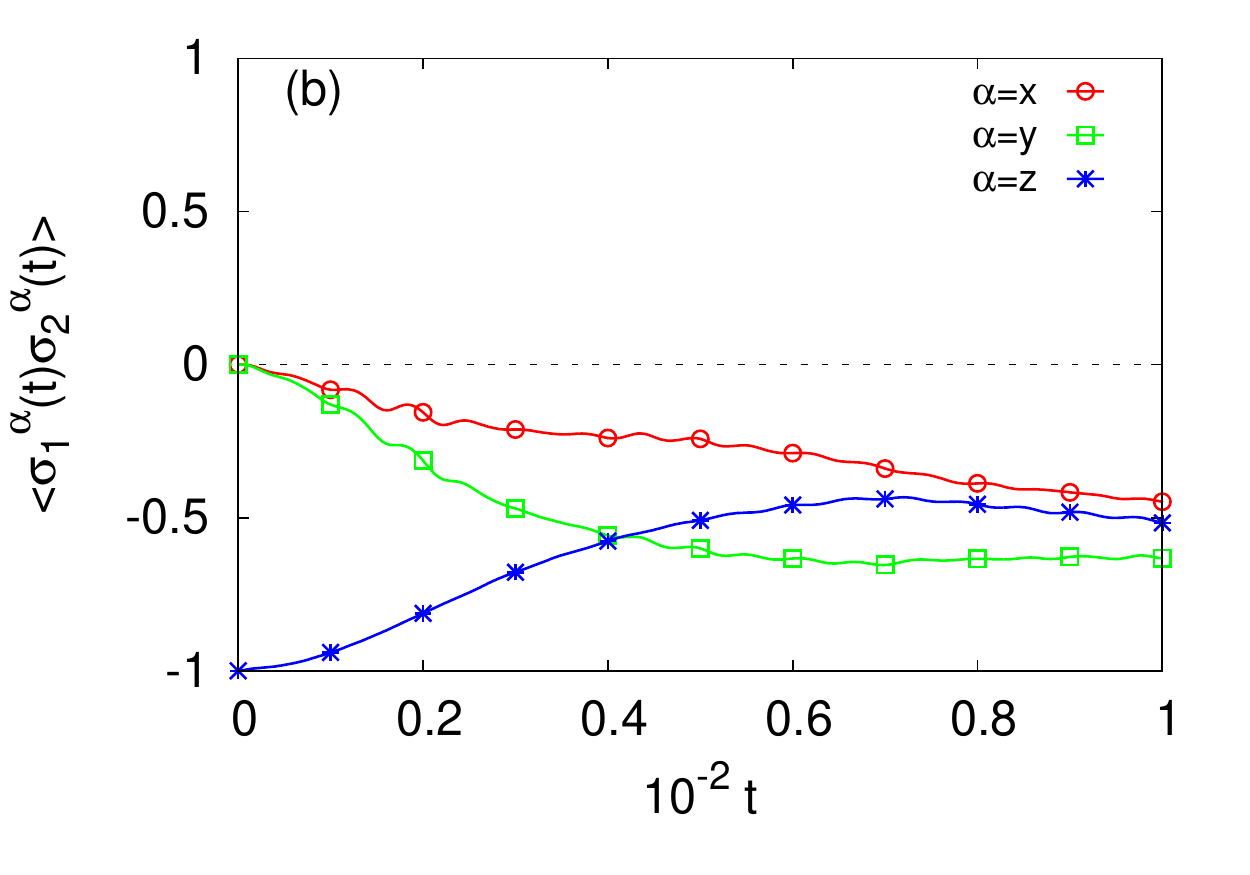}
\includegraphics[width=0.33\hsize]{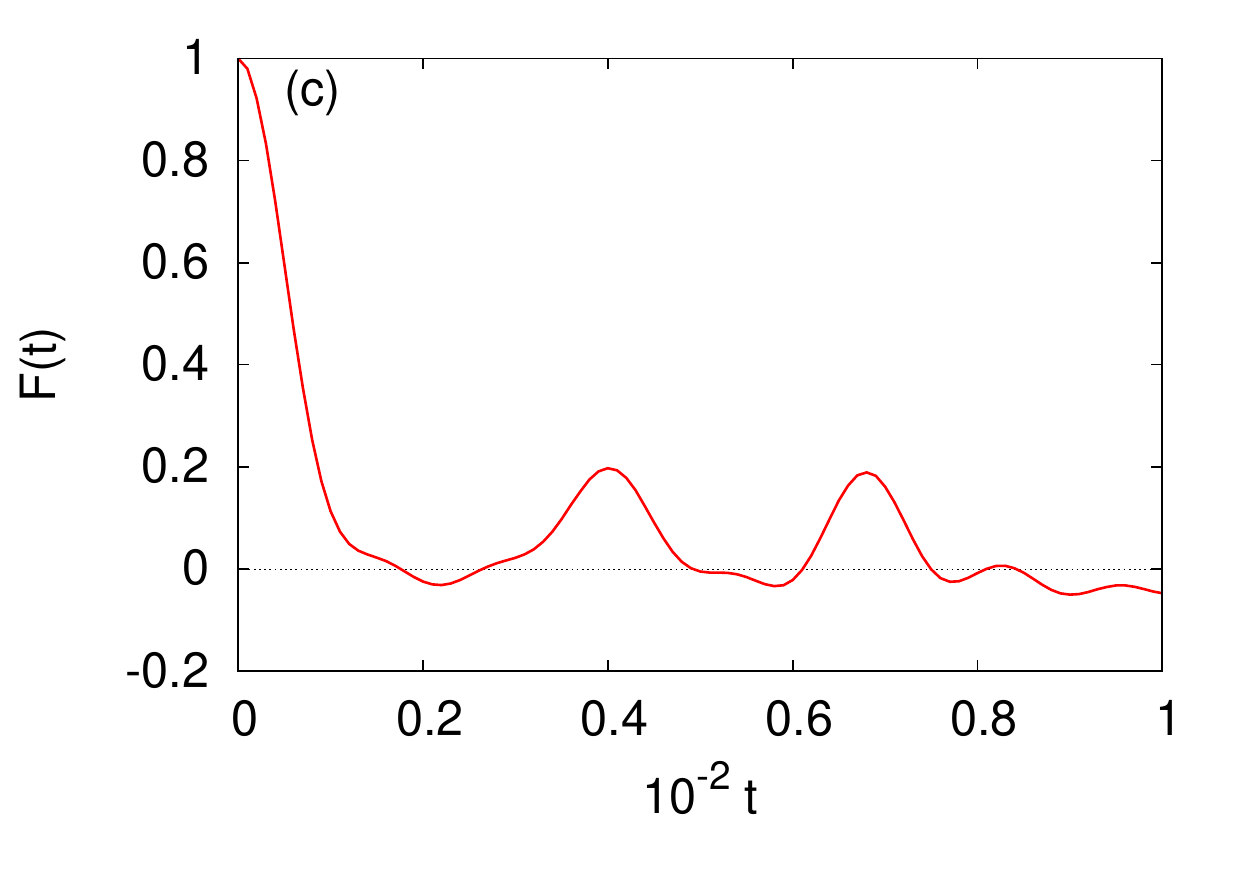}
\caption{(color online) %
(a) The absolute values of three of the nine bath-operator correlations Eq.~(\ref{term1}) as obtained by solving the TDSE
for a bath of $N_\mathrm{B}=20$ spins with the initial state $|\uparrow\downarrow>\otimes|\Phi(\beta=5)\rangle$.
The bath-operator correlations that have absolute values that are too small to be seen on the scale
of the plot have been omitted.
The Hamiltonian of the bath is given by Eq.~(\ref{s42a}) with $K=1/2$ and $h^x_\mathrm{B}=h^z_\mathrm{B}=0$;
(b) Correspondig two-spin averages;
(c) $F(t)$ as obtained from the $20$ lowest eigenvalues of the bath Hamiltonian.
}
\label{figapp3}
\end{center}
\end{figure}

Although it is unlikely that the initial states, Hamiltonians and observables that we consider in this
paper satisfy the conditions to derive the testable theoretical result mentioned earlier~\cite{REIM15},
it may nevertheless be instructive to inquire whether there is a relation between
$F(t)$ and the relaxation time of certain operators.
The most likely candidates for operators that show fast decay are the bath-operators
\begin{eqnarray}
B_x&=&-J^x_{n}\sigma^x_{n}-J^x_{m}\sigma^x_{m} 
\nonumber \\
B_y&=&-J^y_{n}\sigma^y_{n}-J^y_{m}\sigma^y_{m} 
\nonumber \\
B_z&=&-J^z_{n}\sigma^z_{n}-J^z_{m}\sigma^z_{m} 
,
\label{term1}
\end{eqnarray}
which define the coupling between the system and the bath, see Eq.~(\ref{s43b}).
In Fig.~\ref{figapp3})(a,b) we present simulation results for $C(\alpha,\beta)=\langle B_\alpha(t) B_\beta(0)\rangle$ for $\alpha,\beta=x,y,z$
and two-spin expectation values $\langle\sigma_1^\alpha(t) \sigma_2^\alpha(t)\rangle$ for $\alpha=x,y,z$
and $F(t)$ as obtained from the $20$ lowest eigenvalues of the bath Hamiltonian.
Although the bath is prepared at fairly low temperature ($\beta=5$), the bath operator correlations decay very fast,
significantly faster than the two-spin averages while the decay of $F(t)$ is, on these
time-scales neither fast or slow.
Hence, not entirely unexpected, $F(t)$ does not describe the relaxation of the bath-operators or two-spin correlations
because the conditions required to describe the relaxation process in terms of $F(t)$ may not apply in the case at hand.

\subsection{Summary}\label{section3e}
Our simulation data show that due to the interaction with the spin bath, the system relaxes to a stationary state.
For finite $N_\mathrm{B}$, this stationary state depends on the initial state of the system.
If the difference between the initial system energy and the thermal energy of the isolated system
is small, the system relaxes to its thermal state.
Otherwise this difference decreases as the number of spins in the bath $N_\mathrm{B}$ increases.

\section{Quantum Master Equation}\label{section4}

\begin{figure}[t]
\begin{center}
\includegraphics[width=0.33\hsize]{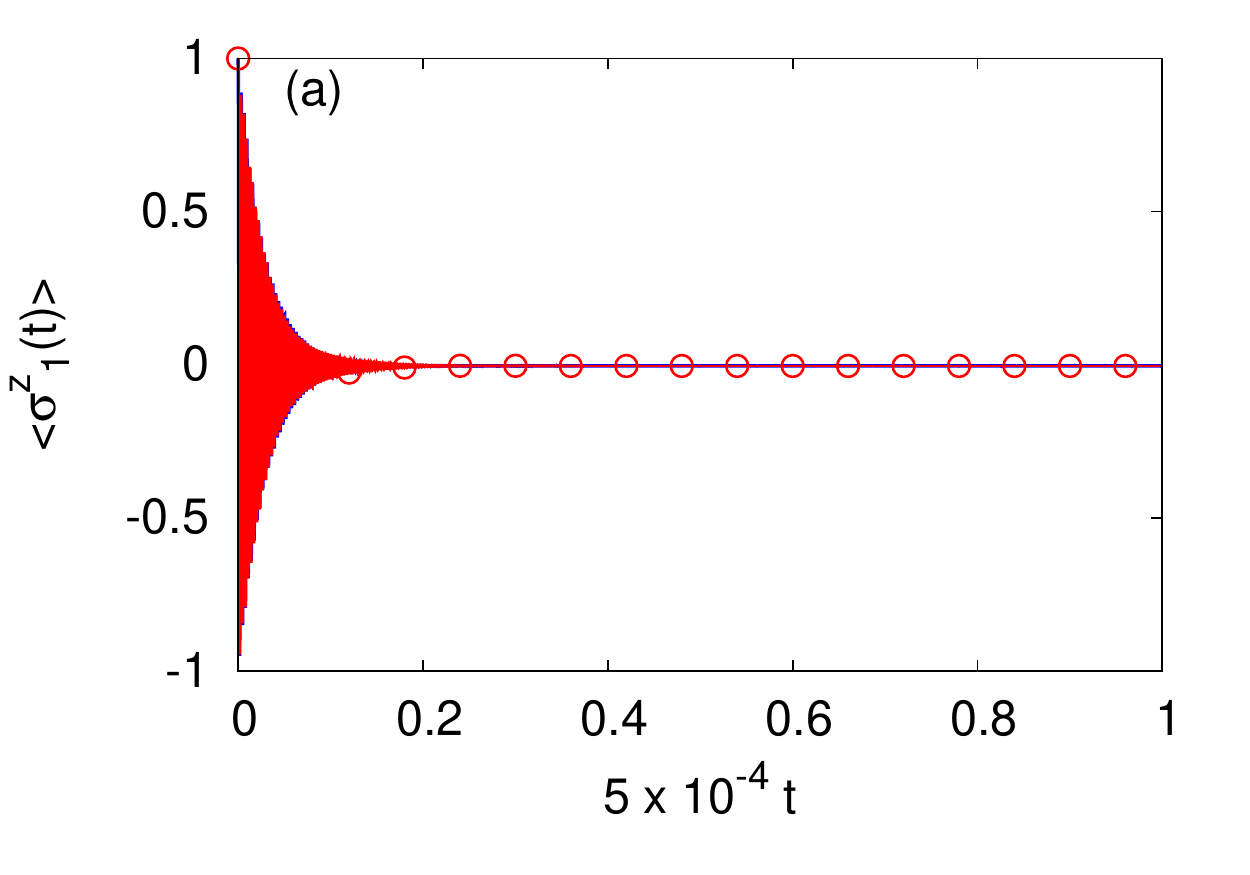}
\includegraphics[width=0.33\hsize]{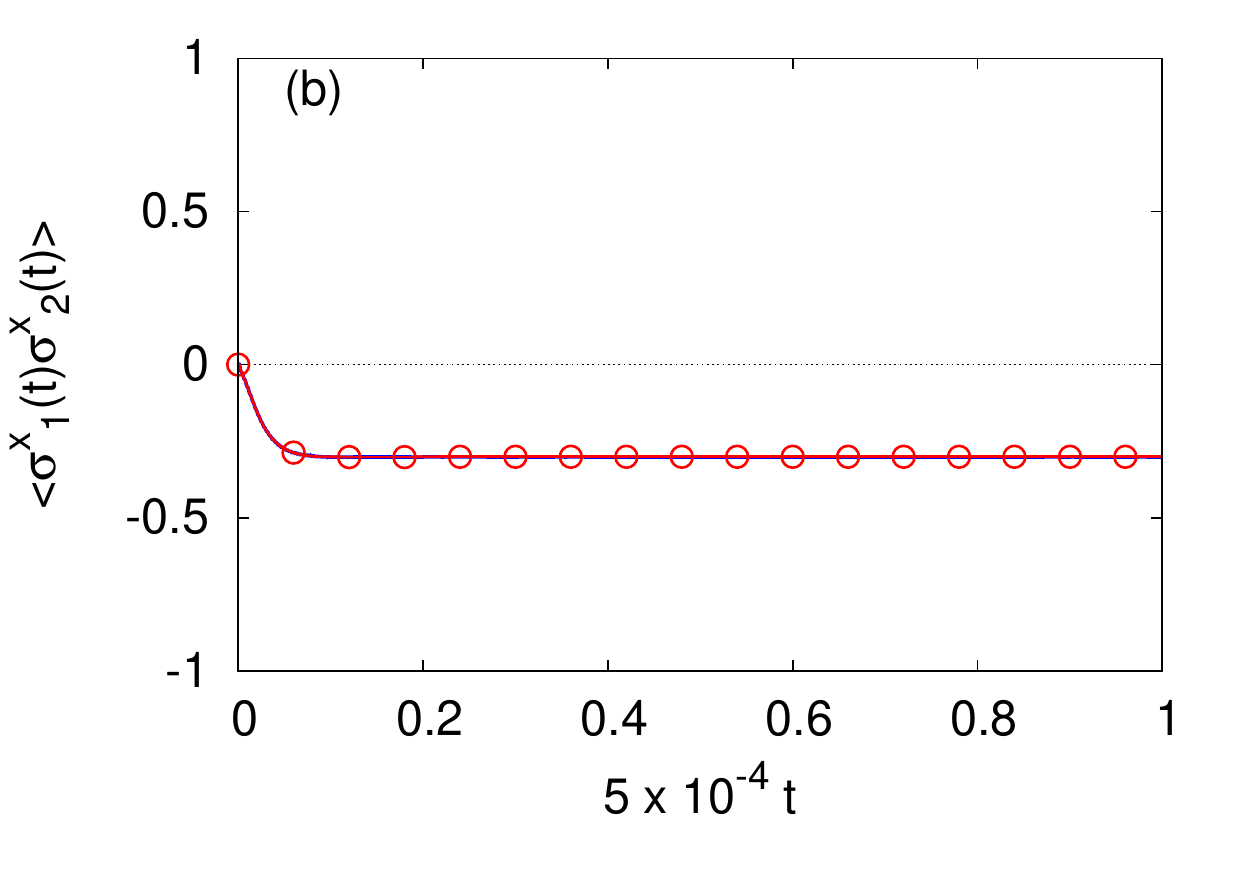}
\includegraphics[width=0.33\hsize]{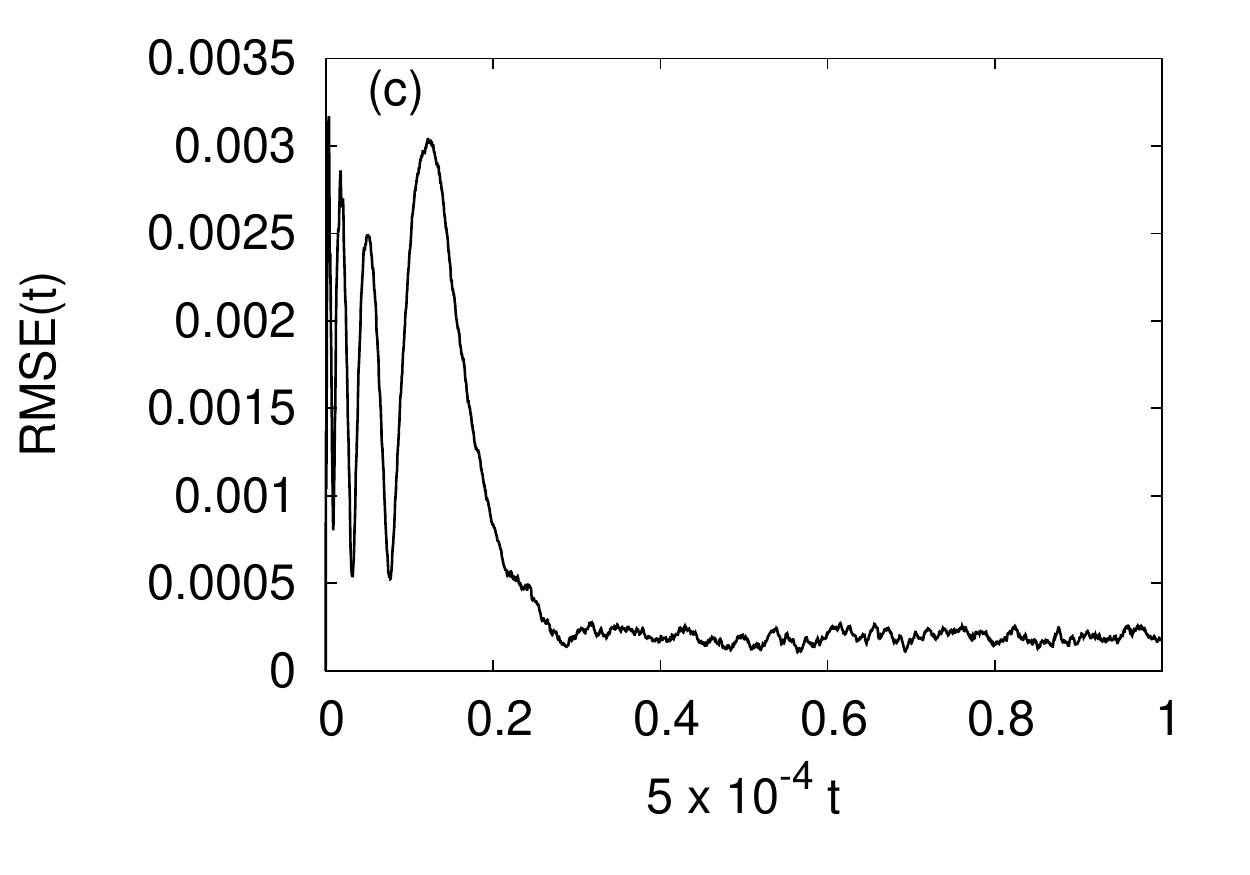}
\includegraphics[width=0.33\hsize]{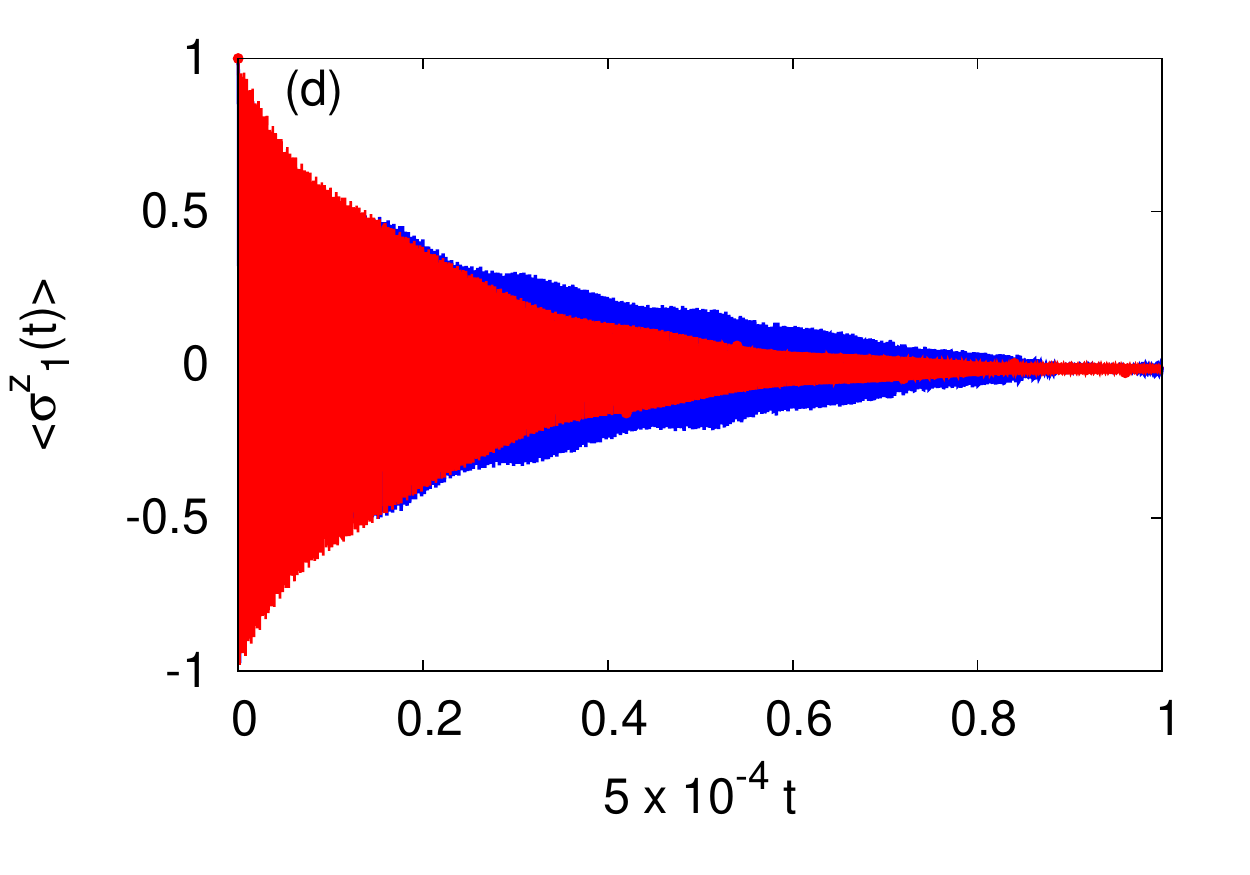}
\includegraphics[width=0.33\hsize]{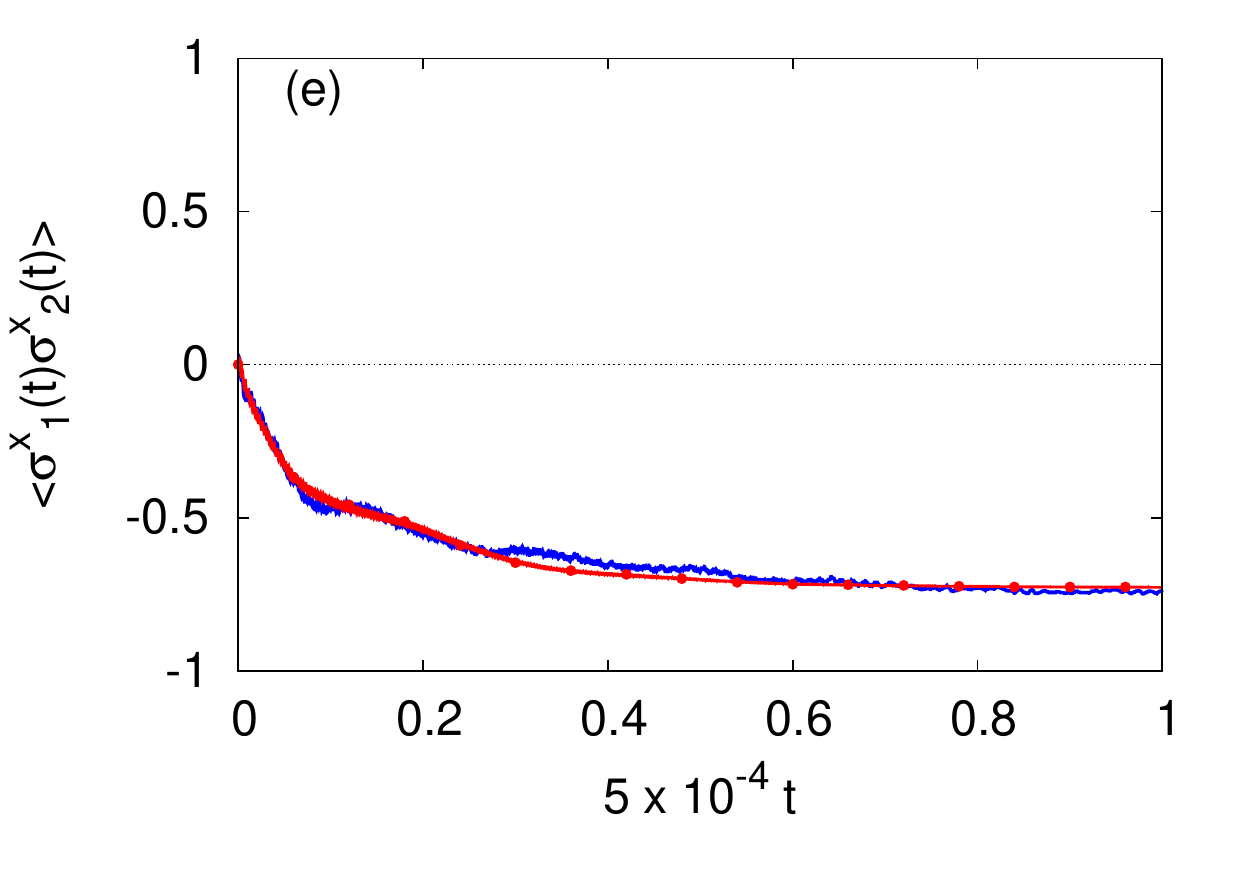}
\includegraphics[width=0.33\hsize]{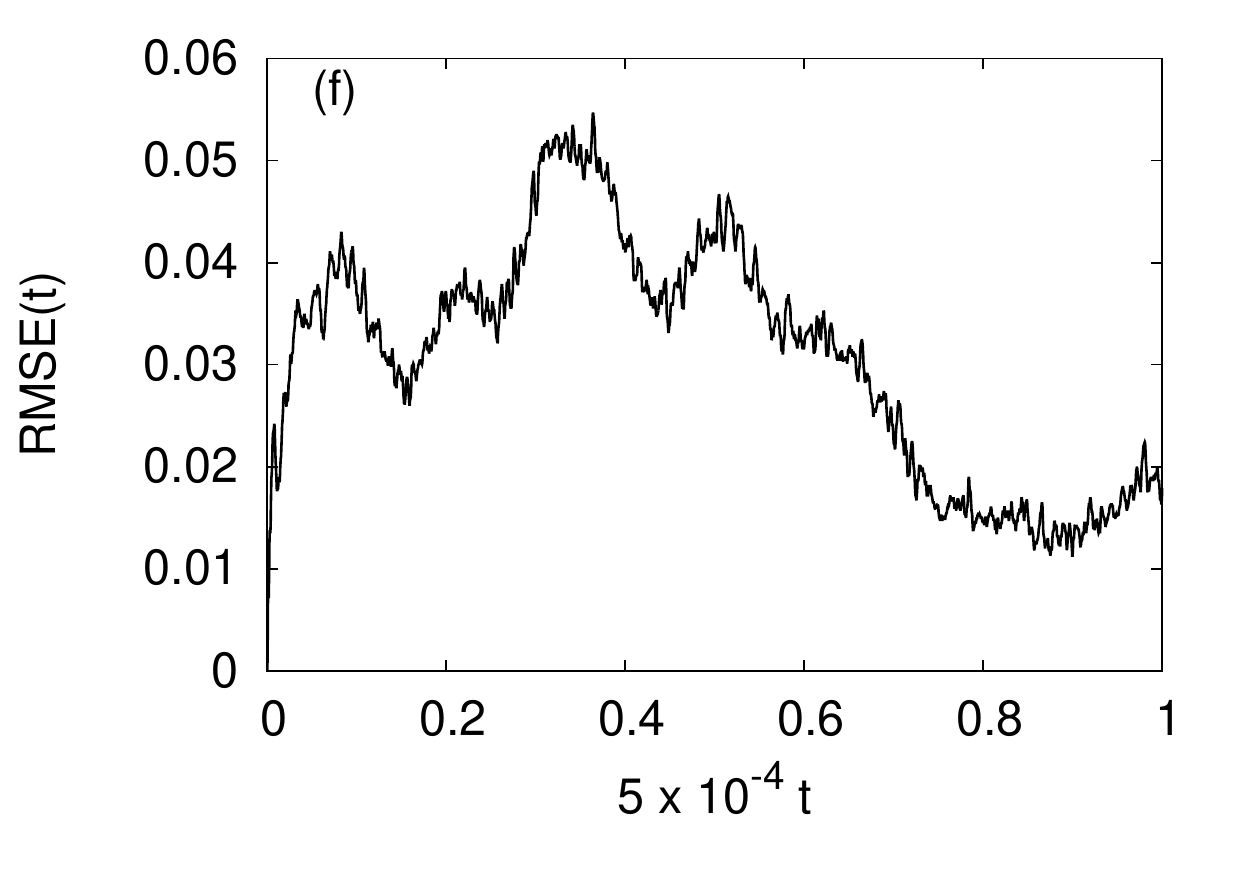}
\caption{(color online) %
A single-spin average (a,d) and a two-spin correlation (b,e) as obtained by solving the TDSE (solid lines)
and the QMEQ (circles) with $e^{\tau\mathbf{A}}$ and $\mathbf{B}$ obtained by least-square fitting to the TDSE data.
The other 13 expectation values show similar agreement and are therefore not shown.
Also show is the RMSE error (c,f) defined by Eq.~(\ref{rmse}).
The system Hamiltonian is given by Eq.~(\ref{s41})
with $J_{\bot}=J_{\parallel}=1/4$ (antiferromagnetic Heisenberg model).
The Hamiltonian of the system-bath interaction and spin bath
are given by Eq.~(\ref{s43b}) and Eq.~(\ref{s42a}), respectively.
The number of bath spins is $N_\mathrm{B}=28$, $K=1$, $h^x_\mathrm{B}=h^z_\mathrm{B}=1/4$.
The system-bath interaction strength is $\lambda=0.25$.
The initial state is $|\uparrow \downarrow\rangle\otimes|\Phi(\beta)\rangle$.
(a,b,c) $\beta=1$;
(d,e,f) $\beta=5$.
}
\label{fig5}
\end{center}
\end{figure}

From the analysis presented in section~\ref{section3}, it follows that
for spin baths of moderate size, the system evolves to a stationary state,
which, depending on the initial state of the system, is close to the thermal state
of the isolated two-spin system.
In this section, we scrutinize how well a Markovian QMEQ of the two-spin system
describes the exact time evolution of the two-spin system coupled to a spin bath.

In general, a Markovian QMEQ can be written as~\cite{BREU02}
\begin{eqnarray}
\frac{\partial \widetilde{\bm{\rho}}(t)}{\partial t}= \mathbf{A}\widetilde{\bm{\rho}}(t) + \mathbf{b}
,
\label{GBE}
\end{eqnarray}
where $\widetilde{\bm{\rho}}$ represents the elements of the reduced density matrix, reshaped as a vector
and the matrix $\mathbf{A}$ and vector $\mathbf{b}$ do not depend on time.
Specifically, for the problem at hand, $\widetilde{\bm{\rho}}(t)=(\widetilde\rho_0(t),\ldots,\widetilde\rho_{15}(t))^T$
with $\widetilde\rho_0(t)=1$ for all $t$,
see Eqs.~(\ref{s48}) and (\ref{s49}).
The formal solution of Eq.~(\ref{GBE}) for a finite time step $\tau$ reads
\begin{eqnarray}
\widetilde{\bm{\rho}}(t+\tau) &=&e^{\tau\mathbf{A}}\widetilde{\bm{\rho}}(t) + \int_{0}^{\tau} e^{(\tau-u)\mathbf{A}}\mathbf{b}\;du
=e^{\tau\mathbf{A}}\widetilde{\bm{\rho}}(t) + \mathbf{B}
,
\label{QMEQ7}
\end{eqnarray}
where
\begin{eqnarray}
\mathbf{B}=\int_{0}^{\tau} e^{(\tau-u)\mathbf{A}}\mathbf{b}\;du
,
\label{QMEQ7a}
\end{eqnarray}
does not depend on the time $t$.

As explained in section~\ref{section3}, solving the TDSE yields the data set
$\Upsilon\equiv\{\rho_i(t)\;|\;i=0,\ldots,15\;;\, t=0,\tau,\ldots,T=m\tau\}$.
This data set can be used as input to a least-square procedure that
determines the matrix $e^{\tau\mathbf{A}}$ and vector $\mathbf{B}$ by minimizing
the root-mean-square-error between the data of the set $\Upsilon$
and the data of the set
$\widetilde\Upsilon\equiv\{\widetilde\rho_i(t)\;|\;i=0,\ldots,15\;;\, t=0,\tau,\ldots,T=m\tau\}$
obtained by solving Eq.~(\ref{QMEQ7}).
A detailed account of this procedure is given in Ref.~\onlinecite{ZHAO16}
and will therefore not be repeated here.
We quantify the difference of the reconstructed data, i.e. the solution of the ``best'' approximation
in terms of the QMEQ, and the original data obtained by solving the TDSE by
the root-mean-square-error (RMSE)
\begin{equation}
\mathrm{RMSE}(t)=\sqrt{\frac{1}{15} \sum_{i=1}^{15} \left(\rho_{i}(t) -\widetilde\rho_{i}(t)\right)^2 }
.
\label{rmse}
\end{equation}
We also check if the approximate density matrix of the system, $\widetilde{\bm{\rho}}(t)$,
is non-negative definite.

In Fig.~\ref{fig5} we present some representative results of fitting
the QMEQ Eq.~(\ref{QMEQ7}) to the TDSE data.
We only show one single-spin average ($\langle\Psi(t)|\sigma^z_1|\Psi(t)\rangle$)
and one two-spin average ($\langle\Psi(t)|\sigma^x_1\sigma^x_2|\Psi(t)\rangle$)
because the other averages show similar good agreement.
For $\beta=1$ the QMEQ describes the TDSE data very well.
For $\beta=5$ the agreement is excellent for the two-spin averages but
apparently, quantitatively, the QMEQ does not describe the decay of the single-spin averages very well.
It overestimates the relaxation time~\cite{ZHAO16}.
The overall, excellent agreement is characteristic for the many data sets that we have analysed.
Therefore, we do not present additional figures.

Although from Fig.~\ref{fig5} the agreement between the ``exact'' TDSE solution for the whole system and the
``fitted'' QMEQ for the two-spin system looks very good, a more detailed analysis reveals that
occasionally, the reduced density matrix obtained by iterating Eq.~(\ref{QMEQ7}) has one negative eigenvalue.
For the data shown in Fig.~(\ref{fig5})(a,b,c) this happens at $t=0.6$ where one eigenvalue
of the reduced density matrix is equal to $-0.00046$ and the three others are positive.
For the data shown in Fig.~(\ref{fig5})(d,e,f) this happens 12 times
in the interval $[0.6,11.4]$, the smallest eigenvalues being larger than $-0.003$ while the other three eigenvalues are positive.
In the course of the project, many data sets (not shown) for different $N_\mathrm{B}$, $\beta$, and model parameters
have been generated. Some of these data sets yield a fitted QMEQ with a density matrix
that has one small negative eigenvalue for a few particular times $t$.
We have not been able to detect any systematics in when and why such small negative eigenvalues occur.

The fact that Markovian QMEQs may lead to density matrices that are not always non-negative definite is well-known~\cite{SUAR92,PECH94}.
In particular, when the characteristic time scale of the system is comparable to that of the thermal bath,
the effect of the finite correlation time of the thermal bath may become important.
Then, the Markovian approximation used to derive the QMEQ may no longer be adequate and
it becomes necessary to account for non-Markovian aspects and treat the
initial condition correctly~\cite{SASS90,GASP99,WEIS99,BREU02,TANI06,BREU06,MORI08,SAEK08,UCHI09,MORI14,CHEN15}.
One exception is the Lindblad QMEQ, which is also of the form Eq.~(\ref{GBE}) and therefore Markovian~\cite{LIND76,BREU02}.
By construction the Lindblad QMEQ preserves positivity (a non-negative definite density matrix)
during the time evolution~\cite{LIND76}.

In contrast to the common procedure of deriving a Markovian QMEQ,
the least-square procedure used to extract the matrix $e^{\tau\mathbf{A}}$ and vector $\mathbf{B}$ from
the TDSE data does not rely on perturbation theory: it simply
finds the QMEQ Eq.~(\ref{QMEQ7}) that fits the TDSE data best (in the least-square sense).
There is no a-priori reason why this procedure should yield a non-negative definite density matrix
but apparently, with some exceptions, it does.
However, in all these exceptions the violation non-negative definiteness is rather small
and may be due to the use of the random-state technology, the finite time step used to fit the QMEQ etc.

\subsection{Relation to Markovian quantum master equations}\label{section4a}

\begin{figure}[t]
\begin{center}
\includegraphics[width=0.33\hsize]{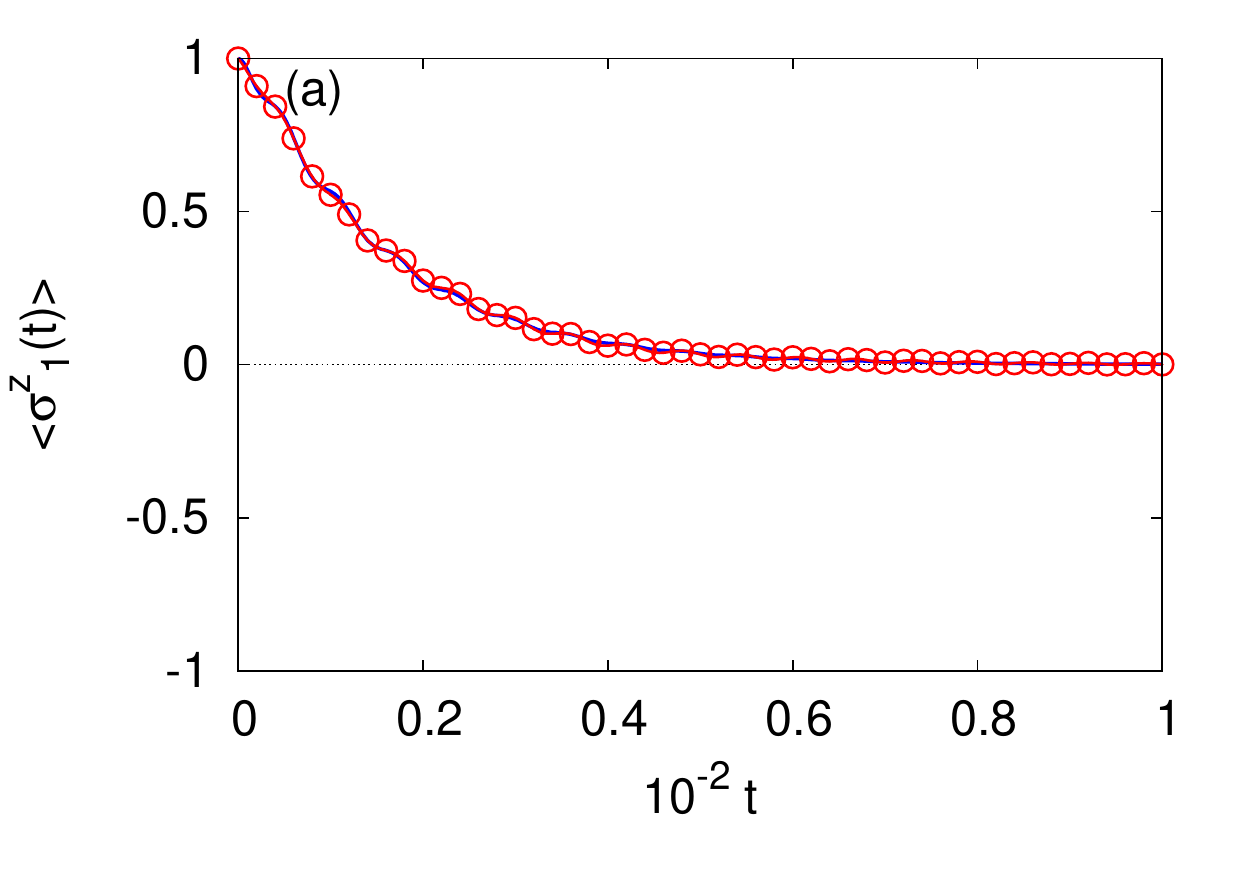}
\includegraphics[width=0.33\hsize]{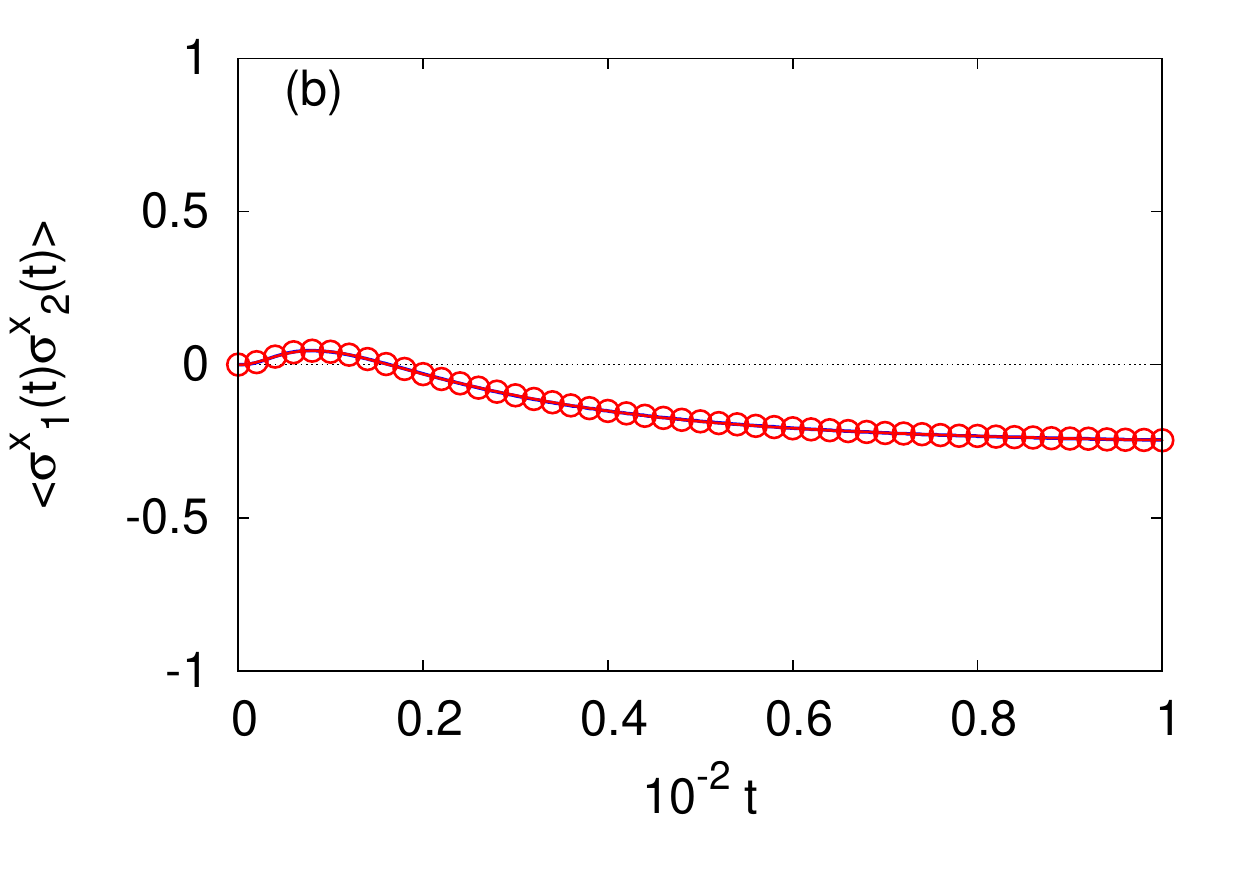}
\includegraphics[width=0.33\hsize]{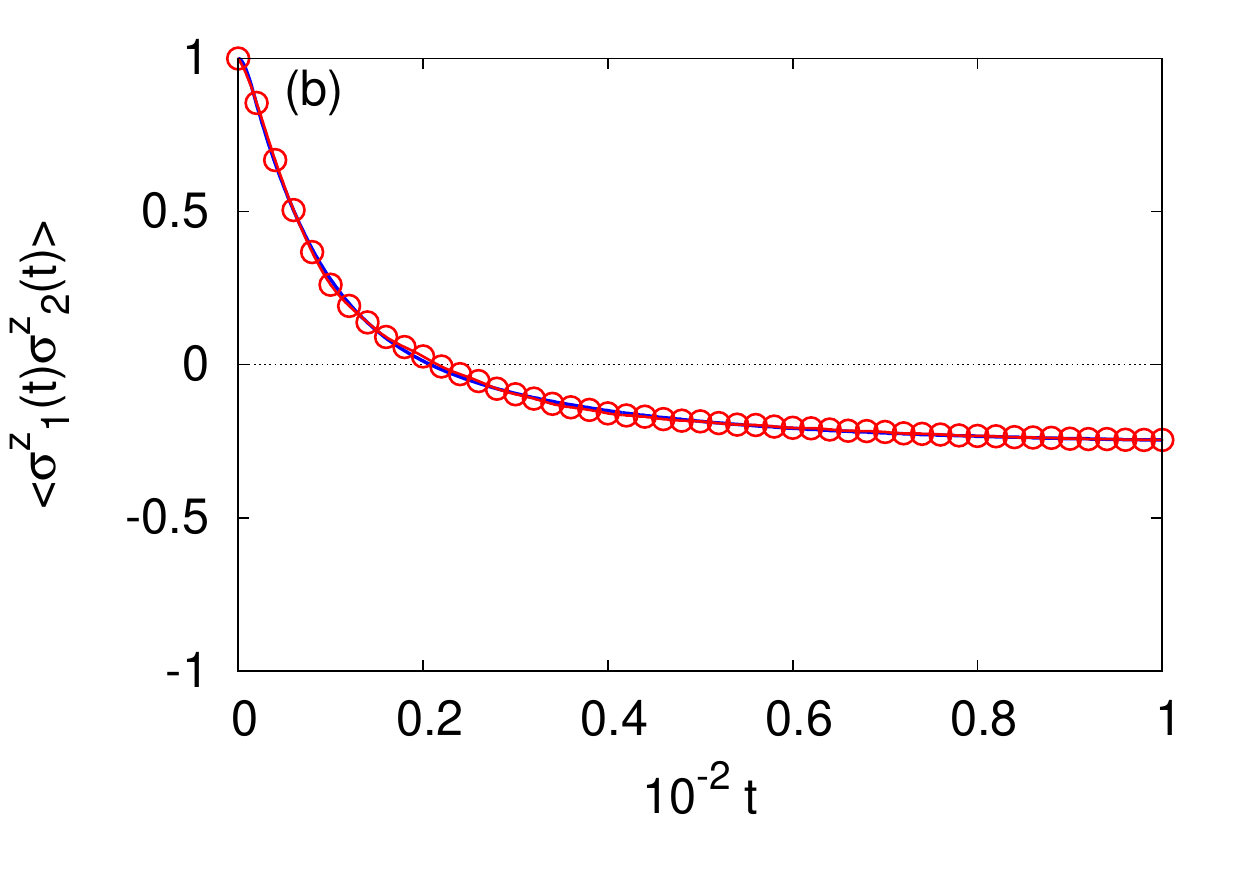}
\caption{(color online) %
A single-spin average (a) and two-spin correlations (b,c) as obtained by solving the TDSE (solid lines)
and the QMEQ (circles) with $e^{\tau\mathbf{A}}$ and $\mathbf{B}$ obtained by least-square fitting to the TDSE data.
The other expectation values show similar agreement and are therefore not shown.
The system Hamiltonian is given by Eq.~(\ref{s41})
with $J_{\bot}=J_{\parallel}=1/4$ (antiferromagnetic Heisenberg model).
The Hamiltonian of the system-bath interaction and spin bath
are given by Eq.~(\ref{s43b}) and the spin-glass model Eq.~(\ref{s42b}), respectively.
The number of bath spins is $N_\mathrm{B}=32$, $K=1/2$ and $h^x_\mathrm{B}=h^z_\mathrm{B}=0$.
The system-bath interaction strength is $\lambda=0.5$.
The initial state is $|\uparrow \uparrow\rangle\otimes|\Phi(\beta=1)\rangle$.
}
\label{fig6}
\end{center}
\end{figure}

It is now of interest to relate the Markovian description Eq.~(\ref{QMEQ7}) to the standard theory
of the Markovian QMEQ~\cite{BREU02}.
In the following, we closely follow Ref.~\onlinecite{BREU02} (chapter 3).

In this paper, we only consider initial states of the whole system that can be represented as
$\rho(0)=\rho_{\mathrm{S}}(0)\otimes \rho_{\mathrm{B}}(0)$, i.e. as a state in which the system and bath degrees of freedom are uncorrelated.
Note that as a result of the unitary time evolution of the whole
system we have $\rho(t)\not=\rho_{\mathrm{S}}(t)\otimes \rho_{\mathrm{B}}(t)$ for $t>0$,
except in the uninteresting case where there is no interaction between system and bath.
Hence, the $4\times4$ density matrix $\rho_{\mathrm{S}}(\tau)$ can be written as~\cite{BREU02}
\begin{eqnarray}
\rho_{\mathrm{S}}(\tau)=V(\tau)\rho_{\mathrm{S}}(0)&=&\mathbf{Tr}_{\mathrm{B}} e^{-i\tau H}\; \rho_{\mathrm{S}}(0)\otimes \rho_{\mathrm{B}}(0)\; e^{+i\tau H}
,
\label{BREU0}
\end{eqnarray}
where $\rho_{\mathrm{B}}(0)$ is the density matrix of the bath at time $t=0$, not necessarily the thermal equilibrium state.
Writing $\rho_{\mathrm{B}}(0)$ in terms of its (non-negative) eigenvalues $\mu_i$ and eigenvectors $|\varphi_j\rangle$, we have~\cite{BREU02}
\begin{eqnarray}
\rho_{\mathrm{S}}(\tau)&=&V(\tau)\rho_{\mathrm{S}}(0)=\sum_{i,j=1}^{D_{\mathbf{B}}}
\mathbf{W}_{i,j}^{\phantom{\dagger}}(\tau) \rho_{\mathrm{S}}(0) \mathbf{W}_{i,j}^{\dagger}(\tau)
,
\label{BREU1}
\end{eqnarray}
where the $(D_{\mathbf{B}})^2$, $4\times4$ matrices $\mathbf{W}_{i,j}(\tau)$ are given by
\begin{eqnarray}
\mathbf{W}_{i,j}(\tau)&=& \sqrt{\mu_j}\langle \varphi_i|e^{-i\tau H}|\varphi_j\rangle
=\frac{1}{4}\sum_{k=0}^{15} \mathbf{e}_{k} \big[\mathbf{Tr}_{\mathrm{B}} \mathbf{e}_{k}\mathbf{W}_{i,j}(\tau)\big]
.
\label{BREU2}
\end{eqnarray}
The last equality in Eq.~(\ref{BREU2}) follows from writing the $4\times4$ matrix $\mathbf{W}_{i,j}(\tau)$ in terms
of the basis vectors $\{\mathbf{e}_0,\ldots,\mathbf{e}_{15} \}$ introduced in section~\ref{section3a}.
Combining Eq.~(\ref{BREU1}) and Eq.~(\ref{BREU2}), we have
\begin{eqnarray}
\rho_{\mathrm{S}}(\tau)&=&V(\tau)\rho_{\mathrm{S}}(0)=\sum_{k,l=0}^{15} \mathrm{c}_{k,l}(\tau) \mathbf{e}_k\rho_{\mathrm{S}}(0)\mathbf{e}_l
,
\label{BREU3}
\end{eqnarray}
where the matrix $\mathrm{c}(\tau)$ with elements
\begin{eqnarray}
\mathrm{c}_{k,l}(\tau)&=& \frac{1}{16}\sum_{i,j=1}^{D_{\mathbf{B}}} \big[\mathbf{Tr}_{\mathrm{B}} \mathbf{e}_{k}\mathbf{W}_{i,j}(\tau)\big]
\big[\mathbf{Tr}_{\mathrm{B}} \mathbf{e}_{l}\mathbf{W}_{i,j}(\tau)\big]^\dagger
,
\label{BREU4}
\end{eqnarray}
is Hermitian and non-negative~\cite{BREU02}.
Using the expansion
$\rho_{\mathrm{S}}(\tau)=4^{-1}\sum_{i=0}^{15} \rho_i(\tau) \mathbf{e}_i$, see Eq.~(\ref{s48}),
and $\mathbf{Tr}_{\mathrm{S}} \mathbf{e}_k\;\mathbf{e}_l=4\delta_{k,l}$,
we obtain
\begin{eqnarray}
\rho_k(\tau)&=&\frac{1}{4}
\sum_{i,j,l=0}^{15} \mathrm{c}_{i,j}(\tau) \rho_l(0) \mathbf{Tr}_{\mathrm{S}} \mathbf{e}_k\mathbf{e}_i\mathbf{e}_l\mathbf{e}_j
=\sum_{i,j=0}^{15} \mathrm{M}_{k,l}(\tau) \rho_l(0)
,
\label{BREU5}
\end{eqnarray}
where
\begin{eqnarray}
\mathrm{M}_{k,l}(\tau)&=&\frac{1}{4}
\sum_{i,j=0}^{15} \mathrm{c}_{i,j}(\tau) \mathbf{Tr}_{\mathrm{S}} \mathbf{e}_k\mathbf{e}_i\mathbf{e}_l\mathbf{e}_j
\equiv \sum_{i,j=0}^{15}  \mathrm{X}_{k,l;i,j} \mathrm{c}_{i,j}(\tau)
\quad,\quad  \mathrm{X}_{k,l;i,j}=\frac{1}{4} \mathbf{Tr}_{\mathrm{S}} \mathbf{e}_k\mathbf{e}_i\mathbf{e}_l\mathbf{e}_j
.
\label{BREU6}
\end{eqnarray}
Regarding the pairs $(i,j)$ and $(k,l)$ as single indices, Eq.~(\ref{BREU6}) takes the form of a linear set of equations.
From the numerical calculation of the matrix elements $\mathrm{X}_{k,l;i,j}$, it follows that the matrix $\mathrm{X}$ is invertible.
The matrix $\mathrm{X}$ relates the elements of the
$4\times4$ density matrix $\rho_{\mathrm{S}}(\tau)$ in representation Eq.~(\ref{BREU3})
to the expectation values $\rho_k(\tau)= \mathbf{Tr}_{\mathrm{S}} \rho_{\mathrm{S}}(\tau)\;\mathbf{e}_k $
for $k=0,\ldots,15$ and vice versa.

The formal relations Eq.~(\ref{BREU0}) -- (\ref{BREU6}) hold for any value of $\tau$.
In particular, we have
$\rho_{\mathrm{S}}(t)=V(t)\rho_{\mathrm{S}}(0)$ and
$\rho_{\mathrm{S}}(t+\tau)=V(t+\tau)\rho_{\mathrm{S}}(0)$.
The assumption of Markovian behavior is often formalized by requiring that $V(t+\tau)=V(\tau)V(t)$ for $t,\tau\ge0$,
i.e. $V(t)$ satisfies the semigroup property~\cite{BREU02}.
Then, we have $\rho_{\mathrm{S}}(t+\tau)=V(\tau)\rho_{\mathrm{S}}(t)$ and Eq.~(\ref{BREU5}) generalizes to
\begin{eqnarray}
\rho_k(t+\tau)&=&\sum_{i,j=0}^{15} \mathrm{M}_{k,l}(\tau) \rho_l(t)
.
\label{BREU5b}
\end{eqnarray}

Now, we are in the position to relate the matrix $e^{\tau\mathbf{A}}$ and vector $\mathbf{B}$
that we obtain from least-square fitting to the TDSE data to the matrix $M(\tau)$.
To this end, we rewrite Eq.~(\ref{QMEQ7}) as
\begin{eqnarray}
\left(\begin{array}{r}
\widetilde\rho_0(t+\tau)\\
\widetilde\rho_1(t+\tau)\\
\ldots\\
\widetilde\rho_{15}(t+\tau)
\end{array}\right)
&=&
\left(\begin{array}{cccc}
1 & 0 &\ldots& 0\\
B_1 & (e^{\tau\mathbf{A}})_{1,1} & \ldots & (e^{\tau\mathbf{A}})_{1,15}\\
\ldots\\
B_{15} & (e^{\tau\mathbf{A}})_{15,1} & \ldots & (e^{\tau\mathbf{A}})_{15,15}
\end{array}\right)
\left(\begin{array}{r}
\widetilde\rho_0(t)\\
\widetilde\rho_1(t)\\
\ldots\\
\widetilde\rho_{15}(t)
\end{array}\right)
\equiv \widetilde{\mathbf{M}}(\tau)
\left(\begin{array}{r}
\widetilde\rho_0(t)\\
\widetilde\rho_1(t)\\
\ldots\\
\widetilde\rho_{15}(t)
\end{array}\right)
,
\label{QMEQ7b}
\end{eqnarray}
where we used the fact that $\widetilde\rho_0(t)=1$ for all $t$.
Assuming that $\rho_k(t)=\widetilde\rho_k(t)$ for $k=0,\ldots,15$ and for all $t\ge0$ it follows
from Eqs.~(\ref{BREU5b}) and (\ref{QMEQ7b}) by inspection that $\mathbf{M}(\tau)=\widetilde{\mathbf{M}}(\tau)$.

From our simulation results, it is an empirical fact that Eq.~(\ref{QMEQ7b}), which clearly is of the Markovian type,
describes the TDSE data of the two-spin system rather well.
On the other hand, the exact relations Eq.~(\ref{BREU0}) -- (\ref{BREU6})
and the assumption that $V(t)$ satisfies the semigroup property also
leads to a Markovian QMEQ~\cite{BREU02}.
Therefore it is of interest to inquire to what extent the theoretical arguments that lead
to Eq.~(\ref{BREU5b}) support our empirical findings.

We address this question by considering a representative example.
In Fig.~\ref{fig6}, we present simulation results of some system-spin averages,
as obtained from a simulation of 34 spins with a spin-glass bath of $N_\mathrm{B}=32$ spins.
From Fig.~\ref{fig6}, it is clear that the QMEQ Eq.~(\ref{QMEQ7b}), with
$e^{\tau\mathbf{A}}$ and $\mathbf{B}$ obtained by least-square fitting to the TDSE data,
describes the TDSE data very well.
In this cases (as in many others), the density matrix reconstructed
from the data $(\widetilde\rho_0(t),\ldots,\widetilde\rho_{15}(t))$ obtained
by iterating Eq.~(\ref{QMEQ7b}) is non-negative definite, for all $t$ multiples of $\tau$.
However, if we compute $\mathrm{c}(\tau)=\mathrm{X}^{-1} \mathrm{M}(\tau) $, we find
that the matrix $\mathrm{c}$ is Hermitian but has eigenvalues in the interval $[-1.0771,1.4809]$,
in conflict with the theoretical treatment in which the Hermitian matrix $\mathrm{c}$ is non-negative definite by construction
(see Eq.~(\ref{BREU4})).
This is the case for all the data that we have analysed.
Recall that the matrix $\mathrm{c}$ given by Eq.~(\ref{BREU4}) being non-negative definite
is a direct consequence of the assumption that at $t=0$ the density matrix of the whole system is a product
state of the system and bath density matrices.
However, in the case at hand, the matrix $\mathrm{c}$ is obtained from the matrix
$\mathrm{M}$ which in turn is determined by least-square fitting to the TDSE data of the whole time interval
and hence there is no theoretical argument that supports the hypothesis that in this case the
matrix $\mathrm{c}$ should be non-negative definite, and indeed it is not.
The coefficients that enter the Lindblad QMEQ are related to
$\lim_{\tau\rightarrow0}\mathrm{c}(\tau)/\tau$~\cite{BREU02}.
As the matrix $\mathrm{c}$ is found to be non-negative definite, we cannot extract a Lindblad QMEQ from the TDSE data.
This suggests that our empirical finding that the Markovian QMEQ  Eq.~(\ref{QMEQ7b})
describes the TDSE data of the two-spin system rather well does not find an explanation
in the standard theory of open quantum systems.

\section{Summary}\label{section5}

Data obtained by the solution of the time-dependent Schr\"odinger equation
of a system of two spin-1/2 particles interacting with a bath of up to 34 spin-1/2 particles
has been used to (i) study the relaxation and thermalization of the two-spin system
and (ii) make a quantitative assessment of the Markovian quantum master equation description
of the two-spin system dynamics.
It was found that the two-spin system relaxes to a stationary state and that
under certain conditions, the two-spin system thermalizes.
We also studied the effect of the finite size of the bath on the thermalization process.

We demonstrated that a least-square fit of a Markovian quantum master equation to the
time-dependent Schr\"odinger equation data of the reduced density matrix,
yields a very good description of the true Schr\"odinger dynamics of the two-spin system,
even though this Markovian quantum master equation seems
mathematically incompatible with the Lindblad equation.
The resolution of this apparent conflict is left for future research.

\section*{Acknowledgements}
The authors gratefully acknowledge the computing time granted by the JARA-HPC
Vergabegremium and provided on the JARA-HPC Partition part of the supercomputer JUQUEEN~\cite{JUQUEEN}
at Forschungszentrum J{\"u}lich.
The work of MIK is supported by European Research Council (ERC) Advanced Grant No. 338957 FEMTO/NANO.

\appendix
\section{Estimate of the fluctuations}\label{APP1}
A simple method to estimate the statistical errors on the averages obtained from the random thermal state
is to make use of the multivariate Taylor expansion for the average
\begin{eqnarray}
\mathrm{E}\left[
\frac{x}{y}\right]
\approx
\frac{\mathrm{E}[x]}{\mathrm{E}[y]}-\frac{\mathrm{Cov}[x,y]}{\mathrm{E}^2[y]}+\frac{\mathrm{E}[x]\mathrm{Var}[y]}{\mathrm{E}^3[y]}
,
\label{AE0}
\end{eqnarray}
where $\mathrm{Cov}[x,y]=\mathrm{E}[xy]-\mathrm{E}[x]\mathrm{E}[y]$
and use the corresponding approximation for the variance
\begin{eqnarray}
\mathrm{Var}\left[\frac{x}{y}\right]
\approx
\frac{\mathrm{Var}[x]}{\mathrm{E}^2[y]}-2\frac{\mathrm{E}[x]\mathrm{Cov}[x,y]}{\mathrm{E}^3[y]}+\frac{\mathrm{E^2}[x]\mathrm{Var}[y]}{\mathrm{E}^4[y]}
.
\label{AE1}
\end{eqnarray}

As explained in Section~\ref{section3b}, the first step in the construction of a random thermal state
is to generate a Gaussian random state $|\Phi\rangle=\sum_{n=1}^D \xi_n |n\rangle$ where
the $\xi_n$'s are complex-valued Gaussian random variables and the set $\{|n\rangle\}$ can be any complete
set of orthonormal states for the Hilbert space of dimension $D$ (in our case, the states
$|\uparrow \ldots \uparrow\rangle, \ldots,  |\downarrow \ldots \downarrow\rangle$).
We denote the expectation with respect to the multivariate Gaussian probability distribution $P(\xi_{1},\ldots,\xi_{D})$
of the $\xi$'s by $\mathrm{E}[ . ]$.
We have
\begin{eqnarray}
P(\xi_{1},\ldots,\xi_{D})&=&\prod_{a=1}^D \left[ \frac{1}{2\pi\sigma^2}e^{-|\xi_{a}|^2/2\sigma^2}\right]
d(\mathrm{Re\;}\xi_a)\,d(\mathrm{Im\;}\xi_a)
\nonumber \\
\mathrm{E}[\xi_{a}^\ast ]&=&\mathrm{E}[\xi_{a}^{\phantom{^\ast}} ]
=\mathrm{E}[\xi_{a}^{\phantom{^\ast}}\xi_{b}^{\phantom{^\ast}} ]
=\mathrm{E}[\xi_{a}^{{^\ast}}\xi_{b}^{{^\ast}} ]
\nonumber \\
\mathrm{E}[\xi_{a}^\ast \xi_{p}^{\phantom{^\ast}}]&=&2\sigma^2\delta_{a,p}
\nonumber \\
\mathrm{E}[\xi_{a}^\ast \xi_{b}^\ast \xi_{p}^{\phantom{^\ast}}\xi_{q}^{\phantom{^\ast}}]&=&
\mathrm{E}[\xi_{a}^\ast \xi_{p}^{\phantom{^\ast}}] \mathrm{E}[\xi_{b}^\ast \xi_{q}^{\phantom{^\ast}}]
+
\mathrm{E}[\xi_{n}^\ast \xi_{q}^{\phantom{^\ast}}] \mathrm{E}[\xi_{m}^\ast \xi_{p}^{\phantom{^\ast}}]
=4\sigma^4\left(\delta_{a,p}\delta_{b,q}+ \delta_{a,q}\delta_{b,p}\right)
.
\label{RS30}
\end{eqnarray}
For the application of interest, we may, without loss of generality, simplify the writing
by choosing $\sigma=1/\sqrt{2}$, hence we will do so in the following.

Making use of the properties Eq.~(\ref{RS30}) of Gaussian random variables, it readily follows that
for any matrix $X$ we have
\begin{eqnarray}
\mathrm{E}[\langle\Phi|X|\Phi\rangle]
&=&
\sum_{a,p=1}^D \mathrm{E}[ \xi_a^\ast \xi_p^{\phantom{\ast}} ] \langle a|X|p\rangle
=\sum_{a}^D \mathrm{E}[ \xi_a^\ast \xi_a^{\phantom{\ast}} ] \langle a|X|a\rangle
=\sum_{a}^D \mathrm{E}[ \xi_a^\ast \xi_a^{\phantom{\ast}} ] \langle a|X|a\rangle
=\mathbf{Tr\;} X
,
\label{RS31}
\end{eqnarray}
and because $\langle\Phi|X|\Phi\rangle=\langle\Phi|X^\dagger|\Phi\rangle^\ast$, the corresponding variance is given by
\begin{eqnarray}
\mathrm{Var}\left(\langle\Phi|X|\Phi\rangle\right)
&=&\mathrm{E}[|\langle\Phi|X|\Phi\rangle|^2]-\left|\mathrm{E}[\langle\Phi|X|\Phi\rangle]\right|^2
\nonumber \\
&=&\sum_{a,p,b,q=1}^D \mathrm{E}[ \xi_a^\ast \xi_p^{\phantom{\ast}} \xi_b^{\phantom{\ast}} \xi_q^\ast ]
\langle a|X|p\rangle
\langle b|X|q\rangle^\ast - \left|\mathbf{Tr\;} X\right|^2
\nonumber \\
&=&\sum_{a,b=1}^D
\bigg(\langle a|X|a\rangle
\langle b|X|b\rangle^\ast
+
\langle a|X|b\rangle
\langle a|X|b\rangle^\ast
\bigg) - \left|\mathbf{Tr\;} X\right|^2
\nonumber \\
&=&\mathbf{Tr\;} XX^\dagger
.
\label{RS32}
\end{eqnarray}

In the case at hand, we use Eqs.~(\ref{AE0})--(\ref{RS32}) as follows.
We set $Z=e^{-\beta H}$ and $X=e^{-\beta H/2} Y e^{-\beta H/2}$ with $H=H^\dagger$ and $Y=Y^\dagger$.
From Eq.~(\ref{RS32}) it follows that
$\mathrm{Var}\left[\langle\Phi|Z|\Phi\rangle\right]=\mathbf{Tr\;} Z^2$.
Furthermore, we have
\begin{eqnarray}
\mathrm{E}[\langle\Phi|X|\Phi\rangle\langle\Phi|Z|\Phi\rangle]
&=&\sum_{a,p,b,q=1}^D \mathrm{E}[ \xi_a^\ast \xi_p^{\phantom{\ast}} \xi_b^\ast \xi_q^{\phantom{\ast}} ]
\langle a|X|p\rangle \langle b|Z|q\rangle
=\left(\mathbf{Tr\;}X \right)\left(\mathbf{Tr\;} Z \right) + \mathbf{Tr\;} XZ
\nonumber \\
&=&\left(\mathbf{Tr\;}ZY \right)\left(\mathbf{Tr\;} Z\right) + \mathbf{Tr\;} e^{-2\beta H} Y
=\left(\mathbf{Tr\;}ZY \right)\left(\mathbf{Tr\;} Z\right) + \mathbf{Tr\;} Z^2 Y
,
\label{AV3}
\end{eqnarray}
from which it follows that
$\mathrm{Cov}[\langle\Phi|X|\Phi\rangle,\langle\Phi|Z|\Phi\rangle]=\mathbf{Tr\;} Z^2 Y$.
Collecting all contributions, we find
\begin{eqnarray}
\mathrm{E}\left[
\frac{\langle\Phi|X|\Phi\rangle}{\langle\Phi|Z|\Phi\rangle}
\right]
&\approx&
\frac{\mathbf{Tr\;} Z Y}{\mathbf{Tr\;} Z}
-
\frac{\mathbf{Tr\;} Z^2 Y}{\left(\mathbf{Tr\;} Z\right)^2}
+
\frac{\mathbf{Tr\;} Z Y}{\mathbf{Tr\;} Z}
\frac{\mathbf{Tr\;} Z^2}{\left(\mathbf{Tr\;} Z\right)^2}
\nonumber \\
&&=
\langle Y \rangle
+\frac{\mathbf{Tr\;} Z^2}{\left(\mathbf{Tr\;} Z\right)^2}
\left\{
\frac{\mathbf{Tr\;} Z Y}{\mathbf{Tr\;} Z}
-
\frac{\mathbf{Tr\;} Z^2 Y}{\mathbf{Tr\;} Z^2}
\right\}
,
\label{AE3}
\end{eqnarray}
and
\begin{eqnarray}
\mathrm{Var}\left[
\frac{\langle\Phi|X|\Phi\rangle}{\langle\Phi|Z|\Phi\rangle}
\right]
&\approx&
\frac{\mathbf{Tr\;} (Z Y)^2}{\left(\mathbf{Tr\;} Z\right)^2}
-2
\frac{\mathbf{Tr\;} Z Y}{\mathbf{Tr\;} Z}
\frac{\mathbf{Tr\;} Z^2 Y}{\left(\mathbf{Tr\;} Z\right)^2}
+
\left(\frac{\mathbf{Tr\;} Z Y}{\mathbf{Tr\;} Z}\right)^2
\frac{\mathbf{Tr\;} Z^2}{\left(\mathbf{Tr\;} Z\right)^2}
\nonumber \\
&&=
\frac{\mathbf{Tr\;} Z^2}{\left(\mathbf{Tr\;} Z\right)^2}
\bigg\{
\frac{\mathbf{Tr\;} (Z Y)^2}{\mathbf{Tr\;} Z^2}
-2
\frac{\mathbf{Tr\;} Z Y}{\mathbf{Tr\;} Z}
\frac{\mathbf{Tr\;} Z^2 Y}{\mathbf{Tr\;} Z^2}
+
\left(\frac{\mathbf{Tr\;} Z Y}{\mathbf{Tr\;} Z}\right)^2
\bigg\}
.
\label{AE5}
\end{eqnarray}

Using the definition $F(\beta)=-(1/\beta)\mathbf{Tr\;} Z$ of the free energy
of the whole system (described by $H$), we may write
\begin{equation}
\frac{\mathbf{Tr\;} Z^2}{\left(\mathbf{Tr\;} Z\right)^2} = e^{-2\beta[F(2\beta)-F(\beta)]}
,
\label{AE4}
\end{equation}
and it is easy to show that
\begin{eqnarray}
\left|
\frac{\mathbf{Tr\;} Z Y}{\mathbf{Tr\;} Z}
-
\frac{\mathbf{Tr\;} Z^2 Y}{\mathbf{Tr\;} Z^2}
\right|
&\le&2\Vert Y \Vert
,
\label{AE6a}
\\
\bigg|
\frac{\mathbf{Tr\;} (Z Y)^2}{\mathbf{Tr\;} Z^2}
-2
\frac{\mathbf{Tr\;} Z Y}{\mathbf{Tr\;} Z}
\frac{\mathbf{Tr\;} Z^2 Y}{\mathbf{Tr\;} Z^2}
+
\left(\frac{\mathbf{Tr\;} Z Y}{\mathbf{Tr\;} Z}\right)^2
\bigg|
&\le&4\Vert Y \Vert^2
,
\label{AE6}
\end{eqnarray}
where $\Vert Y \Vert$ denotes the largest (in absolute value) eigenvalue of $Y$.

For finite values of $\beta$, $F(2\beta)-F(\beta)={\cal O}(N)$ where $N=\log_2 D$ denotes the number of spins
of the whole system. Therefore, for finite values of $\beta$, the correction term
in Eq.~(\ref{AE3}) and the variance Eq.~(\ref{AE5}) vanish exponentially with the number of spins.
In particular, in the limit of infinite temperatures we have
$\lim_{\beta\rightarrow0}\mathbf{Tr\;} Z^2/\left(\mathbf{Tr\;} Z\right)^2=1/D$.

In the zero temperature limit, it is expedient to write Eqs.~(\ref{AE3}) and (\ref{AE5})
in terms of the eigenvalues $E_0<E_1\le\ldots$ of $H$.
We have $\mathbf{Tr\;} Z=e^{-\beta E_0}\left[1+e^{-\beta (E_1-E_0)}+\ldots\right]$ and find that
\begin{eqnarray}
\lim_{\beta\rightarrow\infty}\frac{\mathbf{Tr\;} Z^2}{\left(\mathbf{Tr\;} Z\right)^2} &=& 1
\nonumber \\
\lim_{\beta\rightarrow\infty}
\left\{ \frac{\mathbf{Tr\;} Z Y}{\mathbf{Tr\;} Z}
- \frac{\mathbf{Tr\;} Z^2 Y}{\mathbf{Tr\;} Z^2} \right\} &=&0
\nonumber \\
\lim_{\beta\rightarrow\infty}
\bigg\{ \frac{\mathbf{Tr\;} (Z Y)^2}{\mathbf{Tr\;} Z^2}
-2 \frac{\mathbf{Tr\;} Z Y}{\mathbf{Tr\;} Z}\frac{\mathbf{Tr\;} Z^2 Y}{\mathbf{Tr\;} Z^2}
+\left(\frac{\mathbf{Tr\;} Z Y}{\mathbf{Tr\;} Z}\right)^2 \bigg\} &=&0
,
\label{AE7}
\end{eqnarray}
showing that in this limit, the random thermal state approach yields the exact average $\langle Y \rangle$.

\section{Harmonic oscillators}\label{OSC}

In this section, we present some simulation results of Bogolyubov's model
of a collection of classical oscillators~\cite{BOGU94,CHIR86,STRO07}.
The Hamiltonian of this model takes the generic form Eq.~(\ref{s40})
with each term being defined by
\begin{eqnarray}
H_S &=& \frac{p^2}{2m} +\frac{1}{2}m\omega^2 q^2,
\label{OSC0a}
\\
H_B &=& \sum_{n=1}^{N_\mathrm{B}} \left( \frac{p_n^2}{2m_n}+\frac{1}{2}m_n\omega_n^2 q_n^2 \right) ,
\label{OSC0b}
\\
\lambda H_{SB} &=& \lambda \sum_{n=1}^{N_\mathrm{B}} \alpha_n q_n q,
\label{OSC0c}
\end{eqnarray}
where $m$, $p$, $q$, $\omega$ and $m_n$, $p_n$, $q_n$, $\omega_n$ are masses, momenta, coordinates,
and frequencies of the oscillator in the system and bath, respectively.
The $\alpha_n$'s represent the system-bath coupling constants and $\lambda$ sets the scale of the latter.
For simplicity, we take $m=m_n=1$.

Bogolyubov proved that the density matrix $\rho_S(t,q,p)$ of the system
approaches the canonical distribution
if the following two conditions are satisfied:
(i) the thermal state of the bath of oscillators is described by the canonical distribution
\begin{equation}
\rho_B =e^{-\beta H_B}/Z_B,
\label{OSC1}
\end{equation}
where $\beta$ is the inverse temperature and $Z_B$ is the partition function of the bath,
and (ii) in the limit $N\rightarrow \infty$, the relation
\begin{equation}
\sum_n \frac{\alpha_n^2}{\omega_n^2} \rightarrow \int d\omega J(\omega),
\label{OSC2}
\end{equation}
holds.
Bogolyubov's original result only concerns the asymptotic $t\rightarrow \infty$
behavior of the ensemble- and time-averaged trajectories of the system.
To the best of our knowledge, Bogolyubov's result is the only rigorous result about the thermalization of
a classical system interacting with a thermostat.
Therefore, in the light of the finite quantum spin systems studied in this paper,
it is of interest to investigate finite-size effects for the classical model Eqs.~(\ref{OSC0a})--(\ref{OSC0c}) by simulation.

The simulation is most conveniently carried out by numerical diagonalization of the sum of the quadratic forms Eqs.~(\ref{OSC0a})--(\ref{OSC0c})
and yields results which are, for all practical purposes, exact.
Writing $P=(p,p_1,p_2,\ldots,p_N)$ and $Q=(q,q_1,q_2,\ldots,q_N)$, the Hamiltonian reads
\begin{equation}
H=\frac{1}{2}P^TP+\frac{1}{2}Q^T M Q ,
\label{OSC3}
\end{equation}
where the matrix $M$ is given by
\begin{equation}
M=\left (
		\begin{tabular} {ccccc}
			$\omega^2$ & $\lambda \alpha_1/2$ & $\lambda \alpha_2/2$& \ldots &  $\lambda \alpha_N/2$ \cr
			$\lambda \alpha_1/2$ & $\omega_1^2$ & 0 &\ldots & 0 \cr
			$\lambda \alpha_2/2$ & 0 & $\omega_2^2$ & \ldots & 0 \cr
			$\vdots$ & $\vdots$ & $\vdots$  & $\ddots$ & $\vdots$ \cr
			$\lambda \alpha_N/2$ & 0 & 0& \ldots & $ \omega_N^2$
		\end{tabular}
\right ).
\label{OSC4}
\end{equation}
The determinant of the matrix $M$ is easily found to be
$\mathbf{det}(M)=\omega_1^2\omega_2^2\cdots \omega_N^2(\omega^2-\lambda^2\sum_{n=1}^{N_\mathrm{B}}\alpha_n^2/4\omega_n^2)$.
For the whole system to be stable, the matrix $M$ should be positive-definite,
implying that the value of the global coupling $\lambda$ should satisfy
$\lambda^2 < 4\omega^2/\sum_{n=1}^{N_\mathrm{B}} \alpha_n^2 /\omega_n^2$.
By diagonalizing the matrix $M=UDU^+$ and setting $P'=U^+ P$ and $Q'=U^+Q$,
the Hamiltonian changes into $H=(P'^TP'+Q'^T D Q')/2$,
i.e. a set of independent harmonic oscillators for which a closed-form analytical solution is known.
The solution in terms of the original coordinates is obtained by application of the transformation $P=UP'$ and $Q=UQ'$.

For finite systems, the above condition (ii) is not easy to fulfil and therefore, it is important to make a judicious choice
of the model parameters.
Inspired by suggestions made in Ref.~\onlinecite{STRO07},
we choose $\omega=1$, $\alpha_n=1$, and $\omega_n^2= a+(b(n-1))^2$ with $a=0.5$ and $b=0.01$.
With this particular choice of the parameters, a bath of ${N_\mathrm{B}}=511$ oscillators
was found to be large enough to mimic the infinite thermostat.
Note that as ${N_\mathrm{B}}\rightarrow \infty$, it is necessary to let $b\rightarrow 0$ in order to
have a well-defined thermodynamic limit.
In the simulation, the initial state (the values of $q_n$ and $p_n$) of the bath are
chosen randomly from the canonical distribution with $\beta=1$ (see condition (i)).
The initial state of the system is chosen to be $q=p=\sqrt{T_{\mathrm{S}}}$, where $T_{\mathrm{S}}$
plays the roles of a fictitious temperature of the isolated system.

\begin{figure}[t]
\includegraphics[width=0.40\columnwidth]{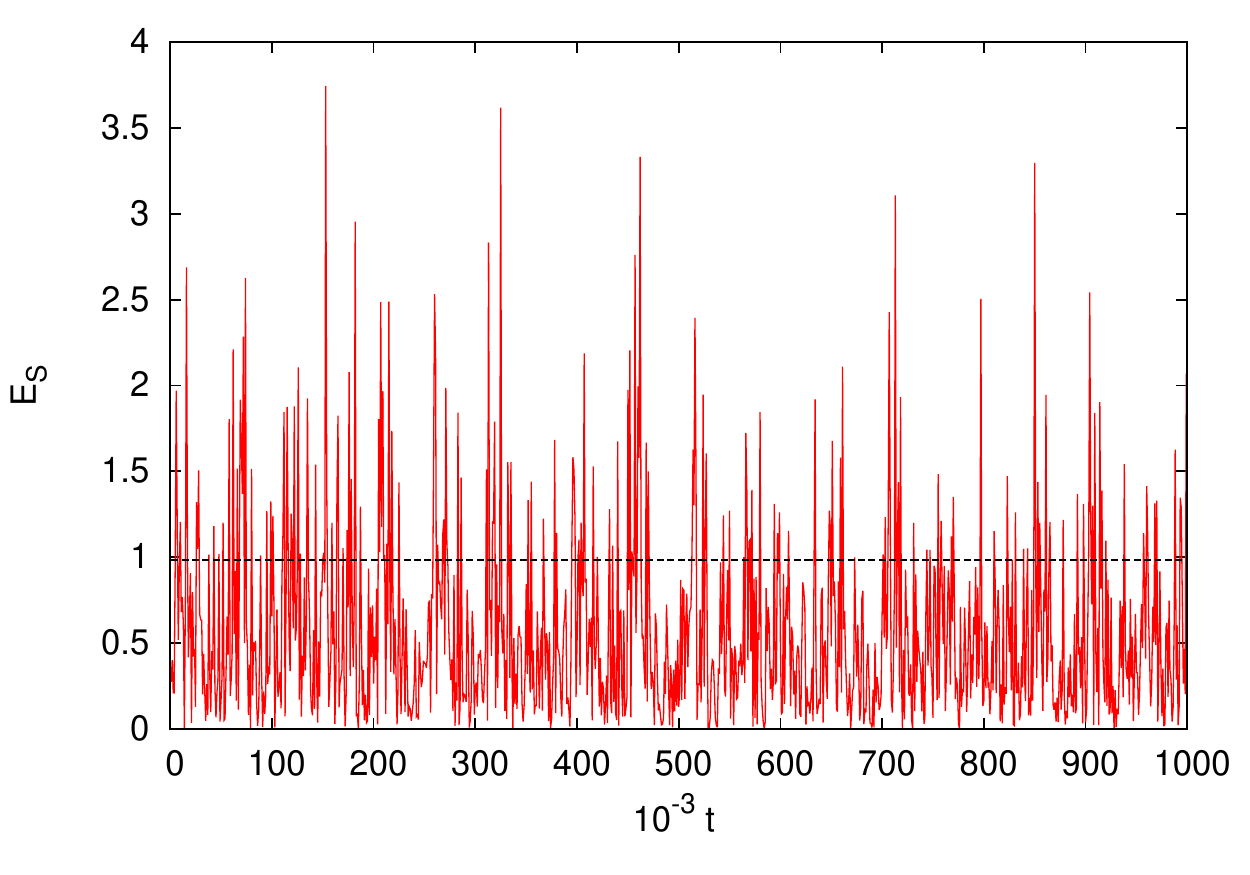}
\caption{The time evolution of the system energy $E_S$
with the model parameters $a=0.5$, $b=0.01$, and $\lambda=0.01$.
The number of bath oscillators is ${N_\mathrm{B}}=511$.
The initial state of the system is $T_S=0$.
The initial state of the bath is drawn randomly from the canonical distribution with $\beta=1$.
The straight line denotes the energy per oscillator of the total system.
}
\label{figure_a1}
\end{figure}

First, we consider a single realization of the initial state and let this state evolve in time.
The quantity of interest is the system energy $E_S=(p^2+q^2)/2$.
If the system thermalizes in the course of following a single trajectory, then the time-averaged system energy $E_S$ should be approximately equal to
the total energy per particle.
In Fig.~{\ref{figure_a1}}, we present the time evolution of the system energy $E_S$
for one particular trajectory up to $t=10^6$.
The total energy per particle is about $E=0.98$.
The time average of the system energy $E_S$ is about $\bar{E}_S=0.54$.
Therefore, it is clear from the simulation results
that in Bogolyubov's model, one trajectory is not enough for the system to thermalize,
in strong contrast with models of coupled harmonic oscillators (integrable system)
or magnetic moments (nonintegrable system) in which the system, defined as a part of the whole system,
thermalizes for one single trajectory~\cite{JIN13b}.

\begin{figure}[t]
\includegraphics[width=0.40\columnwidth]{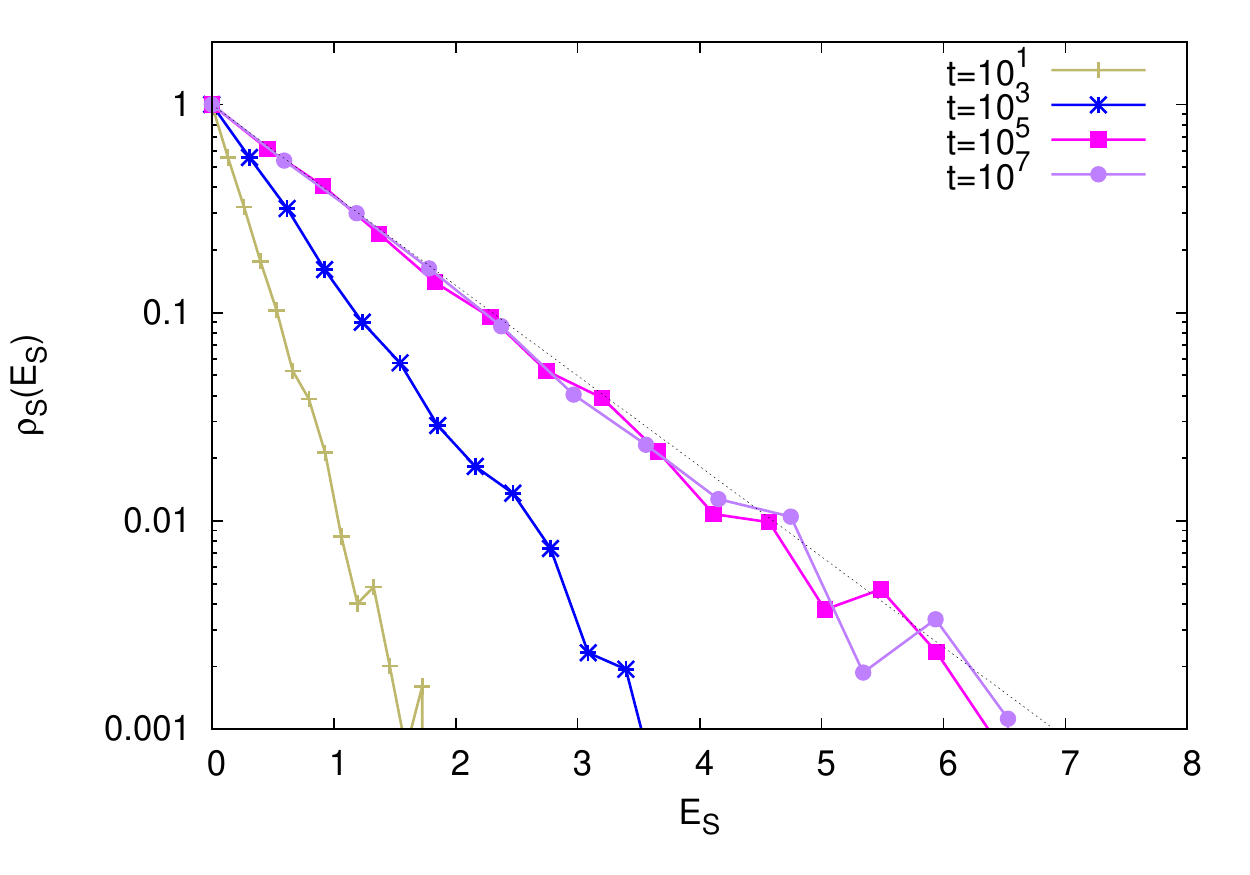}
\includegraphics[width=0.40\columnwidth]{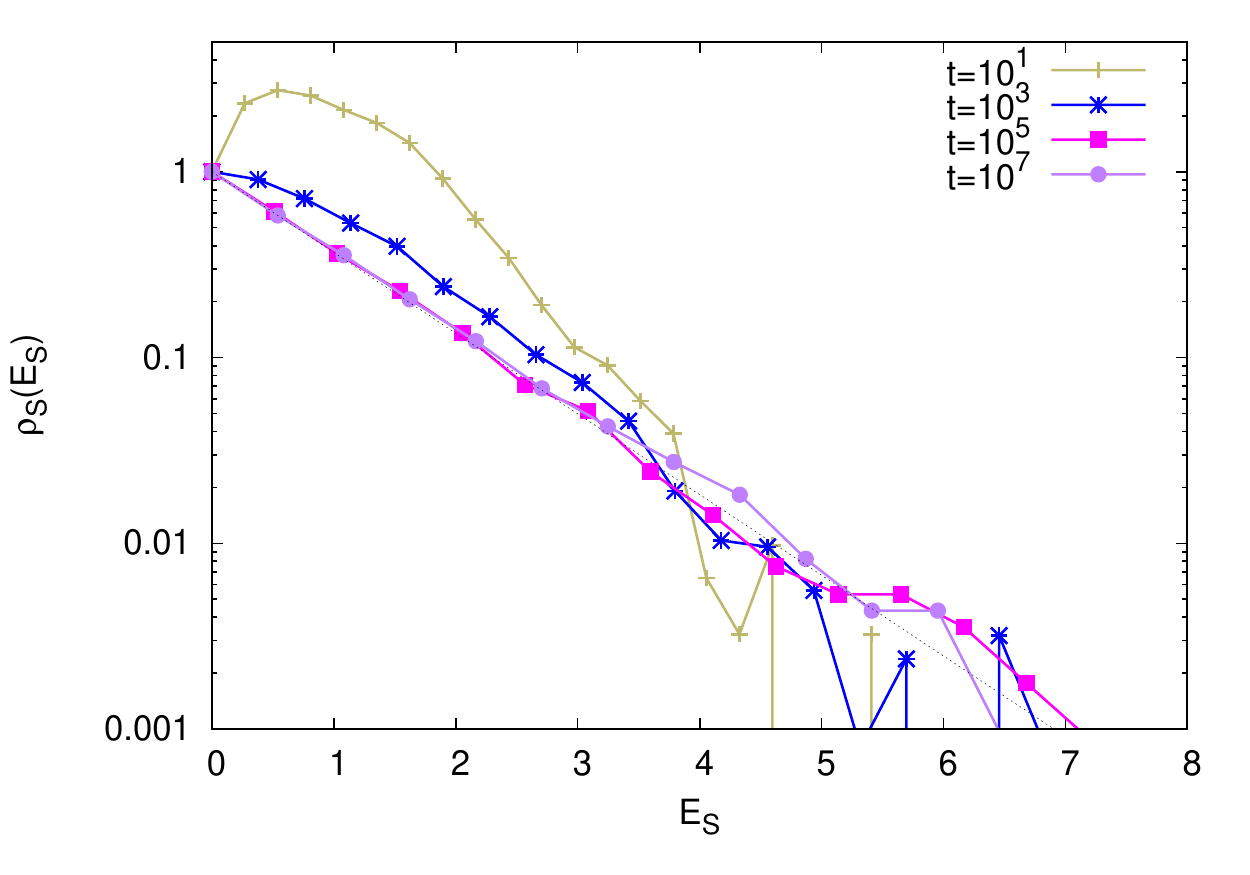}
\caption{The histogram $\rho_S(E_S)$ at different times and two different initial states of the system, $T_S=0$ (left)
and $T_S=1.1$ (right). The model parameters are the same as in Fig.~{\ref{figure_a1}}.
The histogram represents an ensemble of $10^4$ random realizations of initial thermal states of the bath at $\beta=1$.
The straight line denotes the expected distribution $\exp(-\beta E_S)$ with $\beta=1$.
}
\label{figure_a2}
\end{figure}

Next, we investigate the properties of an ensemble of many trajectories, obtained by starting from many different initial states.
As mentioned earlier, the initial state of the bath degrees of freedom are drawn randomly from the canonical distribution.
Now the quantity of interest is the distribution of the system energy $E_S$ at specific times.
We build a histogram $\rho_S(E_S)$ by recording, at specific times, the number of trajectories with system energies in the range $[E_S, E_S+\Delta E_S]$.
Several representative results of $\rho_S(E_S)$ are presented in Fig.~{\ref{figure_a2}}.
This figure shows results obtained from simulations with two different initial states of the system as a function of time $t$
and clearly demonstrates that by taking the ensemble average, the system thermalizes at long times, i.e.,
$\rho_S(E_S)\rightarrow e^{-E_S}$ as $t\rightarrow \infty$.

Summarizing: for the mentioned special choice of the model parameters, we have verified Bogolyubov's result by numerical simulations.
By taking ensemble averages and for sufficiently long times, the system is described by the canonical distribution.
We also show that the ensemble averaging is necessary to recover Bogolyubov's result,
very much unlike in our previous study on the coupled harmonic oscillators and magnetic moments~\cite{JIN13b}.
We have tried out quite a few other choices of the model parameters (data not shown)
but we rarely observed nice thermalization of the system, at least not with the number of bath oscillators
for which exact diagonalization is possible.

\bibliography{../../../all17}

\begin{thebibliography}{67}%
\makeatletter
\providecommand \@ifxundefined [1]{%
 \@ifx{#1\undefined}
}%
\providecommand \@ifnum [1]{%
 \ifnum #1\expandafter \@firstoftwo
 \else \expandafter \@secondoftwo
 \fi
}%
\providecommand \@ifx [1]{%
 \ifx #1\expandafter \@firstoftwo
 \else \expandafter \@secondoftwo
 \fi
}%
\providecommand \natexlab [1]{#1}%
\providecommand \enquote  [1]{``#1''}%
\providecommand \bibnamefont  [1]{#1}%
\providecommand \bibfnamefont [1]{#1}%
\providecommand \citenamefont [1]{#1}%
\providecommand \href@noop [0]{\@secondoftwo}%
\providecommand \href [0]{\begingroup \@sanitize@url \@href}%
\providecommand \@href[1]{\@@startlink{#1}\@@href}%
\providecommand \@@href[1]{\endgroup#1\@@endlink}%
\providecommand \@sanitize@url [0]{\catcode `\\12\catcode `\$12\catcode
  `\&12\catcode `\#12\catcode `\^12\catcode `\_12\catcode `\%12\relax}%
\providecommand \@@startlink[1]{}%
\providecommand \@@endlink[0]{}%
\providecommand \url  [0]{\begingroup\@sanitize@url \@url }%
\providecommand \@url [1]{\endgroup\@href {#1}{\urlprefix }}%
\providecommand \urlprefix  [0]{URL }%
\providecommand \Eprint [0]{\href }%
\providecommand \doibase [0]{http://dx.doi.org/}%
\providecommand \selectlanguage [0]{\@gobble}%
\providecommand \bibinfo  [0]{\@secondoftwo}%
\providecommand \bibfield  [0]{\@secondoftwo}%
\providecommand \translation [1]{[#1]}%
\providecommand \BibitemOpen [0]{}%
\providecommand \bibitemStop [0]{}%
\providecommand \bibitemNoStop [0]{.\EOS\space}%
\providecommand \EOS [0]{\spacefactor3000\relax}%
\providecommand \BibitemShut  [1]{\csname bibitem#1\endcsname}%
\let\auto@bib@innerbib\@empty
\bibitem [{\citenamefont {Redfield}(1957)}]{REDF57}%
  \BibitemOpen
  \bibfield  {author} {\bibinfo {author} {\bibfnamefont {A.}~\bibnamefont
  {Redfield}},\ }\bibfield  {title} {\enquote {\bibinfo {title} {On the theory
  of relaxation processes},}\ }\href@noop {} {\bibfield  {journal} {\bibinfo
  {journal} {IBM J. Res. Develop.}\ }\textbf {\bibinfo {volume} {1}},\ \bibinfo
  {pages} {19--31} (\bibinfo {year} {1957})}\BibitemShut {NoStop}%
\bibitem [{\citenamefont {Nakajima}(1958)}]{NAKA58}%
  \BibitemOpen
  \bibfield  {author} {\bibinfo {author} {\bibfnamefont {S.}~\bibnamefont
  {Nakajima}},\ }\bibfield  {title} {\enquote {\bibinfo {title} {On quantum
  theory of transport phenomena},}\ }\href@noop {} {\bibfield  {journal}
  {\bibinfo  {journal} {Prog. Theor. Phys.}\ }\textbf {\bibinfo {volume}
  {20}},\ \bibinfo {pages} {948 -- 959} (\bibinfo {year} {1958})}\BibitemShut
  {NoStop}%
\bibitem [{\citenamefont {Zwanzig}(1960)}]{ZWAN60}%
  \BibitemOpen
  \bibfield  {author} {\bibinfo {author} {\bibfnamefont {R.}~\bibnamefont
  {Zwanzig}},\ }\bibfield  {title} {\enquote {\bibinfo {title} {Ensemble method
  in the theory of irreversibility},}\ }\href@noop {} {\bibfield  {journal}
  {\bibinfo  {journal} {J. Chem. Phys.}\ }\textbf {\bibinfo {volume} {33}},\
  \bibinfo {pages} {1338 -- 1341} (\bibinfo {year} {1960})}\BibitemShut
  {NoStop}%
\bibitem [{\citenamefont {Breuer}\ and\ \citenamefont
  {Petruccione}(2002)}]{BREU02}%
  \BibitemOpen
  \bibfield  {author} {\bibinfo {author} {\bibfnamefont {H.-P.}\ \bibnamefont
  {Breuer}}\ and\ \bibinfo {author} {\bibfnamefont {F.}~\bibnamefont
  {Petruccione}},\ }\href@noop {} {\emph {\bibinfo {title} {{The Theory of Open
  Quantum Systems}}}}\ (\bibinfo  {publisher} {Oxford University Press},\
  \bibinfo {address} {Oxford},\ \bibinfo {year} {2002})\BibitemShut {NoStop}%
\bibitem [{\citenamefont {Sassetti}\ and\ \citenamefont
  {Weiss}(1990)}]{SASS90}%
  \BibitemOpen
  \bibfield  {author} {\bibinfo {author} {\bibfnamefont {M.}~\bibnamefont
  {Sassetti}}\ and\ \bibinfo {author} {\bibfnamefont {U.}~\bibnamefont
  {Weiss}},\ }\bibfield  {title} {\enquote {\bibinfo {title} {Correlation
  functions for dissipative two-state systems: Effects of the initial
  preparation},}\ }\href@noop {} {\bibfield  {journal} {\bibinfo  {journal}
  {Phys. Rev. A}\ }\textbf {\bibinfo {volume} {41}},\ \bibinfo {pages}
  {5383--5393} (\bibinfo {year} {1990})}\BibitemShut {NoStop}%
\bibitem [{\citenamefont {Weiss}(1999)}]{WEIS99}%
  \BibitemOpen
  \bibfield  {author} {\bibinfo {author} {\bibfnamefont {U.}~\bibnamefont
  {Weiss}},\ }\href@noop {} {\emph {\bibinfo {title} {Quantum Dissipative
  Systems}}},\ \bibinfo {edition} {2nd}\ ed.\ (\bibinfo  {publisher} {World
  Scientific},\ \bibinfo {address} {Singapore},\ \bibinfo {year}
  {1999})\BibitemShut {NoStop}%
\bibitem [{\citenamefont {Tanimura}(2006)}]{TANI06}%
  \BibitemOpen
  \bibfield  {author} {\bibinfo {author} {\bibfnamefont {Y.}~\bibnamefont
  {Tanimura}},\ }\bibfield  {title} {\enquote {\bibinfo {title} {Stochastic
  {Liouville}, {Langevin}, {Fokker}–-{Planck}, and master equation approaches
  to quantum dissipative systems},}\ }\href@noop {} {\bibfield  {journal}
  {\bibinfo  {journal} {J. Phys. Soc. Jpn.}\ }\textbf {\bibinfo {volume}
  {75}},\ \bibinfo {pages} {082001} (\bibinfo {year} {2006})}\BibitemShut
  {NoStop}%
\bibitem [{\citenamefont {Breuer}\ \emph {et~al.}(2006)\citenamefont {Breuer},
  \citenamefont {Gemmer},\ and\ \citenamefont {Michel}}]{BREU06}%
  \BibitemOpen
  \bibfield  {author} {\bibinfo {author} {\bibfnamefont {H.-P.}\ \bibnamefont
  {Breuer}}, \bibinfo {author} {\bibfnamefont {J.}~\bibnamefont {Gemmer}}, \
  and\ \bibinfo {author} {\bibfnamefont {M.}~\bibnamefont {Michel}},\
  }\bibfield  {title} {\enquote {\bibinfo {title} {{Non-Markovian} quantum
  dynamics: Correlated projection superoperators and {Hilbert} space
  averaging},}\ }\href@noop {} {\bibfield  {journal} {\bibinfo  {journal}
  {Phys. Rev. E}\ }\textbf {\bibinfo {volume} {73}},\ \bibinfo {pages} {016139}
  (\bibinfo {year} {2006})}\BibitemShut {NoStop}%
\bibitem [{\citenamefont {Mori}\ and\ \citenamefont
  {Miyashita}(2008)}]{MORI08}%
  \BibitemOpen
  \bibfield  {author} {\bibinfo {author} {\bibfnamefont {T.}~\bibnamefont
  {Mori}}\ and\ \bibinfo {author} {\bibfnamefont {S.}~\bibnamefont
  {Miyashita}},\ }\bibfield  {title} {\enquote {\bibinfo {title} {Dynamics of
  the density matrix in contact with a thermal bath and the quantum master
  equation},}\ }\href@noop {} {\bibfield  {journal} {\bibinfo  {journal} {J.
  Phys. Soc. Jpn.}\ }\textbf {\bibinfo {volume} {77}},\ \bibinfo {pages}
  {124005} (\bibinfo {year} {2008})}\BibitemShut {NoStop}%
\bibitem [{\citenamefont {Saeki}(2008)}]{SAEK08}%
  \BibitemOpen
  \bibfield  {author} {\bibinfo {author} {\bibfnamefont {M.}~\bibnamefont
  {Saeki}},\ }\bibfield  {title} {\enquote {\bibinfo {title} {Relaxation method
  and {TCLE} method of linear response in terms of thermo-field dynamics},}\
  }\href@noop {} {\bibfield  {journal} {\bibinfo  {journal} {Physica A}\
  }\textbf {\bibinfo {volume} {387}},\ \bibinfo {pages} {1827--1850} (\bibinfo
  {year} {2008})}\BibitemShut {NoStop}%
\bibitem [{\citenamefont {Uchiyama}\ \emph {et~al.}(2009)\citenamefont
  {Uchiyama}, \citenamefont {Aihara}, \citenamefont {Saeki},\ and\
  \citenamefont {Miyashita}}]{UCHI09}%
  \BibitemOpen
  \bibfield  {author} {\bibinfo {author} {\bibfnamefont {C.}~\bibnamefont
  {Uchiyama}}, \bibinfo {author} {\bibfnamefont {M.}~\bibnamefont {Aihara}},
  \bibinfo {author} {\bibfnamefont {M.}~\bibnamefont {Saeki}}, \ and\ \bibinfo
  {author} {\bibfnamefont {S.}~\bibnamefont {Miyashita}},\ }\bibfield  {title}
  {\enquote {\bibinfo {title} {Master equation approach to line shape in
  dissipative systems},}\ }\href@noop {} {\bibfield  {journal} {\bibinfo
  {journal} {Phys. Rev. E}\ }\textbf {\bibinfo {volume} {80}},\ \bibinfo
  {pages} {021128} (\bibinfo {year} {2009})}\BibitemShut {NoStop}%
\bibitem [{\citenamefont {Mori}(2014)}]{MORI14}%
  \BibitemOpen
  \bibfield  {author} {\bibinfo {author} {\bibfnamefont {T.}~\bibnamefont
  {Mori}},\ }\bibfield  {title} {\enquote {\bibinfo {title} {Natural
  correlation between a system and a thermal reservoir},}\ }\href@noop {}
  {\bibfield  {journal} {\bibinfo  {journal} {Phys. Rev. A}\ }\textbf {\bibinfo
  {volume} {89}},\ \bibinfo {pages} {040101(R)} (\bibinfo {year}
  {2014})}\BibitemShut {NoStop}%
\bibitem [{\citenamefont {Chen}\ \emph {et~al.}(2015)\citenamefont {Chen},
  \citenamefont {Lambert}, \citenamefont {Cheng}, \citenamefont {Chen},\ and\
  \citenamefont {Nori}}]{CHEN15}%
  \BibitemOpen
  \bibfield  {author} {\bibinfo {author} {\bibfnamefont {H.-B.}\ \bibnamefont
  {Chen}}, \bibinfo {author} {\bibfnamefont {N.}~\bibnamefont {Lambert}},
  \bibinfo {author} {\bibfnamefont {Y.-C.}\ \bibnamefont {Cheng}}, \bibinfo
  {author} {\bibfnamefont {Y.-N.}\ \bibnamefont {Chen}}, \ and\ \bibinfo
  {author} {\bibfnamefont {F.}~\bibnamefont {Nori}},\ }\bibfield  {title}
  {\enquote {\bibinfo {title} {Using non-{Markovian} measures to evaluate
  quantum master equations for photosynthesis},}\ }\href@noop {} {\bibfield
  {journal} {\bibinfo  {journal} {Sci. Rep.}\ }\textbf {\bibinfo {volume}
  {5}},\ \bibinfo {pages} {12753} (\bibinfo {year} {2015})}\BibitemShut
  {NoStop}%
\bibitem [{\citenamefont {Lindblad}(1976)}]{LIND76}%
  \BibitemOpen
  \bibfield  {author} {\bibinfo {author} {\bibfnamefont {G.}~\bibnamefont
  {Lindblad}},\ }\bibfield  {title} {\enquote {\bibinfo {title} {On the
  generators of quantum dynamical semigroups},}\ }\href@noop {} {\bibfield
  {journal} {\bibinfo  {journal} {Commun. Math. Phys.}\ }\textbf {\bibinfo
  {volume} {48}},\ \bibinfo {pages} {119--130} (\bibinfo {year}
  {1976})}\BibitemShut {NoStop}%
\bibitem [{\citenamefont {Fonda}\ \emph {et~al.}(1978)\citenamefont {Fonda},
  \citenamefont {Ghirardi},\ and\ \citenamefont {Rimini}}]{FOND78}%
  \BibitemOpen
  \bibfield  {author} {\bibinfo {author} {\bibfnamefont {L.}~\bibnamefont
  {Fonda}}, \bibinfo {author} {\bibfnamefont {G.~C.}\ \bibnamefont {Ghirardi}},
  \ and\ \bibinfo {author} {\bibfnamefont {A.}~\bibnamefont {Rimini}},\
  }\bibfield  {title} {\enquote {\bibinfo {title} {Decay theory of unstable
  quantum systems},}\ }\href@noop {} {\bibfield  {journal} {\bibinfo  {journal}
  {Rep. Prog. Phys.}\ }\textbf {\bibinfo {volume} {41}},\ \bibinfo {pages} {587
  -- 631} (\bibinfo {year} {1978})}\BibitemShut {NoStop}%
\bibitem [{\citenamefont {{Jin}}\ \emph {et~al.}(2010)\citenamefont {{Jin}},
  \citenamefont {{De Raedt}}, \citenamefont {{Yuan}}, \citenamefont
  {{Katsnelson}}, \citenamefont {{Miyashita}},\ and\ \citenamefont
  {{Michielsen}}}]{JIN10x}%
  \BibitemOpen
  \bibfield  {author} {\bibinfo {author} {\bibfnamefont {F.}~\bibnamefont
  {{Jin}}}, \bibinfo {author} {\bibfnamefont {H.}~\bibnamefont {{De Raedt}}},
  \bibinfo {author} {\bibfnamefont {S.}~\bibnamefont {{Yuan}}}, \bibinfo
  {author} {\bibfnamefont {M.~I.}\ \bibnamefont {{Katsnelson}}}, \bibinfo
  {author} {\bibfnamefont {S.}~\bibnamefont {{Miyashita}}}, \ and\ \bibinfo
  {author} {\bibfnamefont {K.}~\bibnamefont {{Michielsen}}},\ }\bibfield
  {title} {\enquote {\bibinfo {title} {{Approach to Equilibrium in Nano-scale
  Systems at Finite Temperature}},}\ }\href@noop {} {\bibfield  {journal}
  {\bibinfo  {journal} {J. Phys. Soc. Jpn.}\ }\textbf {\bibinfo {volume}
  {79}},\ \bibinfo {pages} {124005} (\bibinfo {year} {2010})}\BibitemShut
  {NoStop}%
\bibitem [{\citenamefont {Donker}\ \emph {et~al.}(2017)\citenamefont {Donker},
  \citenamefont {{De Raedt}},\ and\ \citenamefont {Katsnelson}}]{DONK17}%
  \BibitemOpen
  \bibfield  {author} {\bibinfo {author} {\bibfnamefont {H.C.}\ \bibnamefont
  {Donker}}, \bibinfo {author} {\bibfnamefont {H.}~\bibnamefont {{De Raedt}}},
  \ and\ \bibinfo {author} {\bibfnamefont {M.I.}\ \bibnamefont {Katsnelson}},\
  }\bibfield  {title} {\enquote {\bibinfo {title} {Decoherence and pointer
  states in small antiferromagnets: A benchmark test},}\ }\href@noop {}
  {\bibfield  {journal} {\bibinfo  {journal} {SciPost Phys.}\ }\textbf
  {\bibinfo {volume} {2}},\ \bibinfo {pages} {010} (\bibinfo {year}
  {2017})}\BibitemShut {NoStop}%
\bibitem [{\citenamefont {Zhao}\ \emph {et~al.}(2016)\citenamefont {Zhao},
  \citenamefont {{De Raedt}}, \citenamefont {Miyashita}, \citenamefont {Jin},\
  and\ \citenamefont {Michielsen}}]{ZHAO16}%
  \BibitemOpen
  \bibfield  {author} {\bibinfo {author} {\bibfnamefont {P.}~\bibnamefont
  {Zhao}}, \bibinfo {author} {\bibfnamefont {H.}~\bibnamefont {{De Raedt}}},
  \bibinfo {author} {\bibfnamefont {S.}~\bibnamefont {Miyashita}}, \bibinfo
  {author} {\bibfnamefont {F.}~\bibnamefont {Jin}}, \ and\ \bibinfo {author}
  {\bibfnamefont {K.}~\bibnamefont {Michielsen}},\ }\bibfield  {title}
  {\enquote {\bibinfo {title} {{Dynamics of open quantum spin systems: An
  assessment of the quantum master equation approach}},}\ }\href@noop {}
  {\bibfield  {journal} {\bibinfo  {journal} {Phys. Rev. E}\ }\textbf {\bibinfo
  {volume} {94}},\ \bibinfo {pages} {022126} (\bibinfo {year}
  {2016})}\BibitemShut {NoStop}%
\bibitem [{\citenamefont {Bloch}(1946)}]{BLOC46}%
  \BibitemOpen
  \bibfield  {author} {\bibinfo {author} {\bibfnamefont {F.}~\bibnamefont
  {Bloch}},\ }\bibfield  {title} {\enquote {\bibinfo {title} {Nuclear
  induction},}\ }\href@noop {} {\bibfield  {journal} {\bibinfo  {journal}
  {Phys. Rev.}\ }\textbf {\bibinfo {volume} {70}},\ \bibinfo {pages} {460--474}
  (\bibinfo {year} {1946})}\BibitemShut {NoStop}%
\bibitem [{\citenamefont {Kubo}(1957)}]{KUBO57}%
  \BibitemOpen
  \bibfield  {author} {\bibinfo {author} {\bibfnamefont {R.}~\bibnamefont
  {Kubo}},\ }\bibfield  {title} {\enquote {\bibinfo {title}
  {Statistical-mechanical theory of irreversible processes. {I.}}}\ }\href@noop
  {} {\bibfield  {journal} {\bibinfo  {journal} {J. Phys. Soc. Jpn.}\ }\textbf
  {\bibinfo {volume} {12}},\ \bibinfo {pages} {570--586} (\bibinfo {year}
  {1957})}\BibitemShut {NoStop}%
\bibitem [{\citenamefont {Abragam}(1961)}]{ABRA61}%
  \BibitemOpen
  \bibfield  {author} {\bibinfo {author} {\bibfnamefont {A.}~\bibnamefont
  {Abragam}},\ }\href@noop {} {\emph {\bibinfo {title} {Principles of Nuclear
  Magnetism}}}\ (\bibinfo  {publisher} {Oxford University Press},\ \bibinfo
  {address} {London},\ \bibinfo {year} {1961})\BibitemShut {NoStop}%
\bibitem [{\citenamefont {Slichter}(1990)}]{SLIC90}%
  \BibitemOpen
  \bibfield  {author} {\bibinfo {author} {\bibfnamefont {S.~P.}\ \bibnamefont
  {Slichter}},\ }\href@noop {} {\emph {\bibinfo {title} {Principles of Magnetic
  Resonance}}}\ (\bibinfo  {publisher} {Spinger},\ \bibinfo {address}
  {Berlin},\ \bibinfo {year} {1990})\BibitemShut {NoStop}%
\bibitem [{\citenamefont {Abragam}\ and\ \citenamefont
  {Bleeney}(1970)}]{ABRA70}%
  \BibitemOpen
  \bibfield  {author} {\bibinfo {author} {\bibfnamefont {A.}~\bibnamefont
  {Abragam}}\ and\ \bibinfo {author} {\bibfnamefont {B.}~\bibnamefont
  {Bleeney}},\ }\href@noop {} {\emph {\bibinfo {title} {Electron Paramagnetic
  Resonance of Transition Ions}}}\ (\bibinfo  {publisher} {Clarendon Press},\
  \bibinfo {address} {Oxford},\ \bibinfo {year} {1970})\BibitemShut {NoStop}%
\bibitem [{\citenamefont {Nielsen}\ and\ \citenamefont
  {Chuang}(2010)}]{NIEL10}%
  \BibitemOpen
  \bibfield  {author} {\bibinfo {author} {\bibfnamefont {M.}~\bibnamefont
  {Nielsen}}\ and\ \bibinfo {author} {\bibfnamefont {I.}~\bibnamefont
  {Chuang}},\ }\href@noop {} {\emph {\bibinfo {title} {Quantum Computation and
  Quantum Information}}},\ \bibinfo {edition} {10th}\ ed.\ (\bibinfo
  {publisher} {Cambridge University Press},\ \bibinfo {address} {Cambridge},\
  \bibinfo {year} {2010})\BibitemShut {NoStop}%
\bibitem [{\citenamefont {Johnson}\ \emph {et~al.}(2011)\citenamefont
  {Johnson}, \citenamefont {Amin}, \citenamefont {Gildert}, \citenamefont
  {Lanting}, \citenamefont {Hamze}, \citenamefont {Dickson}, \citenamefont
  {Harris}, \citenamefont {Berkley}, \citenamefont {Johansson}, \citenamefont
  {Bunyk}, \citenamefont {Chapple}, \citenamefont {Enderud}, \citenamefont
  {Hilton}, \citenamefont {Karimi}, \citenamefont {Ladizinsky}, \citenamefont
  {Ladizinsky}, \citenamefont {Oh}, \citenamefont {Perminov}, \citenamefont
  {Rich}, \citenamefont {Thom}, \citenamefont {Tolkacheva}, \citenamefont
  {Truncik}, \citenamefont {Uchaikin}, \citenamefont {Wang}, \citenamefont
  {Wilson},\ and\ \citenamefont {Rose}}]{JOHN11}%
  \BibitemOpen
  \bibfield  {author} {\bibinfo {author} {\bibfnamefont {M.~W.}\ \bibnamefont
  {Johnson}}, \bibinfo {author} {\bibfnamefont {M.~H.~S.}\ \bibnamefont
  {Amin}}, \bibinfo {author} {\bibfnamefont {S.}~\bibnamefont {Gildert}},
  \bibinfo {author} {\bibfnamefont {T.}~\bibnamefont {Lanting}}, \bibinfo
  {author} {\bibfnamefont {F.}~\bibnamefont {Hamze}}, \bibinfo {author}
  {\bibfnamefont {N.}~\bibnamefont {Dickson}}, \bibinfo {author} {\bibfnamefont
  {R.}~\bibnamefont {Harris}}, \bibinfo {author} {\bibfnamefont {A.~J.}\
  \bibnamefont {Berkley}}, \bibinfo {author} {\bibfnamefont {J.}~\bibnamefont
  {Johansson}}, \bibinfo {author} {\bibfnamefont {P.}~\bibnamefont {Bunyk}},
  \bibinfo {author} {\bibfnamefont {E.~M.}\ \bibnamefont {Chapple}}, \bibinfo
  {author} {\bibfnamefont {C.}~\bibnamefont {Enderud}}, \bibinfo {author}
  {\bibfnamefont {J.~P.}\ \bibnamefont {Hilton}}, \bibinfo {author}
  {\bibfnamefont {K.}~\bibnamefont {Karimi}}, \bibinfo {author} {\bibfnamefont
  {E.}~\bibnamefont {Ladizinsky}}, \bibinfo {author} {\bibfnamefont
  {N.}~\bibnamefont {Ladizinsky}}, \bibinfo {author} {\bibfnamefont
  {T.}~\bibnamefont {Oh}}, \bibinfo {author} {\bibfnamefont {I.}~\bibnamefont
  {Perminov}}, \bibinfo {author} {\bibfnamefont {C.}~\bibnamefont {Rich}},
  \bibinfo {author} {\bibfnamefont {M.~C.}\ \bibnamefont {Thom}}, \bibinfo
  {author} {\bibfnamefont {E.}~\bibnamefont {Tolkacheva}}, \bibinfo {author}
  {\bibfnamefont {C.~J.~S.}\ \bibnamefont {Truncik}}, \bibinfo {author}
  {\bibfnamefont {S.}~\bibnamefont {Uchaikin}}, \bibinfo {author}
  {\bibfnamefont {J.}~\bibnamefont {Wang}}, \bibinfo {author} {\bibfnamefont
  {B.}~\bibnamefont {Wilson}}, \ and\ \bibinfo {author} {\bibfnamefont
  {G.}~\bibnamefont {Rose}},\ }\bibfield  {title} {\enquote {\bibinfo {title}
  {Quantum annealing with manufactured spins},}\ }\href@noop {} {\bibfield
  {journal} {\bibinfo  {journal} {Nature}\ }\textbf {\bibinfo {volume} {473}},\
  \bibinfo {pages} {194--198} (\bibinfo {year} {2011})}\BibitemShut {NoStop}%
\bibitem [{\citenamefont {{IBM}}(2016)}]{IBMQE}%
  \BibitemOpen
  \bibfield  {author} {\bibinfo {author} {\bibnamefont {{IBM}}},\ }\href@noop
  {} {\enquote {\bibinfo {title} {The quantum experience},}\ } (\bibinfo {year}
  {2016}),\ \bibinfo {note}
  {\url{http://www.research.ibm.com/quantum/}}\BibitemShut {NoStop}%
\bibitem [{\citenamefont {Bethe}(1931)}]{BETH31}%
  \BibitemOpen
  \bibfield  {author} {\bibinfo {author} {\bibfnamefont {H.~A.}\ \bibnamefont
  {Bethe}},\ }\bibfield  {title} {\enquote {\bibinfo {title} {{Zur Theorie der
  metalle: I. Eigenwerte und Eigenfunktionen der linearen Atomketter}},}\
  }\href@noop {} {\bibfield  {journal} {\bibinfo  {journal} {Z. Phys.}\
  }\textbf {\bibinfo {volume} {71}},\ \bibinfo {pages} {205 -- 226} (\bibinfo
  {year} {1931})}\BibitemShut {NoStop}%
\bibitem [{\citenamefont {Hulthen}(1938)}]{HULT38}%
  \BibitemOpen
  \bibfield  {author} {\bibinfo {author} {\bibfnamefont {L.}~\bibnamefont
  {Hulthen}},\ }\bibfield  {title} {\enquote {\bibinfo {title} {{\"Uber das
  austauschproblem eines Kristalles}},}\ }\href@noop {} {\bibfield  {journal}
  {\bibinfo  {journal} {Arkiv Math. Astron. Fysik}\ }\textbf {\bibinfo {volume}
  {26A}},\ \bibinfo {pages} {1 -- 106} (\bibinfo {year} {1938})}\BibitemShut
  {NoStop}%
\bibitem [{\citenamefont {Gaudin}(1983)}]{GAUD83}%
  \BibitemOpen
  \bibfield  {author} {\bibinfo {author} {\bibfnamefont {M.}~\bibnamefont
  {Gaudin}},\ }\href@noop {} {\emph {\bibinfo {title} {{La Fonction D'onde De
  Bethe}}}}\ (\bibinfo  {publisher} {Masson},\ \bibinfo {address} {Paris},\
  \bibinfo {year} {1983})\BibitemShut {NoStop}%
\bibitem [{\citenamefont {Lages}\ \emph {et~al.}(2005)\citenamefont {Lages},
  \citenamefont {Dobrovitski}, \citenamefont {Katsnelson}, \citenamefont
  {De~Raedt},\ and\ \citenamefont {Harmon}}]{LAGE05}%
  \BibitemOpen
  \bibfield  {author} {\bibinfo {author} {\bibfnamefont {J.}~\bibnamefont
  {Lages}}, \bibinfo {author} {\bibfnamefont {V.~V.}\ \bibnamefont
  {Dobrovitski}}, \bibinfo {author} {\bibfnamefont {M.~I.}\ \bibnamefont
  {Katsnelson}}, \bibinfo {author} {\bibfnamefont {H.~A.}\ \bibnamefont
  {De~Raedt}}, \ and\ \bibinfo {author} {\bibfnamefont {B.~N.}\ \bibnamefont
  {Harmon}},\ }\bibfield  {title} {\enquote {\bibinfo {title} {Decoherence by a
  chaotic many-spin bath},}\ }\href@noop {} {\bibfield  {journal} {\bibinfo
  {journal} {Phys. Rev. E}\ }\textbf {\bibinfo {volume} {72}},\ \bibinfo
  {pages} {026225} (\bibinfo {year} {2005})}\BibitemShut {NoStop}%
\bibitem [{\citenamefont {Yuan}\ \emph {et~al.}(2009)\citenamefont {Yuan},
  \citenamefont {Katsnelson},\ and\ \citenamefont {{De Raedt}}}]{YUAN09}%
  \BibitemOpen
  \bibfield  {author} {\bibinfo {author} {\bibfnamefont {S.}~\bibnamefont
  {Yuan}}, \bibinfo {author} {\bibfnamefont {M.I.}\ \bibnamefont {Katsnelson}},
  \ and\ \bibinfo {author} {\bibfnamefont {H.}~\bibnamefont {{De Raedt}}},\
  }\bibfield  {title} {\enquote {\bibinfo {title} {Origin of the canonical
  ensemble: Thermalization with decoherence},}\ }\href@noop {} {\bibfield
  {journal} {\bibinfo  {journal} {J. Phys. Soc. Jpn.}\ }\textbf {\bibinfo
  {volume} {78}},\ \bibinfo {pages} {094003} (\bibinfo {year}
  {2009})}\BibitemShut {NoStop}%
\bibitem [{\citenamefont {Yuan}\ \emph {et~al.}(2006)\citenamefont {Yuan},
  \citenamefont {Katsnelson},\ and\ \citenamefont {{De Raedt}}}]{YUAN06}%
  \BibitemOpen
  \bibfield  {author} {\bibinfo {author} {\bibfnamefont {S.}~\bibnamefont
  {Yuan}}, \bibinfo {author} {\bibfnamefont {M.I.}\ \bibnamefont {Katsnelson}},
  \ and\ \bibinfo {author} {\bibfnamefont {H.}~\bibnamefont {{De Raedt}}},\
  }\bibfield  {title} {\enquote {\bibinfo {title} {Giant enhancement of quantum
  decoherence by frustrated environments},}\ }\href@noop {} {\bibfield
  {journal} {\bibinfo  {journal} {JETP Lett.}\ }\textbf {\bibinfo {volume}
  {84}},\ \bibinfo {pages} {99} (\bibinfo {year} {2006})}\BibitemShut {NoStop}%
\bibitem [{\citenamefont {Yuan}\ \emph {et~al.}(2007)\citenamefont {Yuan},
  \citenamefont {Katsnelson},\ and\ \citenamefont {{De Raedt}}}]{YUAN07}%
  \BibitemOpen
  \bibfield  {author} {\bibinfo {author} {\bibfnamefont {S.}~\bibnamefont
  {Yuan}}, \bibinfo {author} {\bibfnamefont {M.I.}\ \bibnamefont {Katsnelson}},
  \ and\ \bibinfo {author} {\bibfnamefont {H.}~\bibnamefont {{De Raedt}}},\
  }\bibfield  {title} {\enquote {\bibinfo {title} {Evolution of a quantum spin
  system to its ground state: Role of entanglement and interaction symmetry},}\
  }\href@noop {} {\bibfield  {journal} {\bibinfo  {journal} {Phys. Rev. A}\
  }\textbf {\bibinfo {volume} {75}},\ \bibinfo {pages} {052109} (\bibinfo
  {year} {2007})}\BibitemShut {NoStop}%
\bibitem [{\citenamefont {Yuan}\ \emph {et~al.}(2008)\citenamefont {Yuan},
  \citenamefont {Katsnelson},\ and\ \citenamefont {{De Raedt}}}]{YUAN08}%
  \BibitemOpen
  \bibfield  {author} {\bibinfo {author} {\bibfnamefont {S.}~\bibnamefont
  {Yuan}}, \bibinfo {author} {\bibfnamefont {M.I.}\ \bibnamefont {Katsnelson}},
  \ and\ \bibinfo {author} {\bibfnamefont {H.}~\bibnamefont {{De Raedt}}},\
  }\bibfield  {title} {\enquote {\bibinfo {title} {Decoherence by a spin
  thermal bath: Role of spin-spin interactions and initial state of the
  bath},}\ }\href@noop {} {\bibfield  {journal} {\bibinfo  {journal} {Phys.
  Rev. B}\ }\textbf {\bibinfo {volume} {77}},\ \bibinfo {pages} {184301}
  (\bibinfo {year} {2008})}\BibitemShut {NoStop}%
\bibitem [{\citenamefont {{Jin}}\ \emph
  {et~al.}(2013{\natexlab{a}})\citenamefont {{Jin}}, \citenamefont
  {{Michielsen}}, \citenamefont {Novotny}, \citenamefont {{Miyashita}},
  \citenamefont {{Yuan}},\ and\ \citenamefont {{De Raedt}}}]{JIN13a}%
  \BibitemOpen
  \bibfield  {author} {\bibinfo {author} {\bibfnamefont {F.}~\bibnamefont
  {{Jin}}}, \bibinfo {author} {\bibfnamefont {K.}~\bibnamefont {{Michielsen}}},
  \bibinfo {author} {\bibfnamefont {M.~A.}\ \bibnamefont {Novotny}}, \bibinfo
  {author} {\bibfnamefont {S.}~\bibnamefont {{Miyashita}}}, \bibinfo {author}
  {\bibfnamefont {S.}~\bibnamefont {{Yuan}}}, \ and\ \bibinfo {author}
  {\bibfnamefont {H.}~\bibnamefont {{De Raedt}}},\ }\bibfield  {title}
  {\enquote {\bibinfo {title} {{Quantum decoherence scaling with bath size:
  Importance of dynamics, connectivity, and randomness}},}\ }\href@noop {}
  {\bibfield  {journal} {\bibinfo  {journal} {Phys. Rev. A}\ }\textbf {\bibinfo
  {volume} {87}},\ \bibinfo {pages} {022117} (\bibinfo {year}
  {2013}{\natexlab{a}})}\BibitemShut {NoStop}%
\bibitem [{\citenamefont {Novotny}\ \emph {et~al.}(2016)\citenamefont
  {Novotny}, \citenamefont {Jin}, \citenamefont {Yuan}, \citenamefont
  {Miyashita}, \citenamefont {{De Raedt}},\ and\ \citenamefont
  {Michielsen}}]{NOVO16}%
  \BibitemOpen
  \bibfield  {author} {\bibinfo {author} {\bibfnamefont {M.~A.}\ \bibnamefont
  {Novotny}}, \bibinfo {author} {\bibfnamefont {F.}~\bibnamefont {Jin}},
  \bibinfo {author} {\bibfnamefont {S.}~\bibnamefont {Yuan}}, \bibinfo {author}
  {\bibfnamefont {S.}~\bibnamefont {Miyashita}}, \bibinfo {author}
  {\bibfnamefont {H.}~\bibnamefont {{De Raedt}}}, \ and\ \bibinfo {author}
  {\bibfnamefont {K.}~\bibnamefont {Michielsen}},\ }\bibfield  {title}
  {\enquote {\bibinfo {title} {Quantum decoherence and thermalization at finite
  temperatures within the canonical-thermal-state ensemble},}\ }\href@noop {}
  {\bibfield  {journal} {\bibinfo  {journal} {Phys. Rev. A}\ }\textbf {\bibinfo
  {volume} {93}},\ \bibinfo {pages} {032110} (\bibinfo {year}
  {2016})}\BibitemShut {NoStop}%
\bibitem [{\citenamefont {{Tal-Ezer}}\ and\ \citenamefont
  {Kosloff}(1984)}]{TALE84}%
  \BibitemOpen
  \bibfield  {author} {\bibinfo {author} {\bibfnamefont {H.}~\bibnamefont
  {{Tal-Ezer}}}\ and\ \bibinfo {author} {\bibfnamefont {R.}~\bibnamefont
  {Kosloff}},\ }\href@noop {} {\bibfield  {journal} {\bibinfo  {journal} {J.
  Chem. Phys.}\ }\textbf {\bibinfo {volume} {81}},\ \bibinfo {pages}
  {3967--3971} (\bibinfo {year} {1984})}\BibitemShut {NoStop}%
\bibitem [{\citenamefont {Leforestier}\ \emph {et~al.}(1991)\citenamefont
  {Leforestier}, \citenamefont {Bisseling}, \citenamefont {Cerjan},
  \citenamefont {Feit}, \citenamefont {Friesner}, \citenamefont {Guldberg},
  \citenamefont {Hammerich}, \citenamefont {Jolicard}, \citenamefont
  {Karrlein}, \citenamefont {Meyer}, \citenamefont {Lipkin}, \citenamefont
  {Roncero},\ and\ \citenamefont {Kosloff}}]{LEFO91}%
  \BibitemOpen
  \bibfield  {author} {\bibinfo {author} {\bibfnamefont {C.}~\bibnamefont
  {Leforestier}}, \bibinfo {author} {\bibfnamefont {R.~H.}\ \bibnamefont
  {Bisseling}}, \bibinfo {author} {\bibfnamefont {C.}~\bibnamefont {Cerjan}},
  \bibinfo {author} {\bibfnamefont {M.~D.}\ \bibnamefont {Feit}}, \bibinfo
  {author} {\bibfnamefont {R.}~\bibnamefont {Friesner}}, \bibinfo {author}
  {\bibfnamefont {A.}~\bibnamefont {Guldberg}}, \bibinfo {author}
  {\bibfnamefont {A.}~\bibnamefont {Hammerich}}, \bibinfo {author}
  {\bibfnamefont {G.}~\bibnamefont {Jolicard}}, \bibinfo {author}
  {\bibfnamefont {W.}~\bibnamefont {Karrlein}}, \bibinfo {author}
  {\bibfnamefont {H.-D.}\ \bibnamefont {Meyer}}, \bibinfo {author}
  {\bibfnamefont {N.}~\bibnamefont {Lipkin}}, \bibinfo {author} {\bibfnamefont
  {O.}~\bibnamefont {Roncero}}, \ and\ \bibinfo {author} {\bibfnamefont
  {R.}~\bibnamefont {Kosloff}},\ }\bibfield  {title} {\enquote {\bibinfo
  {title} {A comparison of different propagation schemes for the time-dependent
  {Schr\"odinger} equation},}\ }\href@noop {} {\bibfield  {journal} {\bibinfo
  {journal} {J. Comput. Phys.}\ }\textbf {\bibinfo {volume} {94}},\ \bibinfo
  {pages} {59--80} (\bibinfo {year} {1991})}\BibitemShut {NoStop}%
\bibitem [{\citenamefont {Iitaka}\ \emph
  {et~al.}(1997{\natexlab{a}})\citenamefont {Iitaka}, \citenamefont {Nomura},
  \citenamefont {Hirayama}, \citenamefont {Zhao}, \citenamefont {Aoyagi},\ and\
  \citenamefont {Sugano}}]{IITA97}%
  \BibitemOpen
  \bibfield  {author} {\bibinfo {author} {\bibfnamefont {T.}~\bibnamefont
  {Iitaka}}, \bibinfo {author} {\bibfnamefont {S.}~\bibnamefont {Nomura}},
  \bibinfo {author} {\bibfnamefont {H.}~\bibnamefont {Hirayama}}, \bibinfo
  {author} {\bibfnamefont {X.}~\bibnamefont {Zhao}}, \bibinfo {author}
  {\bibfnamefont {Y.}~\bibnamefont {Aoyagi}}, \ and\ \bibinfo {author}
  {\bibfnamefont {T.}~\bibnamefont {Sugano}},\ }\bibfield  {title} {\enquote
  {\bibinfo {title} {Calculating the linear response functions of
  noninteracting electrons with a time-dependent {Schr\"odinger} equation},}\
  }\href@noop {} {\bibfield  {journal} {\bibinfo  {journal} {Phys. Rev. E}\
  }\textbf {\bibinfo {volume} {56}},\ \bibinfo {pages} {1222--1229} (\bibinfo
  {year} {1997}{\natexlab{a}})}\BibitemShut {NoStop}%
\bibitem [{\citenamefont {Dobrovitski}\ and\ \citenamefont {{De
  Raedt}}(2003)}]{DOBR03}%
  \BibitemOpen
  \bibfield  {author} {\bibinfo {author} {\bibfnamefont {V.~V.}\ \bibnamefont
  {Dobrovitski}}\ and\ \bibinfo {author} {\bibfnamefont {H.}~\bibnamefont {{De
  Raedt}}},\ }\bibfield  {title} {\enquote {\bibinfo {title} {Efficient scheme
  for numerical simulations of the spin-bath decoherence},}\ }\href@noop {}
  {\bibfield  {journal} {\bibinfo  {journal} {Phys. Rev. E}\ }\textbf {\bibinfo
  {volume} {67}},\ \bibinfo {pages} {056702} (\bibinfo {year}
  {2003})}\BibitemShut {NoStop}%
\bibitem [{\citenamefont {{De Raedt}}\ and\ \citenamefont
  {Michielsen}(2006)}]{RAED06}%
  \BibitemOpen
  \bibfield  {author} {\bibinfo {author} {\bibfnamefont {H.}~\bibnamefont {{De
  Raedt}}}\ and\ \bibinfo {author} {\bibfnamefont {K.}~\bibnamefont
  {Michielsen}},\ }\bibfield  {title} {\enquote {\bibinfo {title}
  {{Computational Methods for Simulating Quantum Computers}},}\ }in\ \href@noop
  {} {\emph {\bibinfo {booktitle} {Handbook of Theoretical and Computational
  Nanotechnology}}},\ \bibinfo {editor} {edited by\ \bibinfo {editor}
  {\bibfnamefont {M.}~\bibnamefont {Rieth}}\ and\ \bibinfo {editor}
  {\bibfnamefont {W.}~\bibnamefont {Schommers}}}\ (\bibinfo  {publisher}
  {American Scientific Publishers},\ \bibinfo {address} {Los Angeles},\
  \bibinfo {year} {2006})\ pp.\ \bibinfo {pages} {2 -- 48}\BibitemShut
  {NoStop}%
\bibitem [{\citenamefont {{De Raedt}}\ \emph {et~al.}(2007)\citenamefont {{De
  Raedt}}, \citenamefont {Michielsen}, \citenamefont {{De Raedt}},
  \citenamefont {Trieu}, \citenamefont {Arnold}, \citenamefont {Richter},
  \citenamefont {Lippert}, \citenamefont {Watanabe},\ and\ \citenamefont
  {Ito}}]{RAED07X}%
  \BibitemOpen
  \bibfield  {author} {\bibinfo {author} {\bibfnamefont {K.}~\bibnamefont {{De
  Raedt}}}, \bibinfo {author} {\bibfnamefont {K.}~\bibnamefont {Michielsen}},
  \bibinfo {author} {\bibfnamefont {H.}~\bibnamefont {{De Raedt}}}, \bibinfo
  {author} {\bibfnamefont {B.}~\bibnamefont {Trieu}}, \bibinfo {author}
  {\bibfnamefont {G.}~\bibnamefont {Arnold}}, \bibinfo {author} {\bibfnamefont
  {M.}~\bibnamefont {Richter}}, \bibinfo {author} {\bibfnamefont {Th.}\
  \bibnamefont {Lippert}}, \bibinfo {author} {\bibfnamefont {H.}~\bibnamefont
  {Watanabe}}, \ and\ \bibinfo {author} {\bibfnamefont {N.}~\bibnamefont
  {Ito}},\ }\bibfield  {title} {\enquote {\bibinfo {title} {Massively parallel
  quantum computer simulator},}\ }\href@noop {} {\bibfield  {journal} {\bibinfo
   {journal} {Comp. Phys. Comm.}\ }\textbf {\bibinfo {volume} {176}},\ \bibinfo
  {pages} {121 -- 136} (\bibinfo {year} {2007})}\BibitemShut {NoStop}%
\bibitem [{\citenamefont {von Neumann}(1955)}]{NEUM55}%
  \BibitemOpen
  \bibfield  {author} {\bibinfo {author} {\bibfnamefont {J.}~\bibnamefont {von
  Neumann}},\ }\href@noop {} {\emph {\bibinfo {title} {Mathematical Foundations
  of Quantum Mechanics}}}\ (\bibinfo  {publisher} {Princeton University
  Press},\ \bibinfo {address} {Princeton},\ \bibinfo {year} {1955})\BibitemShut
  {NoStop}%
\bibitem [{\citenamefont {Ballentine}(2003)}]{BALL03}%
  \BibitemOpen
  \bibfield  {author} {\bibinfo {author} {\bibfnamefont {L.~E.}\ \bibnamefont
  {Ballentine}},\ }\href@noop {} {\emph {\bibinfo {title} {{Quantum Mechanics:
  A Modern Development}}}}\ (\bibinfo  {publisher} {World Scientific},\
  \bibinfo {address} {Singapore},\ \bibinfo {year} {2003})\BibitemShut
  {NoStop}%
\bibitem [{\citenamefont {{Hams}}\ and\ \citenamefont {{De
  Raedt}}(2000)}]{HAMS00}%
  \BibitemOpen
  \bibfield  {author} {\bibinfo {author} {\bibfnamefont {A.}~\bibnamefont
  {{Hams}}}\ and\ \bibinfo {author} {\bibfnamefont {H.}~\bibnamefont {{De
  Raedt}}},\ }\bibfield  {title} {\enquote {\bibinfo {title} {{Fast algorithm
  for finding the eigenvalue distribution of very large matrices}},}\
  }\href@noop {} {\bibfield  {journal} {\bibinfo  {journal} {Phys. Rev. E}\
  }\textbf {\bibinfo {volume} {62}},\ \bibinfo {pages} {4365 -- 4377} (\bibinfo
  {year} {2000})}\BibitemShut {NoStop}%
\bibitem [{\citenamefont {Reimann}(2007)}]{REIM07}%
  \BibitemOpen
  \bibfield  {author} {\bibinfo {author} {\bibfnamefont {P.}~\bibnamefont
  {Reimann}},\ }\bibfield  {title} {\enquote {\bibinfo {title} {{Typicality for
  Generalized Microcanonical Ensembles}},}\ }\href@noop {} {\bibfield
  {journal} {\bibinfo  {journal} {Phys. Rev. Lett.}\ }\textbf {\bibinfo
  {volume} {99}},\ \bibinfo {pages} {160404} (\bibinfo {year}
  {2007})}\BibitemShut {NoStop}%
\bibitem [{\citenamefont {Bartsch}\ and\ \citenamefont
  {Gemmer}(2009)}]{BART09}%
  \BibitemOpen
  \bibfield  {author} {\bibinfo {author} {\bibfnamefont {C.}~\bibnamefont
  {Bartsch}}\ and\ \bibinfo {author} {\bibfnamefont {J.}~\bibnamefont
  {Gemmer}},\ }\bibfield  {title} {\enquote {\bibinfo {title} {Dynamical
  typicality of quantum expectation values},}\ }\href@noop {} {\bibfield
  {journal} {\bibinfo  {journal} {Phys. Rev. Lett}\ }\textbf {\bibinfo {volume}
  {102}},\ \bibinfo {pages} {110403} (\bibinfo {year} {2009})}\BibitemShut
  {NoStop}%
\bibitem [{\citenamefont {Sugiura}\ and\ \citenamefont
  {Shimizu}(2012)}]{SUGI12}%
  \BibitemOpen
  \bibfield  {author} {\bibinfo {author} {\bibfnamefont {S.}~\bibnamefont
  {Sugiura}}\ and\ \bibinfo {author} {\bibfnamefont {A.}~\bibnamefont
  {Shimizu}},\ }\bibfield  {title} {\enquote {\bibinfo {title} {Thermal pure
  quantum states at finite temperature},}\ }\href@noop {} {\bibfield  {journal}
  {\bibinfo  {journal} {Phys. Rev. Lett.}\ }\textbf {\bibinfo {volume} {108}},\
  \bibinfo {pages} {240401} (\bibinfo {year} {2012})}\BibitemShut {NoStop}%
\bibitem [{\citenamefont {Sugiura}\ and\ \citenamefont
  {Shimizu}(2013)}]{SUGI13}%
  \BibitemOpen
  \bibfield  {author} {\bibinfo {author} {\bibfnamefont {S.}~\bibnamefont
  {Sugiura}}\ and\ \bibinfo {author} {\bibfnamefont {A.}~\bibnamefont
  {Shimizu}},\ }\bibfield  {title} {\enquote {\bibinfo {title} {Canonical
  thermal pure quantum state},}\ }\href@noop {} {\bibfield  {journal} {\bibinfo
   {journal} {Phys. Rev. Lett.}\ }\textbf {\bibinfo {volume} {111}},\ \bibinfo
  {pages} {010401} (\bibinfo {year} {2013})}\BibitemShut {NoStop}%
\bibitem [{\citenamefont {Steinigeweg}\ \emph {et~al.}(2014)\citenamefont
  {Steinigeweg}, \citenamefont {Gemmer},\ and\ \citenamefont
  {Brenig}}]{STEI14b}%
  \BibitemOpen
  \bibfield  {author} {\bibinfo {author} {\bibfnamefont {Robin}\ \bibnamefont
  {Steinigeweg}}, \bibinfo {author} {\bibfnamefont {Jochen}\ \bibnamefont
  {Gemmer}}, \ and\ \bibinfo {author} {\bibfnamefont {Wolfram}\ \bibnamefont
  {Brenig}},\ }\bibfield  {title} {\enquote {\bibinfo {title} {Spin-current
  autocorrelations from single pure-state propagation},}\ }\href@noop {}
  {\bibfield  {journal} {\bibinfo  {journal} {Phys. Rev. Lett.}\ }\textbf
  {\bibinfo {volume} {112}},\ \bibinfo {pages} {120601} (\bibinfo {year}
  {2014})}\BibitemShut {NoStop}%
\bibitem [{\citenamefont {Bogolyubov}(1970)}]{BOGU70}%
  \BibitemOpen
  \bibfield  {author} {\bibinfo {author} {\bibfnamefont {N.~N.}\ \bibnamefont
  {Bogolyubov}},\ }\enquote {\bibinfo {title} {{N. N. Bogolyubov, Selected
  Works Vol.2}},}\ \ (\bibinfo  {publisher} {in Russian, Naukova Dumka},\
  \bibinfo {address} {Kiev},\ \bibinfo {year} {1970})\ Chap.\ \bibinfo
  {chapter} {Elementary example of establishment of statistical equilibrium in
  a system coupled to a thermostat}, pp.\ \bibinfo {pages} {77 --
  98}\BibitemShut {NoStop}%
\bibitem [{\citenamefont {(jr)}\ and\ \citenamefont
  {Sankovich}(1994)}]{BOGU94}%
  \BibitemOpen
  \bibfield  {author} {\bibinfo {author} {\bibfnamefont {N.N.~Bogolyubov}\
  \bibnamefont {(jr)}}\ and\ \bibinfo {author} {\bibfnamefont {D.P.}\
  \bibnamefont {Sankovich}},\ }\bibfield  {title} {\enquote {\bibinfo {title}
  {{N.N. Bogolyubov} and statistical mechanics},}\ }\href@noop {} {\bibfield
  {journal} {\bibinfo  {journal} {Russian Math. Surveys}\ }\textbf {\bibinfo
  {volume} {49}},\ \bibinfo {pages} {19 -- 49} (\bibinfo {year}
  {1994})}\BibitemShut {NoStop}%
\bibitem [{\citenamefont {Chirikov}(1986)}]{CHIR86}%
  \BibitemOpen
  \bibfield  {author} {\bibinfo {author} {\bibfnamefont {B.V.}\ \bibnamefont
  {Chirikov}},\ }\bibfield  {title} {\enquote {\bibinfo {title} {Transient
  chaos in quantum and classical mechanics},}\ }\href@noop {} {\bibfield
  {journal} {\bibinfo  {journal} {Found. Phys.}\ }\textbf {\bibinfo {volume}
  {16}},\ \bibinfo {pages} {39 -- 47} (\bibinfo {year} {1986})}\BibitemShut
  {NoStop}%
\bibitem [{\citenamefont {Strokov}(2007)}]{STRO07}%
  \BibitemOpen
  \bibfield  {author} {\bibinfo {author} {\bibfnamefont {V.}~\bibnamefont
  {Strokov}},\ }\bibfield  {title} {\enquote {\bibinfo {title} {{On convergence
  to equilibrium in strongly coupled {Bogoliubov's} oscillator model}},}\
  }\href@noop {} {\bibfield  {journal} {\bibinfo  {journal} {Infinite
  Dimensional Analysis, Quantum Probability and Related Topics}\ }\textbf
  {\bibinfo {volume} {10}},\ \bibinfo {pages} {573 -- 589} (\bibinfo {year}
  {2007})}\BibitemShut {NoStop}%
\bibitem [{\citenamefont {{Jin}}\ \emph
  {et~al.}(2013{\natexlab{b}})\citenamefont {{Jin}}, \citenamefont {Neuhaus},
  \citenamefont {{Michielsen}}, \citenamefont {Novotny}, \citenamefont
  {{Miyashita}}, \citenamefont {Katsnelson},\ and\ \citenamefont {{De
  Raedt}}}]{JIN13b}%
  \BibitemOpen
  \bibfield  {author} {\bibinfo {author} {\bibfnamefont {F.}~\bibnamefont
  {{Jin}}}, \bibinfo {author} {\bibfnamefont {T.}~\bibnamefont {Neuhaus}},
  \bibinfo {author} {\bibfnamefont {K.}~\bibnamefont {{Michielsen}}}, \bibinfo
  {author} {\bibfnamefont {M.~A.}\ \bibnamefont {Novotny}}, \bibinfo {author}
  {\bibfnamefont {S.}~\bibnamefont {{Miyashita}}}, \bibinfo {author}
  {\bibfnamefont {M.I.}\ \bibnamefont {Katsnelson}}, \ and\ \bibinfo {author}
  {\bibfnamefont {H.}~\bibnamefont {{De Raedt}}},\ }\bibfield  {title}
  {\enquote {\bibinfo {title} {Equilibration and thermalization of classical
  systems},}\ }\href@noop {} {\bibfield  {journal} {\bibinfo  {journal} {New J.
  Phys.}\ }\textbf {\bibinfo {volume} {15}},\ \bibinfo {pages} {033009}
  (\bibinfo {year} {2013}{\natexlab{b}})}\BibitemShut {NoStop}%
\bibitem [{\citenamefont {Gemmer}\ and\ \citenamefont {Mahler}(2003)}]{GEMM03}%
  \BibitemOpen
  \bibfield  {author} {\bibinfo {author} {\bibfnamefont {J.}~\bibnamefont
  {Gemmer}}\ and\ \bibinfo {author} {\bibfnamefont {G.}~\bibnamefont
  {Mahler}},\ }\bibfield  {title} {\enquote {\bibinfo {title} {Finite quantum
  environments as thermostats: an analysis based on the {Hilbert} space average
  method},}\ }\href@noop {} {\bibfield  {journal} {\bibinfo  {journal} {Eur.
  Phys. J. B}\ }\textbf {\bibinfo {volume} {31}},\ \bibinfo {pages} {249 --
  257} (\bibinfo {year} {2003})}\BibitemShut {NoStop}%
\bibitem [{\citenamefont {Goldstein}\ \emph {et~al.}(2006)\citenamefont
  {Goldstein}, \citenamefont {Lebowitz}, \citenamefont {Tumulka},\ and\
  \citenamefont {Zangh{\`{i}}}}]{GOLD06}%
  \BibitemOpen
  \bibfield  {author} {\bibinfo {author} {\bibfnamefont {S.}~\bibnamefont
  {Goldstein}}, \bibinfo {author} {\bibfnamefont {J.~L.}\ \bibnamefont
  {Lebowitz}}, \bibinfo {author} {\bibfnamefont {R.}~\bibnamefont {Tumulka}}, \
  and\ \bibinfo {author} {\bibfnamefont {N.}~\bibnamefont {Zangh{\`{i}}}},\
  }\bibfield  {title} {\enquote {\bibinfo {title} {{Canonical typicality}},}\
  }\href@noop {} {\bibfield  {journal} {\bibinfo  {journal} {Phys. Rev. Lett.}\
  }\textbf {\bibinfo {volume} {96}},\ \bibinfo {pages} {050403} (\bibinfo
  {year} {2006})}\BibitemShut {NoStop}%
\bibitem [{\citenamefont {Popescu}\ \emph {et~al.}(2006)\citenamefont
  {Popescu}, \citenamefont {Short},\ and\ \citenamefont {Winter}}]{POPE06}%
  \BibitemOpen
  \bibfield  {author} {\bibinfo {author} {\bibfnamefont {S.}~\bibnamefont
  {Popescu}}, \bibinfo {author} {\bibfnamefont {A.~J.}\ \bibnamefont {Short}},
  \ and\ \bibinfo {author} {\bibfnamefont {A.}~\bibnamefont {Winter}},\
  }\bibfield  {title} {\enquote {\bibinfo {title} {{Entanglement and the
  foundations of statistical mechanics}},}\ }\href@noop {} {\bibfield
  {journal} {\bibinfo  {journal} {Nature Phys.}\ }\textbf {\bibinfo {volume}
  {2}},\ \bibinfo {pages} {754 -- 758} (\bibinfo {year} {2006})}\BibitemShut
  {NoStop}%
\bibitem [{\citenamefont {Iitaka}\ \emph
  {et~al.}(1997{\natexlab{b}})\citenamefont {Iitaka}, \citenamefont {Nomura},
  \citenamefont {Hirayama}, \citenamefont {Zhao},\ and\ \citenamefont
  {Aoyagi}}]{IITA97b}%
  \BibitemOpen
  \bibfield  {author} {\bibinfo {author} {\bibfnamefont {T.}~\bibnamefont
  {Iitaka}}, \bibinfo {author} {\bibfnamefont {S.}~\bibnamefont {Nomura}},
  \bibinfo {author} {\bibfnamefont {H.}~\bibnamefont {Hirayama}}, \bibinfo
  {author} {\bibfnamefont {X.}~\bibnamefont {Zhao}}, \ and\ \bibinfo {author}
  {\bibfnamefont {Y.}~\bibnamefont {Aoyagi}},\ }\bibfield  {title} {\enquote
  {\bibinfo {title} {Linear scaling calculation for optical-absorption spectra
  of large hydrogenated silicon nanocrystallites},}\ }\href@noop {} {\bibfield
  {journal} {\bibinfo  {journal} {Phys. Rev. B}\ }\textbf {\bibinfo {volume}
  {56}},\ \bibinfo {pages} {R4348--R4350} (\bibinfo {year}
  {1997}{\natexlab{b}})}\BibitemShut {NoStop}%
\bibitem [{\citenamefont {Gelman}\ and\ \citenamefont
  {Kosloff}(2003)}]{GELM03}%
  \BibitemOpen
  \bibfield  {author} {\bibinfo {author} {\bibfnamefont {D.}~\bibnamefont
  {Gelman}}\ and\ \bibinfo {author} {\bibfnamefont {R.}~\bibnamefont
  {Kosloff}},\ }\bibfield  {title} {\enquote {\bibinfo {title} {Simulating
  dissipative phenomena with a random thermal phase wavefunctions, high
  temperature application of the surrogate {Hamiltonian} approach},}\
  }\href@noop {} {\bibfield  {journal} {\bibinfo  {journal} {Chem. Phys.
  Lett.}\ }\textbf {\bibinfo {volume} {381}},\ \bibinfo {pages} {129--138}
  (\bibinfo {year} {2003})}\BibitemShut {NoStop}%
\bibitem [{\citenamefont {Gelman}\ \emph {et~al.}(2004)\citenamefont {Gelman},
  \citenamefont {Koch},\ and\ \citenamefont {Kosloff}}]{GELM04}%
  \BibitemOpen
  \bibfield  {author} {\bibinfo {author} {\bibfnamefont {D.}~\bibnamefont
  {Gelman}}, \bibinfo {author} {\bibfnamefont {C.P.}\ \bibnamefont {Koch}}, \
  and\ \bibinfo {author} {\bibfnamefont {R.}~\bibnamefont {Kosloff}},\
  }\bibfield  {title} {\enquote {\bibinfo {title} {Dissipative quantum dynamics
  with the surrogate {Hamiltonian} approach. {A} comparison between spin and
  harmonic baths},}\ }\href@noop {} {\bibfield  {journal} {\bibinfo  {journal}
  {J. Chem. Phys.}\ }\textbf {\bibinfo {volume} {121}},\ \bibinfo {pages}
  {661--671} (\bibinfo {year} {2004})}\BibitemShut {NoStop}%
\bibitem [{\citenamefont {Reimann}(2015)}]{REIM15}%
  \BibitemOpen
  \bibfield  {author} {\bibinfo {author} {\bibfnamefont {P.}~\bibnamefont
  {Reimann}},\ }\bibfield  {title} {\enquote {\bibinfo {title} {Typical fast
  thermalization processes in closed many-body systems},}\ }\href@noop {}
  {\bibfield  {journal} {\bibinfo  {journal} {Nat. Comm.}\ }\textbf {\bibinfo
  {volume} {7}},\ \bibinfo {pages} {10821} (\bibinfo {year}
  {2015})}\BibitemShut {NoStop}%
\bibitem [{\citenamefont {Goldstein}\ \emph {et~al.}(2015)\citenamefont
  {Goldstein}, \citenamefont {Hara},\ and\ \citenamefont {Tasaki}}]{GOLD15}%
  \BibitemOpen
  \bibfield  {author} {\bibinfo {author} {\bibfnamefont {S.}~\bibnamefont
  {Goldstein}}, \bibinfo {author} {\bibfnamefont {T.}~\bibnamefont {Hara}}, \
  and\ \bibinfo {author} {\bibfnamefont {H.}~\bibnamefont {Tasaki}},\
  }\bibfield  {title} {\enquote {\bibinfo {title} {Extremely quick
  thermalization in a macroscopic quantum system for a typical nonequilibrium
  subspace.}}\ }\href@noop {} {\bibfield  {journal} {\bibinfo  {journal} {New.
  J. Phys.}\ }\textbf {\bibinfo {volume} {17}},\ \bibinfo {pages} {045002}
  (\bibinfo {year} {2015})}\BibitemShut {NoStop}%
\bibitem [{\citenamefont {Su{\'a}rez}\ \emph {et~al.}(1992)\citenamefont
  {Su{\'a}rez}, \citenamefont {Silbey},\ and\ \citenamefont
  {Oppenheim}}]{SUAR92}%
  \BibitemOpen
  \bibfield  {author} {\bibinfo {author} {\bibfnamefont {A.}~\bibnamefont
  {Su{\'a}rez}}, \bibinfo {author} {\bibfnamefont {R.}~\bibnamefont {Silbey}},
  \ and\ \bibinfo {author} {\bibfnamefont {I.}~\bibnamefont {Oppenheim}},\
  }\bibfield  {title} {\enquote {\bibinfo {title} {Memory effects in the
  relaxation of quantum systems},}\ }\href@noop {} {\bibfield  {journal}
  {\bibinfo  {journal} {J. Chem. Phys.}\ }\textbf {\bibinfo {volume} {97}},\
  \bibinfo {pages} {5101 -- 5107} (\bibinfo {year} {1992})}\BibitemShut
  {NoStop}%
\bibitem [{\citenamefont {Pechukas}(1994)}]{PECH94}%
  \BibitemOpen
  \bibfield  {author} {\bibinfo {author} {\bibfnamefont {P.}~\bibnamefont
  {Pechukas}},\ }\bibfield  {title} {\enquote {\bibinfo {title} {Reduced
  dynamics need not be completely positive},}\ }\href@noop {} {\bibfield
  {journal} {\bibinfo  {journal} {Phys. Rev. Lett.}\ }\textbf {\bibinfo
  {volume} {73}},\ \bibinfo {pages} {1060--1062} (\bibinfo {year}
  {1994})}\BibitemShut {NoStop}%
\bibitem [{\citenamefont {Gaspard}\ and\ \citenamefont
  {Nagaoka}(1999)}]{GASP99}%
  \BibitemOpen
  \bibfield  {author} {\bibinfo {author} {\bibfnamefont {P.}~\bibnamefont
  {Gaspard}}\ and\ \bibinfo {author} {\bibfnamefont {M.}~\bibnamefont
  {Nagaoka}},\ }\bibfield  {title} {\enquote {\bibinfo {title} {Slippage of
  initial conditions for the {Redfield} master equation},}\ }\href@noop {}
  {\bibfield  {journal} {\bibinfo  {journal} {J. Chem. Phys.}\ }\textbf
  {\bibinfo {volume} {111}},\ \bibinfo {pages} {5668 -- 5675} (\bibinfo {year}
  {1999})}\BibitemShut {NoStop}%
\bibitem [{\citenamefont {Stephan}\ and\ \citenamefont
  {Docter}(2015)}]{JUQUEEN}%
  \BibitemOpen
  \bibfield  {author} {\bibinfo {author} {\bibfnamefont {M.}~\bibnamefont
  {Stephan}}\ and\ \bibinfo {author} {\bibfnamefont {J.}~\bibnamefont
  {Docter}},\ }\bibfield  {title} {\enquote {\bibinfo {title} {{JUQUEEN: IBM
  Blue Gene/Q Supercomputer System at the J{\"u}lich Supercomputing Centre}},}\
  }\href@noop {} {\bibfield  {journal} {\bibinfo  {journal} {J. of Large-Scale
  Res. Facil.}\ }\textbf {\bibinfo {volume} {1}},\ \bibinfo {pages} {A1}
  (\bibinfo {year} {2015})}\BibitemShut {NoStop}%
\end{thebibliography}%
\end{document}